\documentclass[12pt]{article}
\usepackage{amsthm}
\usepackage{amsmath}
\usepackage{natbib}
\usepackage[colorlinks,citecolor=blue,urlcolor=blue,filecolor=blue,backref=page]{hyperref}
\usepackage{graphicx}
\RequirePackage[OT1]{fontenc}
\RequirePackage{amsthm,amsmath}
\RequirePackage{natbib}
\RequirePackage[colorlinks,citecolor=blue,urlcolor=blue]{hyperref}
\usepackage[table]{xcolor}
\RequirePackage[OT1]{fontenc}
\RequirePackage{amsthm,amsmath}
\RequirePackage{graphicx,bm, graphics}

\usepackage{subcaption}
\usepackage{amssymb, url}
\usepackage{subcaption}
\usepackage{tikz}
\usetikzlibrary{matrix}
\usetikzlibrary{arrows,fit,positioning}
\usepackage{lmodern} 
\usepackage[counterclockwise, figuresleft]{rotating}
\usepackage{booktabs}

\newcommand{\B}{\mathcal{B}}
\newcommand{\Y}{\mathbf{Y}}

\newcommand{\M}{\mathbf{M}}
\newcommand{\U}{\mathbf{U}}
\newcommand{\V}{\mathbf{V}}

\renewcommand{\u}{\mathbf{u}}
\renewcommand{\v}{\mathbf{v}}

\newcommand{\x}{\mathbf{x}}

\def\real{\mathbb{R}}

\renewcommand{\S}{\mathbf{S}}
\renewcommand{\P}{\mathbf{P}}

\renewcommand{\b}{\boldsymbol{\beta}}

\newcommand{\T}{\boldsymbol{\Theta}}

\newcommand{\normdist}[2]{\ensuremath{\mathcal{N}(#1,#2)}}

\newcommand{\gamdist}[2]{\ensuremath{\mathcal{G}(#1,#2)}}

\newcommand{\betadist}[2]{\ensuremath{\mathcal{B}eta(#1,#2)}}

\tikzset{
  latentnode/.style  ={draw,minimum width=2.5em, shape=circle,thick, black,fill=white},
  visiblenode/.style ={draw, minimum width=2.5em, shape=circle,thick, black,fill=black!20},
  plate/.style={draw,
    shape=rectangle,
    thick,
    minimum width=4em,
    minimum height=4em,
    align=right,
    inner sep=5em,
    inner ysep=5em,
    label={[xshift=-16pt,yshift=11pt]south east:#1}},
  line/.style={draw, -latex'},
  jump left/.style={draw=none, inner xsep=0pt},
  jump line/.style={line, shorten <=5pt}
}

\begin{document}


\title{Detecting Structural Changes in Longitudinal Network Data}


\author{Jong Hee Park \\
Seoul National University\\ 
\texttt{jongheepark@snu.ac.kr} \\
\and 
Yunkyu Sohn\\
Waseda University\\ 
\texttt{ysohn@aoni.waseda.jp}
}
\maketitle

\begin{abstract}
Dynamic modeling of longitudinal networks has been an increasingly important topic in applied research. While longitudinal network data commonly exhibit dramatic changes in its structures, existing methods have largely focused on modeling smooth topological changes over time. In this paper, we develop a hidden Markov multilinear tensor model (HMTM) that combines the multilinear tensor regression model \citep{Hoff2011} with a hidden Markov model using Bayesian inference. We model changes in network structure as shifts in discrete states yielding particular sets of network generating parameters. Our simulation results demonstrate that the proposed method correctly detects the number, locations, and types of changes in latent node characteristics. We apply the proposed method to international military alliance networks to find structural changes in the coalition structure among nations. 
\end{abstract}





\section{Introduction}
Tensor decomposition is becoming a standard means to analyze longitudinal network datasets \citep{hoff2009_cmot, Hoff2011, Rai2015, Hoff2015, Minhas2016, johndrow2017, Han2018MultiresolutionTD}. Compared to network models for static snapshots or matrix-valued datasets, this approach significantly advances our modeling possibility. A longitudinal network data set can be represented as a tensor $\mathcal{Y}=\{\Y_t\vert t\in\{1,\ldots,T\}\}\in\real^{N \times N \times T}$, which is an array of $N \times N$ square matrices $\Y_t=\{y_{ijt}\vert i,j\in\{1,\ldots,N\}\}$. Here $y_{ijt}$ informs the dyadic relationship between actors $i$ and $j$ at time $t$. While dynamic modeling of tensor-valued longitudinal networks, mainly in a form of reduced-rank decomposition, has been an increasingly important topic in social, biological, and other fields of science, a fully probabilistic treatment of dynamic network process has been a challenging problem due to simultaneous dependence between dyadic and temporal observations that are often associated with fundamental shifts in data generating processes.\footnote{For a range of examples for longitudinal network analysis, see \cite{holme2012temporal} and references therein.}

\begin{figure}[!t]
	\centering
	\begin{tabular}{@{}cc@{}}
		\includegraphics[width=.75\textwidth]{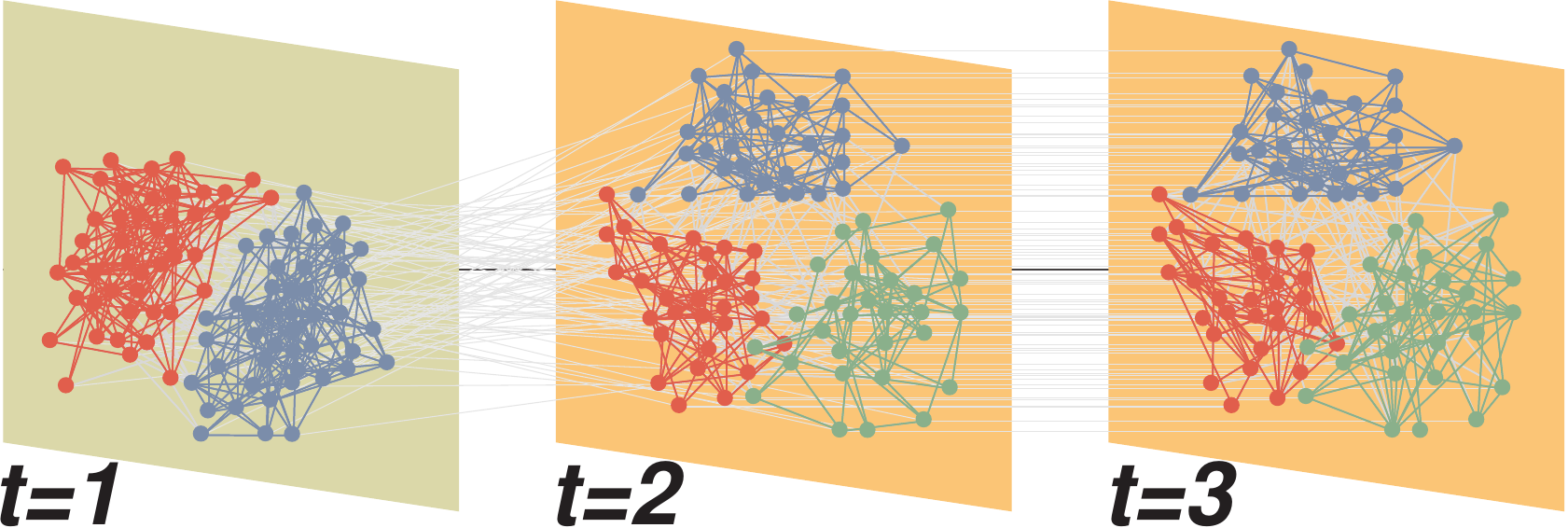} \\
		\includegraphics[width=.75\textwidth]{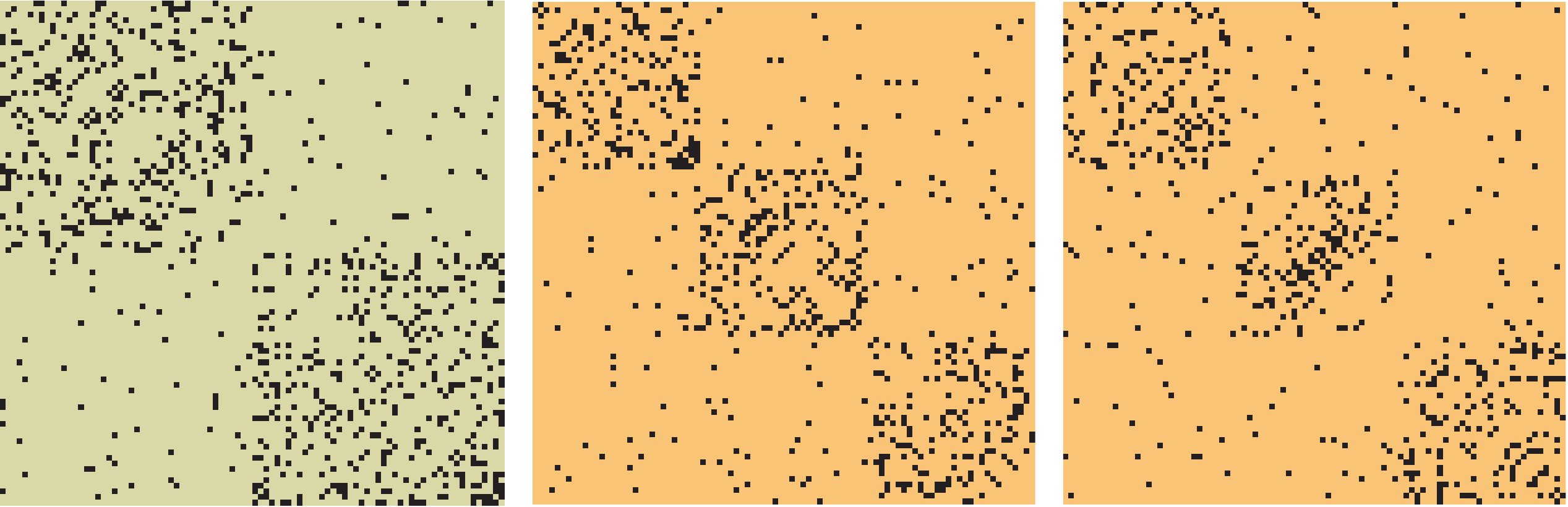} 
	\end{tabular}
	\caption{A dynamic network with 3 temporal snapshots embedded in their {\it{latent}} (\emph{i.e.} unobserved) node positions and {\it{latent}} (colored) regimes of layers that can be recovered using our method (top panel). The same network is represented in a tensor format (bottom panel). Gray lines between layers indicate node identity. Nodes in the same color exhibit dense connections whereas nodes in different colors exhibit sparse connections. Same cluster connections are represented by their node group color and inter-cluster connections are colored in white. The patterns of connections of the three snapshots can be represented as a tensor as shown in the bottom panel. Each network snapshot is shown as a matrix with black dots representing the presence of connections between corresponding node pairs. Olive and yellow colors shown in the backgrounds of top and bottom panels indicate the hidden regimes of latent coordinates shared among layers. $t=1$ network has distinct characteristics, consisting of two clusters, whereas $t=2$ and $t=3$ networks share latent coordinates of three clusters.}
	\label{multinet}
\end{figure}

By employing a Bayesian framework of tensor decomposition by \cite{Hoff2011} and \cite{Hoff2015}, we present a Bayesian method to model change-point process in longitudinal networks. The motivation of our method is firmly based on a common observation in network analysis that longitudinal network datasets frequently exhibit irregular dynamics, implying multiple changes in their data generating processes \cite[e.g.][]{Guo2007, Heard2010, Wang2014, Cribben2016, Barnett2016, Ridder2016}. 

Figure~\ref{multinet} shows an example of longitudinal network data with 3 layers of time series and 90 nodes. The data generating parameters of nodes in each network are depicted by 2 dimensional latent traits at each time point. The distance between each pair of nodes represents their probability of connection, so that proximal nodes are more likely to have connections. As clearly shown by the blocks of matrices at the bottom panel, the clustered patterns of connections are well depicted by the node positions, that can be recovered using our method, on the top panel. Colors in the backgrounds of the layers represent latent regimes inferred using our method.  In addition to the latent traits that are specific to each network, one can easily notice that the two cluster networks at $t=1$ turned into a three cluster network at $t=2$. Contrary to the dramatic change of the overall network structure at $t=2$, the network at $t=3$ exhibits identical node positions to the $t=2$ network. The same color indicates layers sharing latent node positions.  The goal of our method is to  \textit{uncover} 1) \textit{latent traits} representing data generating processes at each regime sharing those traits (colored node positions/groupings in the top panel) and 2) the \textit{timing of unspecified number of changes} ($t = \{1\}$ for regime 1 and $t = \{2, 3\}$ for regime 2). 

Conventional approaches to dynamic network modeling typically extend a static network analysis framework by assuming smooth topological changes over time or applying completely separate models for each time period \citep{Robins2001, Hanneke2010, Desmarais2012, Snijders2006, Snijders2010,  Westveld2011, Ward2013}. These methods rely largely on heuristic approaches to detect structural changes in data generating parameters. Recently, several methods for the  ``network change-point detection'' problem have been proposed, noting the importance of irregular changes in network structures.  For example, \cite{Cribben2016} introduce a two-step approach to network change-point detection in which the cosine distances for the principal eigenvectors of time-specific graph Laplacian matrix are used to find change-points given pre-specified significance thresholds.\footnote{Graph Laplacian is one of the most well-known linear operators for adjacency matrix that is designed to minimize the summed quadratic distances between latent positions of connected (unconnected) node pairs for an assortative (dissortative) network.} Another group of studies \citep{Guo2007, Wang2014} allow parameter values of exponential random graph models (ERGMs) to change over time. However, both models exhibit computational inefficiency. For instance, the maximum size of network analyzed was 11 nodes in \cite{Guo2007} and 6 nodes in \cite{Wang2014}.\footnote{In the framework of the temporal exponential random graph models (TERGM), \cite{Cranmer2014} pre-tested the existence of parametric breaks in global network statistics. Although this type of two-step approaches could be useful in learning specific aspects of network evolution, they are inherently unstable and inefficient by understating uncertainties in each estimation step and hence do not provide principled tools to select the number of parametric breaks.}  By incorporating the stochastic blockmodel (SBM) framework, which presumes the existence of discrete node groups, \cite{Ridder2016} propose a method to identify a single parametric break. \cite{Ridder2016}'s method compares the bootstrapped distribution of the log-likelihood ratio between a null model and an alternative. However, the asymptotic distribution of a SBM with a break approaches to a mixture of $\chi^2$-distributions. Hence it does not meet the regularity condition of the log-likelihood ratio test statistic \citep{drton2009}. A recent approach by \cite{Bartolucci2018} is also restricted to model changes in group membership in the SBM setting when the number of group is fixed. Likewise, existing methods for the  ``network change-point detection'' problem lack the capacity of a fully probabilistic modeling and fail to incorporate uncertainty in the model structure and parameter estimation. %



Our approach diverges from previous methods in two significant ways. First, we build a dynamic model using \cite{Hoff2011, Hoff2015}'s multilinear tensor regression model (MTRM), which is a multilayer (\emph{i.e.} tensor) extension of the latent space approach to network data. MTRM allows us to decompose longitudinal network data into node-specific (or row and column) random effects and time-specific (or layer-specific) random effects.\footnote{These two effects correspond to the node positions at the top panel of Figure~\ref{multinet} and data generating parameters associated with the global patterns of connections respectively.} For example, let $\{z_{i,j,t}\}$ be latent propensities to form a link between $i$ and $j$ observed at time $t$ and $\x_{i,j,t}$ be a vector of known covariates affecting $\{z_{i,j,t}\}$. Then, based on the notion of multilayer exchangeability \citep{hoff2009}, MTRM models the latent propensity of link formation ($\{z_{i,j,t}\}$) as a function of covariates, node-specific random effects ($\{\u_1, \ldots, \u_N\}$) and time-specific random effects ($\V_t$):
\begin{eqnarray}
\Pr(y_{i,j,t} = 1|  \x_{i,j,t}, \u_i, \u_j, \V)&=& \Phi(z_{i,j,t} )\\\label{eq1}
z_{i,j,t} &=& \x_{i,j,t}\b + \u_i^T\V_t \u_j + \epsilon_{i,j, t}\\
\epsilon_{i,j,t} &\sim& \normdist{0}{1}
\end{eqnarray}
where $\{\u_1, \ldots, \u_N\}$ represent (time-constant) $R$-dimensional latent node positions and $\V_t$ is a diagonal matrix of (time-varying) node-connection rules.  As we will explain in details, this multiplicative decomposition is highly useful for the joint estimation of \emph{time-varying network generation rules} in conjunction with \emph{latent node positions} that are constant for the duration of a hidden regime. Different from SBM formulation, the continuous multidimensional node position formulation let us to model any underlying latent structure including both group-structured networks, treated by SBM, and networks without group substructures that are unable to be modeled by SBM. 

The second departure of our approach from existing methods is the use of hidden Markov model (HMM) to characterize the change-point process. As shown in other applications \citep{Baum1970, Chib1998, RobertRydenTitt00, Cappe2005, Scott2005, Sylvia2006, Yee2006}, the conditional independence assumption in HMM turns out to be highly useful to model unknown changes in the latent network traits. More specifically, latent node positions ($\{\u_1, \ldots, \u_N\}$), which are constrained to be constant over time in \cite{Hoff2011}, are allowed to change over time depending on the transition of hidden states.

The resulting model is a hidden Markov multilinear tensor model (HMTM) as it combines \cite{Hoff2011}'s MTRM with HMM. HMTM assumes that a dynamic network process can be modeled as discrete changes in the latent space representation of network layers at each time point. These changes reflect fundamental shifts in structural properties of network under consideration. For example, structural changes in military alliance networks reflect the transformation of the international system such as the balance of power system during the Concert of Europe, the bifurcated system in the run-up to the World War I, and the bipolar structure during the Cold War, as we will see shortly.  

The proposed method has several notable contributions to longitudinal network analysis. First, we show that degree heterogeneity hinders the recovery of meaningful traits in the latent space approach and demonstrate that degree correction formulations \citep{karrer2011stochastic,chaudhuri2012spectral} make a crucial difference in the recovery of ground-truth group structures underlying our example data generation. Second, our method provides an important tool to understand dynamic network processes by allowing researchers to model fundamental changes in factors underlying the evolution of longitudinal networks. Changes in longitudinal network data can take a variety of forms and our method does not restrict the types of network generating models. Finally, we provide an open-source \textbf{\texttt{R}} package,  \textbf{\texttt{NetworkChange}}, that implements all the methods introduced in the paper including Bayesian model diagnostic tools: the approximate log marginal likelihood \citep{Chib1995}, the Watanabe-Akaike Information Criterion (WAIC) \citep{Watanabe2010}. We report the performance test results of these diagnostics.

\section{Understanding Multilinear Tensor Regression Model}
\subsection{Latent Space Model for Tensor}
 Let $\U = (\u_1, \ldots, \u_N)^\top\in\mathbb{R}^{N\times R}$ be the $R$-dimensional latent node positions of $N$ nodes and $\v_t=(v_{1t},\ldots,v_{Rt})\in\mathbb{R}^{R}$ be a vector exhibiting dimension-specific node connection rules at time $t$. In this formulation, network effects are modeled by the product of latent node traits ($\u_i$ for node $i$ and $\u_j$ for node $j$) and layer-specific node-connection rules ($\v_{t}$ at time $t$ or $t$th layer) as follows: 
\begin{eqnarray}\label{mtrm}
	\Pr(y_{i,j,t} = 1|  \x_{i,j,t}, \u_i, \u_j, \v_t)&=& \boldsymbol{\Phi}(\x_{i,j,t}\b + \langle \u_{i}, \v_{t} , \u_{j} \rangle)\\
	\U &\sim& \text{matrix normal}(\M = \mathbf{1}\boldsymbol{\mu}_{U}^T, \mathbf{I}_N, \Psi_U)\\
		\V &\sim& \text{matrix normal}(\M = \mathbf{1}\boldsymbol{\mu}_{V}^T, 
\mathbf{I}_T, \Psi_V)\\
	 \epsilon_{i,j, t}&\sim&\normdist{0}{\sigma^2}
\end{eqnarray}
where $\langle \u_{i}, \v_{t} , \u_{j} \rangle = \sum_{r=1}^R u_{i, r} v_{r, t} u_{j,r}$ and \text{matrix normal}$(\M, \U, \V)$ is a $N \times R$ matrix-variate normal distribution with mean matrix $\M$, row variance $\U$, and column variance $\V$. 

The resulting estimates of node-specific latent variables recover a specific type of similarity between nodes that is easily interpretable \citep{hoff2008}. If $\u_i$ and $\u_j$ exhibit similar values, they will have similar inner product outcomes with node $k$'s latent position vector $\u_k$. This means that the probability of connection with $k$ is analogous for $i$ and $j$. This corresponds to the notion in network theory that nodes $i$ and $j$ are structurally equivalent \citep{wasserman1994social}. In addition, the generation rule parameter $\v_t$ contains the information on what the distance relationships on each dimension of the $\U$ space reveal about their connection probability. For example, $v_{rt}>0$ corresponds to the case when a network generation rule for the $r$th dimension at time $t$ is homophilous (assortative). In words, $v_{rt}>0$ indicates that two nodes on $r$th dimension at time $t$ are more likely to be connected if they are located in the same side of the axis and the magnitude of their product is high. Similarly, $v_{rt}<0$ corresponds to the case when a network generation rule for $r$th dimension at time $t$ is heterophilous (dissortative), so that nodes located on the opposite sides are more likely to be connected than the ones with the same sign.

\subsection{Degree Correction}
One of most important features of the proposed method is detecting changes in \emph{meso-scopic network properties}, grouping of nodes, such as homophilous or heterophilous groups and core-periphery substructures in a network. The emergence (and changes) of group structures is commonly observed in real-world network data \citep{Borgatti1999, nowicki2001estimation,newman2006modularity,fortunato2010community, Xu2015, Ridder2016}. The formulation of MTRM in Equation (\ref{mtrm}), however, is designed to recover consistent regression parameters ($\b$) considering network effects as a nuisance parameter. Hence it entails a critical weakness in uncovering latent meso-scopic network features. 


Except for exogenous covariates ($\x_{i,j,t}$), Equation (\ref{mtrm}) has no treatment to account for degree heterogeneity that has been known to confound the group structure recovery  \cite[e.g.][]{newman2006modularity, newman2010networks, karrer2011stochastic}. The intuition is that the distribution of degrees in empirical networks is highly heterogeneous and skewed following power law or exponential distributions \citep{clauset2009power} 
while the implicit assumption in the group structure recovery is that the expected degree of nodes having a similar role (i.e. proximal in the latent space or belonging to the same group) is similar. 
This problem is well known in the network science literature and various degree-correction methods have been proposed \citep{newman2006modularity, newman2010networks, karrer2011stochastic, chaudhuri2012spectral, zhao2012consistency}. 
 


\begin{figure}[ht]\centering \vspace{0.2cm}
	\includegraphics[width=1 \textwidth]{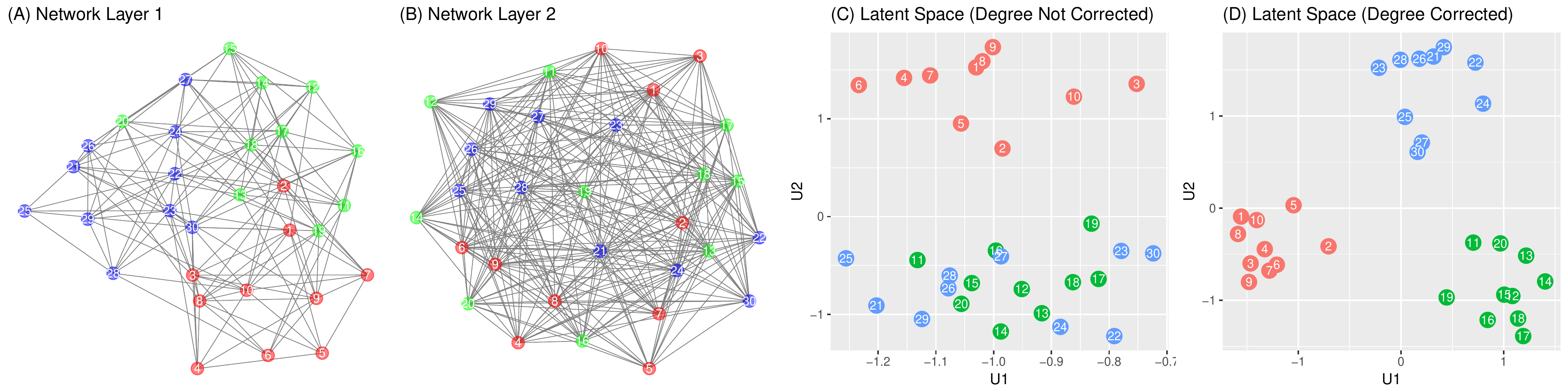}		
	\caption{Degree Correction for Group Structure Discovery: Two undirected networks with 30 nodes and 3 groups are generated. In Layer 1, the within-group link probability of 0.5 and the between-group link probability of 0.2 (homophily) and in Layer 2 the within-group link probability of 0.2 and the between-group link probability of 0.5 (heterophily). }\label{degree}
\end{figure}

Figure \ref{degree} illustrates the problem in a simple setting using a 2 layer network where we assume no change in node positions. We generate two undirected networks consisting 3 groups, each group composed of 10 nodes, in Figure \ref{degree}(A) and Figure \ref{degree}(B). The number of groups and node membership are identical but the connection rules are opposite. In Figure \ref{degree}(A), the within group link probability (0.5) is much larger than the between group link probability (0.2) and the probabilities are flipped in Figure \ref{degree}(B). The node colors indicate group memberships and the lines indicate links. Without assuming a change between 2 layers, our task is to identify 3 hidden groups from the data using their recovered node positions. 

We first fit an probit MTRM shown in Equation (\ref{mtrm}) with an unconditional mean parameter without external covariates. Then, we applied the $k$-means clustering algorithm to the estimated latent node positions to identify group membership. The results are reported in Figure \ref{degree}(C). The MTRM fails to distinguish the green group from the blue group and researchers will conclude erroneously that the data are generated by two groups. 

A simple fix to this problem is to use a null model to control for the expected level of associations among pairs of nodes. One example is an additive null model ($\omega_{ijt}$), consisting of the principal eigenvalue ($\lambda^{princ}_t=\text{max}(\vert\lambda(\mathbf{Y}_t)\vert)$) and its associated eigenvector \citep{peixoto2013eigenvalue}:
\begin{equation}\label{omega}
\omega_{ijt} = \lambda^{princ}_t \tilde{\u}_{it} \tilde{\u}_{it}^{T},
\end{equation}
 where $\tilde{\u}_{it}$ is the $i$th row of the associated eigenvector.\footnote{Alternatively, one can use a modularity matrix ($M_{ijt}$): 
\begin{equation*}\label{modul} M_{ijt}= y_{ijt}-\frac{k_i k_j}{2m} 
\end{equation*}
where $m=\frac{\sum_{i=1}^{N}k_i}{2}$ and $k_i$ is the sum of weights for $i$ \citep{newman2004finding}. Both methods are available in \textbf{\texttt{NetworkChange}}.} In matrix form, we denote the principal eigenmatrix as $\mathbf{\Omega}_t$. 

Figure \ref{degree}(D) shows the results from the linear MTRM on the transformed data ($\mathbf{B}_{t} = \mathbf{Y}_t - \mathbf{\Omega}_t$). Colors are allocated by the $k$-means clustering analysis. The use of a null model allows us to recover three distinct blocs in the data. Of course, one can think of an alternative way of controlling for the expected level of associations by including a  list of external covariates. However, when the goal is to identify hidden groups, not coefficients of covariates, using a null model is more intuitive and computationally less expensive than including a list of covariates. 

\section{The Proposed Method}
In order to develop a dynamic network model for structural changes, we must start from the question, ``What constitutes structural changes in networks?" On the one hand, one can think of a change in summary statistics of \emph{macro-scopic} network properties, such as average shortest path length or network density as a structural change \citep{Cranmer2014}. On the other hand, a change in the population statistics of \emph{micro-scopic} network properties, such as transitivity or node degree, can be considered as a structural change \citep{Heard2010, Lung-Yut-Fong2012,kolar2012estimating}. But global network statistics and local indices cannot fully represent generative processes of dynamic networks as the granularity of the information entailed in such measures is too limited. Instead, studies in network science have paid an increasing amount of attention to \emph{meso-scopic} features of networks (e.g. community structures, stochastic blocks, core-periphery structures) \citep{nowicki2001estimation, newman2006modularity, fortunato2010community, Tibely2011, Sporns2014}. Technically, various approaches for meso-scopic trait discovery locates nodes in a discrete or continuous latent space on the basis of their similarity. In this paper, a structural change in networks is defined as a change in meso-scopic features of networks. We support this claim by using synthetic examples and show that this perspective is effective enough to recover fundamental aspects of changes in network generation. 

\subsection{Hidden Markov Multilinear Tensor Model}
As shown by \cite{Chib1998} and \cite{Park2012}, multiple change-point problems are equivalent to the estimation of a nonergodic (or forward-moving) HMM, which  has advantages in latent state identification and parameter estimation, thanks to the order constraint in latent states. Let us denote $\S$ as a vector of hidden state variables where $S_t$  is an integer-valued hidden state variable at $t$ 
\begin{equation}
\S = \{(S_1,  \ldots, S_T) : S_t \in \{1, \ldots, M\}, t = 1, \ldots, T\}, 
\end{equation}
and $\P$ as a $M \times M$ transition matrix where $p_{i,i}$ is the $i$th diagonal element of $\P$ and $M$ is the number of hidden states. 


Then, the probability distribution of a degree-corrected longitudinal network data with $M-1$ breaks can be modeled as a Markov mixture of $M$-component MTRMs using the conditional independence assumption of the HMM. Suppose $\T$ be a collection of parameters that represent a network generating process of a longitudinal network. Then, 
\begin{eqnarray}
p(\mathcal{B} | \T) &=& \int p(S_1 | \T) p(\mathbf{B}_1| S_1, \T)  \prod_{t=2}^T \sum^{M}_{m=1} p(\mathbf{B}_t | \T_m) \Pr(S_t = m | S_{t-1},  \T) d\S,\nonumber 
\end{eqnarray}
where $p(\mathbf{B}_t | \T_m)$ a generative network model at regime $m$. Here, the duration of hidden state $m$ follows a geometric distribution of $1 - p_{mm}$ where $p_{mm}$ is the $m$th diagonal element of an $M \times M$ transition matrix. The regime change probability can be easily computed using the posterior draws of hidden states (e.g. $\frac{1}{G} \sum_{g=1}^G \mathcal{I}(S_t^{(g)} \neq S_{t-1}^{(g)})$).   

Following the MTRM, the HMTM decomposes the degree corrected network data at $t$ as a bilinear product of latent node positions and dimension weights subject to hidden state changes: 
\begin{eqnarray}\label{hmtm1}
	\mathbf{B}_{t} &=&\U_{S_t} \mathbf{V}_t  \U_{S_t}^T + \mathbf{E}_t\\\label{hmtm2}
	\mathbf{E}_t &\sim&  \left\{ \begin{array}{rcl}\label{hmtm3}
\mathcal{N}_{N \times N}(\mathbf{0}, \sigma_{S_t}^2\mathbf{I}_{N}, \mathbf{I}_N) &\mbox{for}  & \text{Normal Error} \\ 
\mathcal{N}_{N \times N}(\mathbf{0}, \gamma_t^{-1}\sigma_{S_t}^2\mathbf{I}_{N}, \mathbf{I}_N) & \mbox{for} & \text{Student-$t$ Error}
\end{array}\right.
\end{eqnarray}
One may concern that the normal distribution of $\mathbf{E}_t$ does not fit the data very well. In that case, the above model can be modified to include a Student-$t$ distributed error \citep{Carlin1991} where the prior distribution of $\gamma_t$ follows a gamma distribution $(\gamdist{\nu_0/2}{\nu_1/2})$. 

For prior distributions of $\U$ and $\V$, we follow \cite{Hoff2011}'s hierarchical scheme with two major  modifications.  First, we orthogonalize each column of $\mathbf{U}_{S_t}$ using the Gram-Schmidt process \citep{Bjorck1996, Dunson2015} in each simulation step. \cite{Hoff2011}'s hierarchical scheme centers rows of $\mathbf{U}_{S_t}$ around its global mean ($\boldsymbol{\mu}_{u,S_t}$) using a multivariate normal distribution. This does not guarantee the orthogonality of each latent factor in $\mathbf{U}_{S_t}$. The lack of orthogonality makes the model unidentified, causing numerical instability in parameter estimation and model diagnostics \citep{murphy2012machine, Dunson2015}.%

Second, we use independent inverse-gamma distributions instead of inverse-Wishart distribution for the prior distribution of a variance parameter ($\Psi_{u, S_t}, \Psi_{v}$). The use of inverse-Wishart distribution for the prior distribution of a variance parameter ($\Psi_{u, S_t}, \Psi_{v}$) comes at a great cost because choosing informative inverse-Wishart prior distributions for $\Psi_{u, m}$ and $\Psi_{v}$ is not easy \citep{Gelman2015} and a poorly specified inverse-Wishart prior distribution has serious impacts on the marginal likelihood estimation. In our trials, the log posterior  inverse-Wishart density  of $\Psi_{u, S_t}$ and $\Psi_{v}$ often goes to a negative infinity, failing to impose proper penalties. In HMTM, the off-diagonal covariance of $\mathbf{U}_m$ is constrained to be 0, thanks to the Gram-Schmidt process, and the off-diagonal covariance of $\mathbf{V}$ is close to 0 as $\v_t$ measures time-varying weights of independent $\mathbf{U}_m$. Thus, inverse-gamma distributions resolve a computational issue without a loss of information. 
 
The resulting prior distributions of $\U$ and $\V$ are matrix-variate normal distributions in which each column vector ($ \u_{i, S_t}$ and $\v_{t}$) follows a multivariate normal distribution. We first discuss the prior distribution of $\U$: 
\begin{eqnarray}\label{prior1}
\U_{S_t} &\equiv&  (\u_{1, S_t}, \ldots, \u_{N, S_t})^\top\in\mathbb{R}^{N\times R}\\\label{prior2}
 \u_{i, S_t}&\sim& \mathcal{N}_{R}(\boldsymbol{\mu}_{u,S_t}, \Psi_{u, S_t})\\\label{prior3}
\boldsymbol{\mu}_{u, S_t}|\Psi_{u, S_t} &\sim& \mathcal{N}_{R}(\boldsymbol{\mu}_{0, u_{S_t}}, \Psi_{u, S_t})\\ \label{prior4}
\Psi_{u, S_t} &\equiv& \left( \begin{array}{ccc}
 \psi_{1, u, S_t} & \ldots & 0 \\
0 & \psi_{r, u, S_t} & 0 \\
0 &  \ldots & \psi_{R, u, S_t} \end{array} \right)\\\label{prior5}
	\psi_{r, u, S_t} &\sim&\mathcal{IG}\left (\frac{u_0}{2}, \frac{u_{1}}{2}\right).
\end{eqnarray}
	
The prior distributions of $\V$ are similar to $\U$ but one difference is that only diagonal elements of $\V_t$ are modeled as a multivariate normal distribution:   	
\begin{eqnarray}\label{prior6}
\mathbf{V}_t &\equiv&  \left( \begin{array}{ccc}
 v_{1, t} & \ldots & 0 \\
0 & v_{r, t} & 0 \\
0 &  \ldots & v_{R, t} \end{array} \right) \\\label{prior7}
\v_{t} &\equiv&(v_{1, t}, \ldots, v_{R, t})^\top\in\mathbb{R}^{R\times 1} \\
\v_{t}&\sim& \mathcal{N}_{R}(\boldsymbol{\mu}_v, \Psi_v)\\\label{prior8}
\boldsymbol{\mu}_{v}|\Psi_{v} &\sim& \mathcal{N}_{R}(\boldsymbol{\mu}_{0, v}, \Psi_{v})\\ \label{prior9}
\Psi_{v} &=& \left( \begin{array}{ccc}
\psi_{1, v} & \ldots & 0 \\
0 & \psi_{r, v} & 0 \\
0 &  \ldots & \psi_{R, v} \end{array} \right)\\\label{prior10}
	\psi_{r, v} &\sim&\mathcal{IG}\left (\frac{v_0}{2}, \frac{v_{1}}{2}\right).
\end{eqnarray}

Then, we complete the model building by introducing HMM-related prior specifications following  \cite{Chib1998}: 
\begin{eqnarray}\label{eq:cp1}
	S_t|S_{t-1}, \P &\sim&\mathcal{M}arkov(\P, \pi_0)\\\label{eq:cp10}
	\underbrace{\P}_{M \times M} &=& (\mathbf{p}_1, \ldots, \mathbf{p}_M)\\\label{eq:cp11}
	\mathbf{p}_i &\sim& \text{Dirichlet}(\alpha_{i,1}, \ldots, \alpha_{i,M})
\end{eqnarray}
where $\pi_0$ is the initial probability of a non-ergodic Markov chain ($\pi_0 = (1, 0, \ldots, 0)$). 

\subsection{Sampling Algorithm}
Let $ \mathbf{\Theta}$ indicate a parameter vector beside hidden states ($\S$) and a transition matrix ($\P$): $\mathbf{\Theta} = \{ \mathbf{U}, \mathbf{V}, \boldsymbol{\mu}_u, \Psi_u, \boldsymbol{\mu}_v, \Psi_v, \sigma^2\}$. Let $\mathbf{\Theta}_{S_t}$ denote regime-specific $\mathbf{\Theta}$ at $t$. Then, the  joint posterior density $p(\mathbf{\Theta}, \P, \S| \B)$ is
\begin{eqnarray}
	p(\mathbf{\Theta}, \P, \S| \B) &\propto& \mathcal{N}_{N \times N}(\mathbf{B}_1|\mathbf{\Theta}_{1})\prod_{t=2}^T \Big( \mathcal{N}_{N \times N}(\mathbf{B}_t| \B_{t-1}, \mathbf{\Theta}_{S_t}) p(S_t| S_{t-1}, \P) \Big)\\
	&&  \prod_{m=1}^M \Big(  \mathcal{N}_{R}(\boldsymbol{\mu}_{u,m}| \boldsymbol{\mu}_{0, u^m}, \psi_{., u, m})\mathcal{N}_{R}(\boldsymbol{\mu}_v|\boldsymbol{\mu}_{0, v}, \psi_{., v, m}) \Big)\\
	&&  \prod_{m=1}^M \prod_{r=1}^R  \Big( \mathcal{IG}(\psi_{r, u, m}| u_{0, m}, u_{1, m})  \mathcal{IG}(\psi_{r, v, m}| v_{0, m}, v_{1, m}) \Big)\\
	&& \prod_{m=1}^M \Big(\mathcal{IG}(\sigma^{2}_{m}| c_0,  d_0)  \mathcal{B}eta(p_{mm}|a, b)\Big) 
\end{eqnarray}
where $\B_{t-1} = (\mathbf{B}_1, \ldots, \mathbf{B}_{t-1})$.  
Using the conditional independence we decompose the joint posterior distribution into three blocks and marginalize conditional distributions \citep{Liu1994a, vanDyk2008}: 
\begin{eqnarray*}
	p(\mathbf{\Theta}, \P, \S| \B) &=& \underbrace{p(\mathbf{\Theta}| \B, \P, \S)}_{\textrm{Part 1}} \underbrace{p( \P| \B, \S)}_{\textrm{Part 2}} \underbrace{p( \S| \B)}_{\textrm{Part 3}}.
\end{eqnarray*}

The sampling algorithm of the HMTM can be summarized as follows:
\begin{description}
\item[Step 1] The sampling of regime specific $\U, \boldsymbol{\mu}, \Psi_{u}$ consists of the following three steps for each regime $m$.  Let $ \Psi_{u} = \left( \begin{array}{ccc}
\psi_{1, u, m} & \ldots & 0 \\
0 & \psi_{r, u, m} & 0 \\
0 &  \ldots & \psi_{R, u, m} \end{array} \right) $.
	\begin{enumerate}
	\item $p(\psi_{r, u, m} |  \B,  \P, \S, \mathbf{\Theta} ^{-\Psi_{u, m}}) \propto \mathcal{IG}\left (\frac{u_0 + N}{2}, \frac{\U_{r, m}^T\U_{r, m} + u_{1}}{2}\right)$.
	
	\item $p(\boldsymbol{\mu}_{u, m}|  \B,  \P, \S, \mathbf{\Theta} ^{-\boldsymbol{\mu}_{u, m}}) \propto  \textrm{multivariate normal}(\U_m^T\mathbf{1}/(N + 1), \Psi_{u, m}/(N + 1))$.
	\item $p(\U_m|  \B,  \P, \S, \mathbf{\Theta} ^{-\U_m}) \propto \text{matrix normal}_{N \times R}(\tilde{\M}_{u, m}, \mathbf{I}_{N}, \tilde{\Psi}_{u, m})$ 
	where 
	\begin{eqnarray*}
	\tilde{\Psi}_{u, m} &=& (\mathbf{Q}_{u, m}/\sigma_m^2 + \Psi_{u, m}^{-1})^{-1} \\
	\tilde{\M}_{u, m} &=& (\mathbf{L}_{u, m}/\sigma_m^2 + \mathbf{1}\boldsymbol{\mu}_{u, m}^T \Psi_{u, m}^{-1}) \tilde{\Psi}_{u, m}\\
	\mathbf{Q}_{u, m} &=& (\U_m^T\U_m)\circ(\V_m^T\V_m) \\
	\mathbf{L}_{u, m} &=& \sum_{j, t:\; t \in S_t = m}b_{\cdot,j, t} \otimes (\U_{m, j, \cdot} \circ   \V_{m, t, \cdot} )
	\end{eqnarray*}
	\item Orthogonalize $\U_m$ using the Gram-Schmidt  algorithm. 
	\end{enumerate}

\item[Step 2] The sampling of $\V, \boldsymbol{\mu}_v, \Psi_v$ is done for each regime.  Let $ \Psi_{v} = \left( \begin{array}{ccc}
\psi_{1, v, m} & \ldots & 0 \\
0 & \psi_{r, v, m} & 0 \\
0 &  \ldots & \psi_{R, v, m} \end{array} \right) $.
	\begin{enumerate}
	\item $p(\psi_{r, v, m} |  \B,  \P, \S, \mathbf{\Theta} ^{-\Psi_{v, m}}) \propto\mathcal{IG}\left (\frac{v_0 + T}{2}, \frac{\V_{r, m}^T\V_{r, m} + v_{1}}{2} \right)$.
	\item  $p(\boldsymbol{\mu}_{v, m}|  \B,  \P, \S, \mathbf{\Theta} ^{-\boldsymbol{\mu}_{v, m}}) \propto\textrm{multivariate normal}(\V_m^T\mathbf{1}/(T_m + 1), \Psi_{v, m}/(T_m + 1))$.
	\item  $p(\V_m|  \B,  \P, \S, \mathbf{\Theta} ^{-\V_m}) \propto\text{matrix normal}_{T_m \times R}(\tilde{\M}_{v, m}, \mathbf{I}_{T_m}, \tilde{\Psi}_{v, m})$ 
	where 
	\begin{eqnarray*}
	\tilde{\Psi}_{v, m} &=& (\mathbf{Q}_{v, m}/\sigma_m^2 + \Psi_{v, m}^{-1})^{-1} \\
	\tilde{\M}_{v, m} &=& (\mathbf{L}_{v, m}/\sigma_m^2 + \mathbf{1}\boldsymbol{\mu}_{v, m}^{T_m}\Psi_{v, m}^{-1}) \tilde{\Psi}_{v, m}\\
	\mathbf{Q}_{v, m} &=& (\U_m^{T}\U_m)\circ(\U_m^{T}\U_m) \\
	\mathbf{L}_{v, m} &=& \sum_{i, j}b_{i, j, \cdot} \otimes (\U_{m, i, \cdot} \circ   \U_{m, j, \cdot} )
	\end{eqnarray*}
	\end{enumerate}


\item[Step 3] The sampling of $\sigma^2_m$ from $\mathcal{IG}\left (\frac{c_0 +  N_m \cdot N_m \cdot T_m}{2}, \frac{d_{0} + \sum_{i = 1}^N\sum_{j = 1}^N\sum_{t=1}^T b_{i, j, t} - \mu_{i, j, t}}{2} \right)$. 

\item[Step 4] Sample $\S$ recursively using \cite{Chib1998}'s algorithm. The joint conditional distribution of the latent states $p(S_0, \ldots, S_T | \mathbf{\Theta}, \B, \P) $ can be written as the product of $T$ numbers of independent conditional distributions: 
\begin{equation*}
 p(S_0, \ldots, S_T |\mathbf{\Theta}, \B, \P) = p(S_T| \mathbf{\Theta}, \B, \P)\ldots p(S_t|\S^{t+1},  \mathbf{\Theta}, \B, \P) \ldots p(S_0|\S^{1},  \mathbf{\Theta}, \B, \P). 
\end{equation*}

Using Bayes' Theorem, \cite{Chib1998} shows that 
\begin{eqnarray*}
p(S_t|\S^{t+1}, \mathbf{\Theta}, \B, \P) &\propto& \underbrace{p(S_t|\mathbf{\Theta}, \mathbf{B}_{1:t}, \P)}_{\text{State probabilities given all data}} \underbrace{p(S_{t+1}|S_t, \P)}_{\text{Transition probability at $t$}}.
\end{eqnarray*}
The second part on the right hand side is a one-step ahead transition probability at $t$, which can be obtained from a sampled transition matrix ($\P$). The first part on the right hand side is state probabilities given all data, which can be simulated via a forward-filtering-backward-sampling algorithm as shown in \cite{Chib1998}.

During the burn-in iterations, if sampled $\S$ has a state with single observation, randomly sample  $\S$ with replacement using a pre-chosen perturbation weight ($\mathbf{w}_{\mathrm{perturb}} = (w_1, \ldots, w_{M})$).  
\item[Step 5] Sample each row of $\P$ from the following Beta distribution: \\
\begin{equation*}
p_{kk} \sim \betadist{a_0 + j_{k, k} - 1}{b_{0} + j_{k, k+1}}
\end{equation*}
where $p_{kk}$ is the probability of staying when the state is $k$, and $j_{k, k}$ is the number of jumps from state $k$ to $k$, and $j_{k, k+1}$ is the number of jumps from state $k$ to $k+1$.
\end{description}
We provide the sampling details of the HMTM with a Student-$t$ distributed error the supplementary material.

\subsection{Assessing Model Uncertainty}
We provide three metrics for model diagnostics and break number detection: the approximate log marginal likelihood method, WAIC, and the average loss of break points. The first measure is the approximate log marginal likelihood method using the candidate's estimator \citep{Chib1995}. Main advantages of the approximate log marginal likelihood are its direct connection with Bayes' theorem and its consistency when models are well identified and MCMC chains converge to the target distribution.  A disadvantage of the approximate log marginal likelihood is its computational cost arising from additional MCMC runs at each Gibbs sampling block. Using the Rao-Blackwell approximation, the approximate log marginal likelihood of HMTM with $M$ numbers of latent states ($\mathcal{M}_M$) can be computed as follows: 
\begin{eqnarray*}\nonumber
\log \hat{p}(\B^{\text{upper}}|  \mathcal{M}_M) &=& \underbrace{\log p(\B^{\text{upper}}|  \boldsymbol{\mu}_{u}^*, \psi_{., u}^*, \boldsymbol{\mu}_{v}^*, \psi_{., v}^*, \sigma^{2*}, \P^*, \mathcal{M}_M)}_{\textrm{the log likelihood}} \\
&& + \underbrace{\sum_{m=1}^M\log p(\boldsymbol{\mu}_{u,m}^*, \psi_{., u,m}^*, \boldsymbol{\mu}_{v,m}^*, \psi_{., v,m}^*, \sigma_m^{2*}, p_{m,m}^*|\mathcal{M}_M)}_{\textrm{the log prior density of posterior means}}\\
&& - \underbrace{\sum_{m=1}^M\log p(\boldsymbol{\mu}_{u,m}^*, \psi_{., u,m}^*, \boldsymbol{\mu}_{v,m}^*, \psi_{., v,m}^*, \sigma_m^{2*}, p_{m,m}^*|\B^{\text{upper}}, \mathcal{M}_M)}_{\textrm{the log posterior density of posterior means}}
\end{eqnarray*}
where $\{\boldsymbol{\mu}_u^*, \psi_{., u}^*, \boldsymbol{\mu}_v^*, \psi_{., v}^*, \sigma^{2*}, \P^*\}$ are posterior means of MCMC outputs.  The log likelihood is computed by summing log predictive density values evaluated at posterior means across all states and over all upper triangular array elements as follows:
\begin{eqnarray*}\nonumber
 \sum_{t=1}^T\sum_{i=1}^{N} \sum_{j=i+1}^{N-1}\sum_{m=1}^{M}p(b_{i,j,t}| \B_{t-1}^{\text{upper}},   \boldsymbol{\mu}_{u,m}^*, \psi_{., u,m}^*, \boldsymbol{\mu}_{v,m}^*, \psi_{., v,m}^*, \sigma_m^{2*}, \P^*, S_t = m, \mathcal{M}_M)  \\
 p(S_t = m| \B_{t-1}^{\text{upper}},   \boldsymbol{\mu}_{u,m}^*, \psi_{., u,m}^*, \boldsymbol{\mu}_{v, m}^*, \psi_{., v, m}^*, \sigma_m^{2*}, \P^*_m, \mathcal{M}_M).
\end{eqnarray*}

The computation of the log posterior density of posterior means requires a careful blocking in a highly parameterized model as discussed in \cite{Chib1995}. In our HMTM, the log posterior density evaluated at posterior means is decomposed into seven blocs:
\begin{eqnarray*}\nonumber
\log p(\boldsymbol{\mu}_u^*, \psi_{., u}^*, \boldsymbol{\mu}_v^*, \psi_{., v}^*, \sigma^{2*}, \P^*|\B) &=& 
 \log p(\boldsymbol{\mu}^*_u|\B) + \sum_{r=1}^R\log  p(\psi^*_{r, u}|\B, \boldsymbol{\mu}^*_u)\\\nonumber
&& + \log p(\boldsymbol{\mu}_v^*|\B, \boldsymbol{\mu}^*_u, \psi^*_{., u}) + \sum_{r=1}^R \log  p(\psi_{r, v}^*|\B, \boldsymbol{\mu}^*_u, \psi^*_{., u}, \boldsymbol{\mu}_v^*) \\ \label{marg}
 && + \log p(\sigma^{2*}|\B,\boldsymbol{\mu}^*_u, \psi^*_{., u}, \boldsymbol{\mu}_v^*, \psi_{., v}^*) \\\nonumber
 && + \log  p(\P^*|\B, \boldsymbol{\mu}^*_u, \psi^*_{., u}, \boldsymbol{\mu}_v^*, \psi_{., v}^*,  \sigma^{2*}).
\end{eqnarray*}


The second measure of model diagnostics is WAIC \citep{Watanabe2010}. WAIC approximates the expected log pointwise predictive density by subtracting a bias for the effective number of parameters from the sum of log pointwise predictive density.  WAIC approximates leave-one-out cross validation (LOO-CV) in singular models and hence can serve as a metric for out-of-sample predictive accuracy of HMTM \citep{Gelman2014}. Predictive accuracy is a good standard for detecting the number of breaks because overfitting is a major concern in analysis using mixture models and HMMs. Also, the cost of computation is very low as WAIC is computed from MCMC outputs. Note that WAIC of HMTM partitions the data into $T$ pieces of conditional density, and the estimated WAIC scores are dependent upon latent state estimates. The dependence on estimated latent states indicate that our measure of WAIC cannot be used to predict future networks given the sequence of network observation. Instead, we aim to use WAIC to compare predictive accuracies of HMTMs given a varying number of breaks.  

Using the formula suggested by \cite{Gelman2014},  WAIC for HMTM with $M$ number of latent states ($\mathcal{M}_M$) is 
\begin{eqnarray*} 
 \text{WAIC}_{\mathcal{M}_M} &=& -2\Bigg(\underbrace{\sum_{t=1}^{T}\log \left [ \frac{1}{G} \sum_{g=1}^{G}p(\B_t^{\text{upper}}| \boldsymbol{\mu}_{u}^{(g)}, \psi_{., u}^{(g)}, \boldsymbol{\mu}_{v}^{(g)}, \psi_{., v}^{(g)}, \sigma^{2, (g)},  \P^{(g)},  \S^{(g)}, \mathcal{M}_M) \right ]}_{\textrm{the expected log pointwise predictive density}} - \\
 && \underbrace{\sum_{t=1}^{T}V_{g=1}^{G} \left [ \log p(\B_t^{\text{upper}}|\boldsymbol{\mu}_{u}^{(g)}, \psi_{., u}^{(g)}, \boldsymbol{\mu}_{v}^{(g)}, \psi_{., v}^{(g)}, \sigma^{2, (g)}, \P^{(g)},  \S^{(g)}, \mathcal{M}_M) \right ]}_{\textrm{bias for the effective number of parameters}} \Bigg ) 
 \end{eqnarray*}
 where $G$ is the MCMC simulation size, $V[\cdot]$ indicates a variance,  and  $\mathbf{\Theta}^{(g)}, \P^{(g)}$ are the $g$th simulated outputs. 
 Throughout the paper, we report the approximate log marginal likelihood in the deviance scale by multiplying -2 to $\log \hat{p}(\B^{\text{upper}}|  \mathcal{M}_M)$ for easy comparison with WAIC following the advice of \cite{Gelman2014}: The smaller the deviance, the better the accuracy.
 
 The last measure of model diagnostics is the average loss of break points. The inclusion of redundant break points (e.g. imposing two breaks on a single break process) produces an instability in draws of hidden state variables. An easy way to check the existence of redundant breaks is to estimate average variances of simulated break points. This measure is equivalent to the average loss of break points assuming the simulation mean of break points ($\bar{\tau}_m$) as true break points:
\begin{eqnarray*} 
\textrm{Average Loss} &=& \frac{1}{M}\sum_{m=1}^{M}\left(\frac{1}{G}\sum_{g=1}^{G} (\bar{\tau}_m - \tau_{m}^{(g)})^2 \right)
 \end{eqnarray*}
where $G$ is the MCMC simulation size and $M$ is the total number of breaks. The average loss is close to 0 if simulated break points are highly stable. Average Loss becomes larger if at least one of break points swings widely in each simulation.

\section{Simulation Studies}
In this section, we check the performance of the proposed method using simulated group-structured network data. We consider five different cases of group-structured network changes: 
\begin{enumerate}
\item group-structured networks with no break (Table \ref{sim.constant})
\item group-structured networks with a group-splitting break (Table \ref{sim.split})
\item group-structured networks with a group-merging break  (Table 3 in Supplementary Material)\footnote{To save space, we report  the simulation results of a group-merging break and of a group-splitting break followed by a group-merging break in Supplementary Material.}
\item group-structured networks with a group-merging break followed by a group-splitting break (Table \ref{sim.merge-split})
\item group-structured networks with a group-splitting break followed by a group-merging break (Table 4 in Supplementary Material)
\end{enumerate}

\subsection{Simulation Setup}
Blocks in simulated data were generated by an assortative rule in which nodes belonging to the same group had a higher connection probability ($p_{in} = 0.5$) than nodes belonging to different groups ($p_{out} = 0.05$).\footnote{\emph{i.e.} $\forall b_i=b_j$, $p_{ij}=p_{in}$ and $\forall b_i\neq b_j$, $p_{ij}=p_{out}$ where $b_i$ and $b_j$ are group labels for nodes $i$ and $j$ respectively. This simple formulation, with two difference values for the block diagonal connection probability and the off-diagonal connection probability, is called the planted partition model.} In the group merging examples, two groups were merged so that the tie formation probability between the members of the two groups changed from $p_{out}$ to $p_{in}$. In the group splitting examples, an existing group split into two equal size groups so that the connection probability between the members of the two different groups became $p_{out}$ from $p_{in}$. The length of time layers was 40. The planted break occurred at $t = 20$ in the case of the single break examples and $t=10$ and $t=30$ in the case of two breaks. We fit four different HMTMs from no break ($\mathcal{M}_0$) to three breaks ($\mathcal{M}_3$) and compare their model diagnostics, recovered latent spaces, and time-varying network generation rules.

\begin{sidewaystable}\footnotesize 
  \centering
\resizebox{\textwidth}{!}{  \begin{tabular}{p{4cm} cccc c}
    \toprule
    Model Fit& \multicolumn{4}{c}{Latent Space ($\mathbf{U}_m$) Changes}&  \multicolumn{1}{c}{Generation Rule ($\mathbf{v}_t$)} \\  
       \cline{2-5} 
    & Regime 1& Regime 2&Regime 3& Regime 4& \\\midrule
     \begin{tabular}{lcl } Break number &=&0 \\ WAIC &=& 14573  \\ -2*log marginal &=& 14435 \\ -2*log likelihood &=& 14382 \end{tabular} 
  &
    \begin{minipage}{.18\textwidth}
      \includegraphics[width=\linewidth]{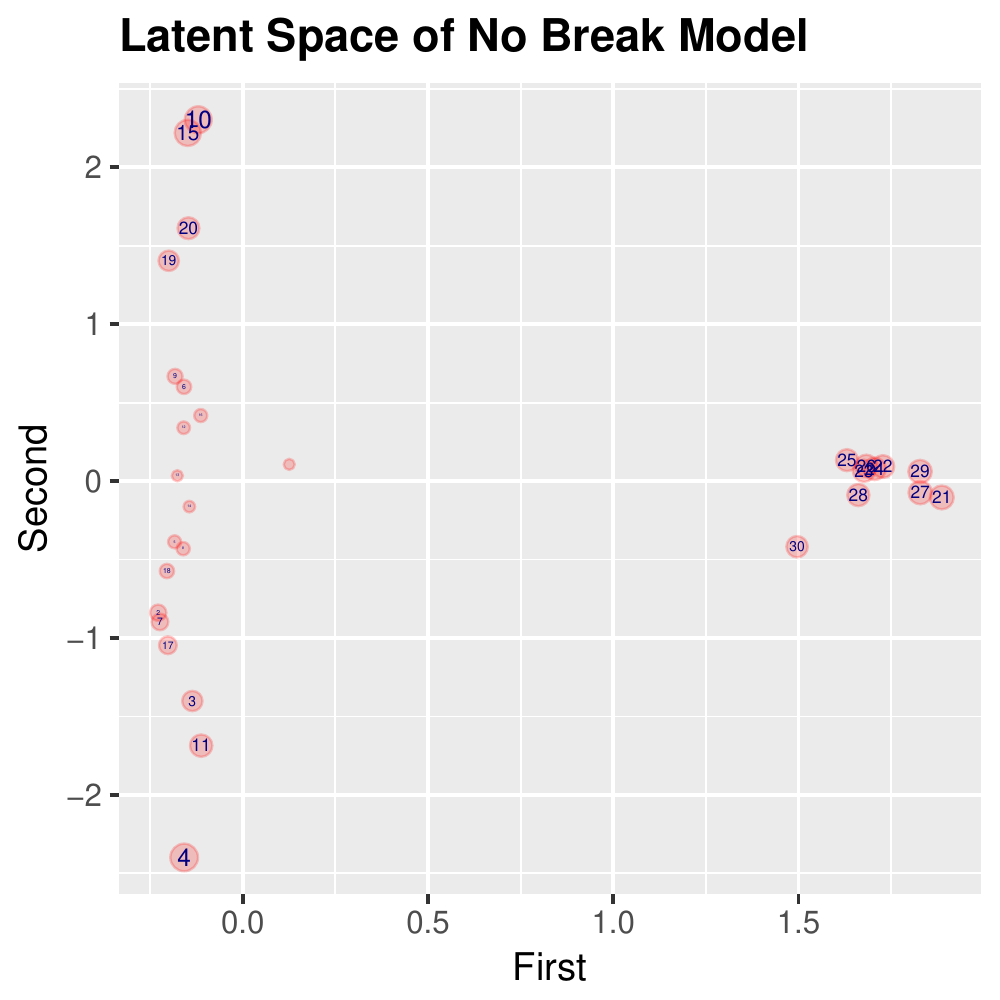}
    \end{minipage}
    &&&&
    \begin{minipage}{.18\textwidth}
      \includegraphics[width=\linewidth]{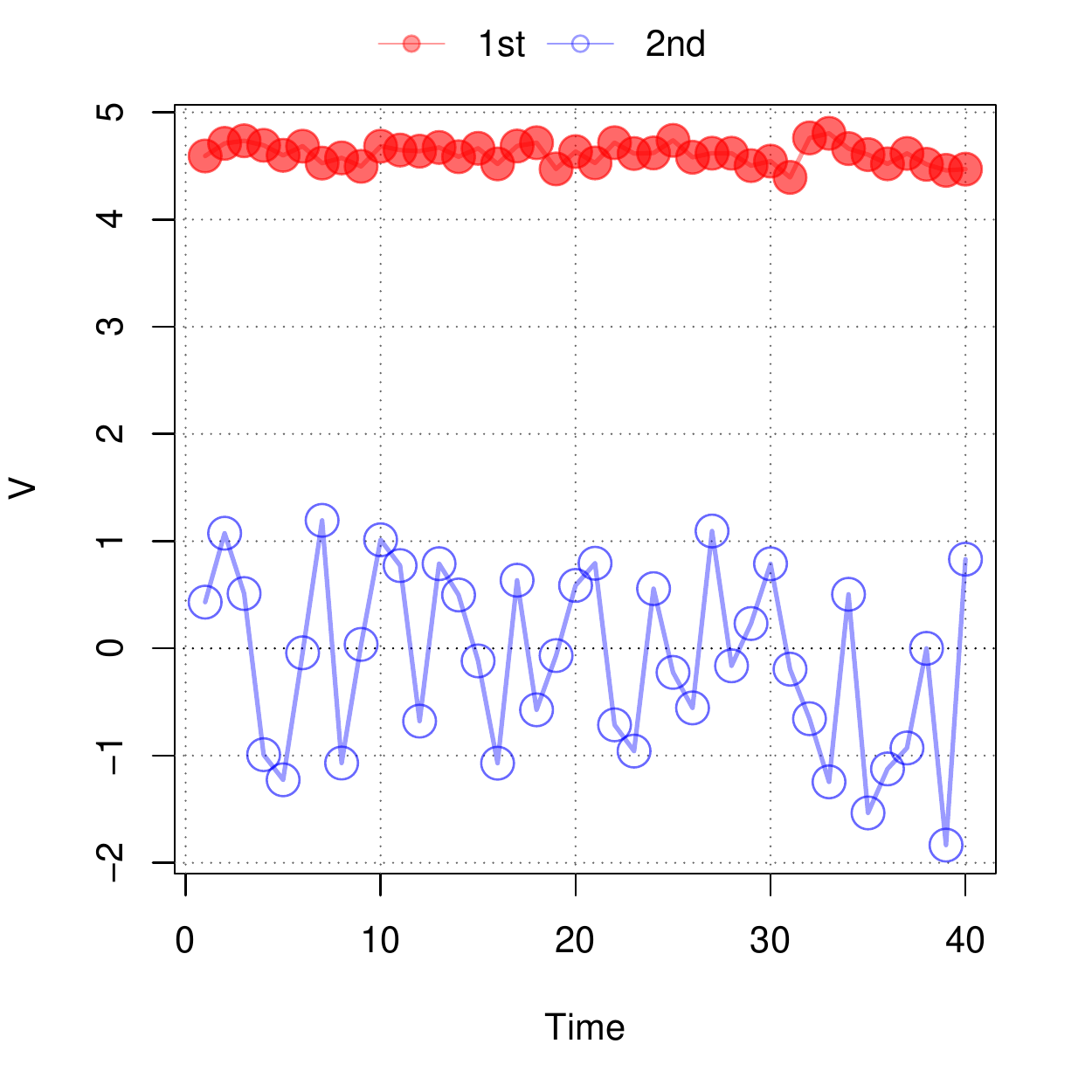}
    \end{minipage} \\
     \begin{tabular}{lcl } Break number &=&1\\ WAIC &=& 14589  \\ -2*log marginal &=& 14404 \\ -2*log likelihood &=& 14310\\ Average Loss &=& 0.08 \end{tabular} 
&
    \begin{minipage}{.18\textwidth}
      \includegraphics[width=\linewidth]{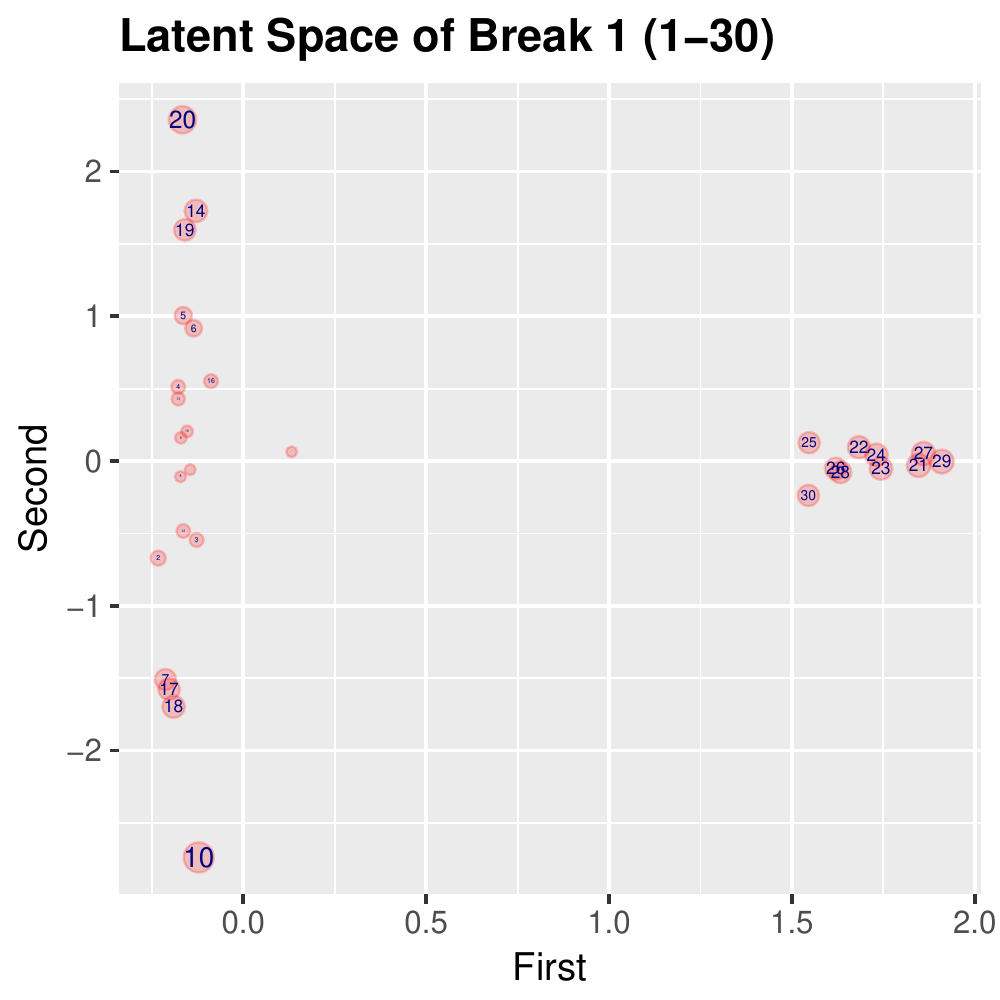}
    \end{minipage}
    &
        \begin{minipage}{.18\textwidth}
      \includegraphics[width=\linewidth]{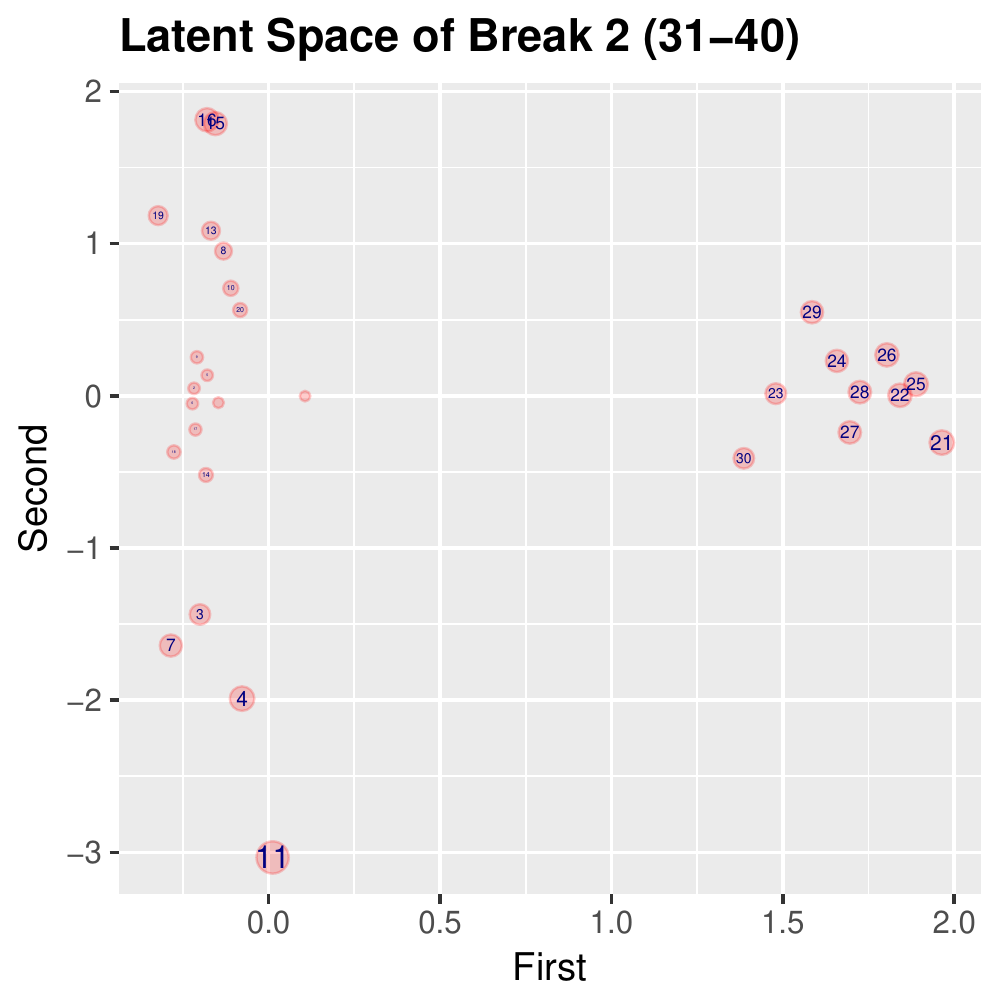}
    \end{minipage}
    &&&
    \begin{minipage}{.18\textwidth}
      \includegraphics[width=\linewidth]{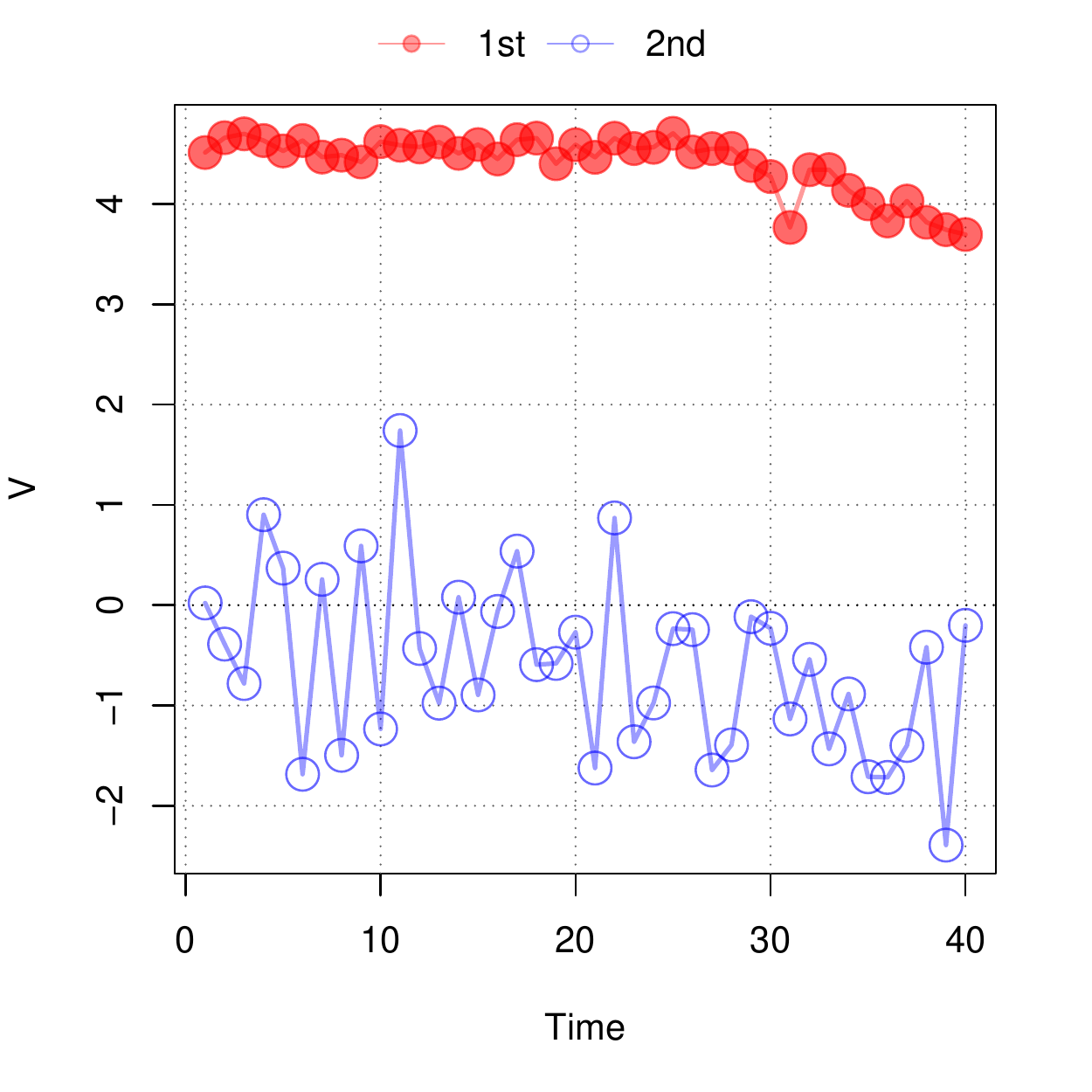}
    \end{minipage} \\
    \begin{tabular}{lcl } Break number &=&2 \\ WAIC &=& 14653\\ -2*log marginal &=&14514  \\ -2*log likelihood &=& 14408 \\ Average Loss &=& 0.55\end{tabular} 
	&
	 \begin{minipage}{.18\textwidth}
      \includegraphics[width=\linewidth]{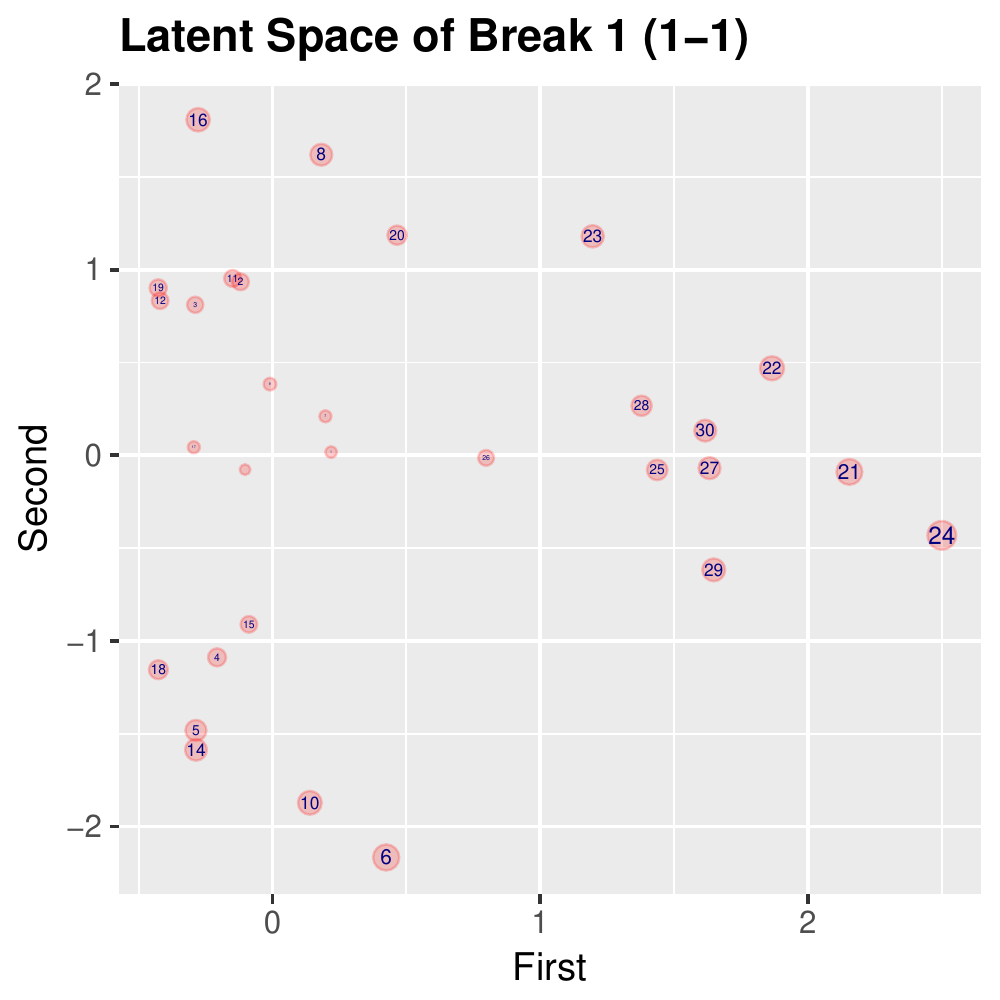}
    \end{minipage}
    &
        \begin{minipage}{.18\textwidth}
      \includegraphics[width=\linewidth]{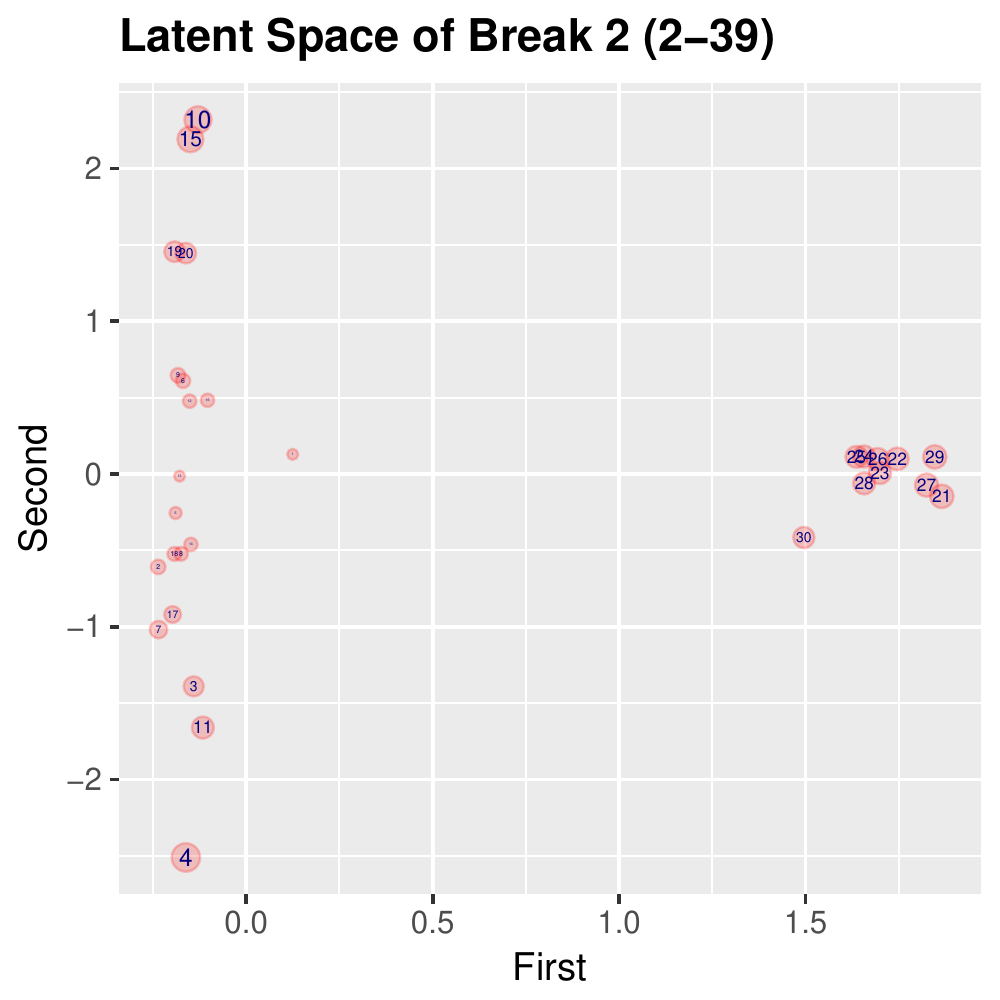}
    \end{minipage}
    &
            \begin{minipage}{.18\textwidth}
      \includegraphics[width=\linewidth]{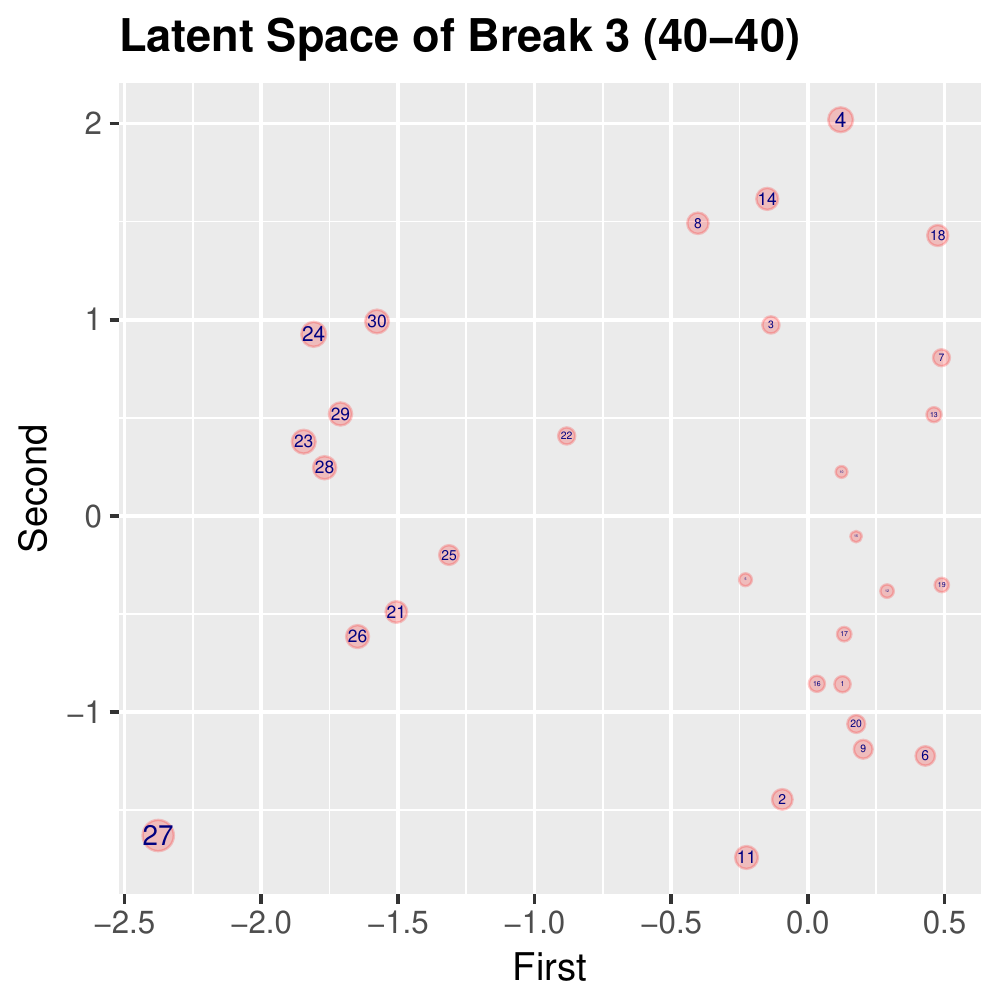}
    \end{minipage}
	&&
    \begin{minipage}{.18\textwidth}
      \includegraphics[width=\linewidth]{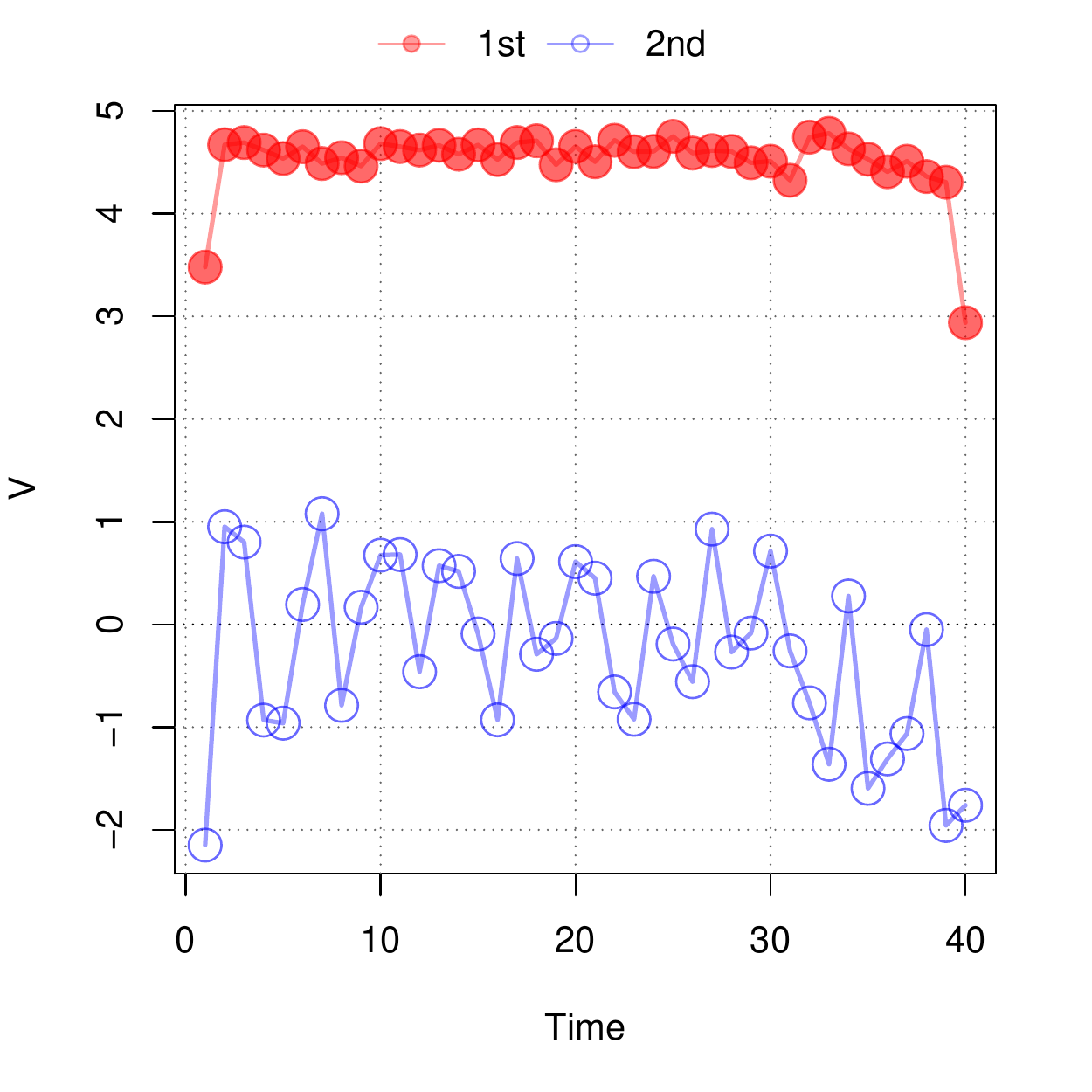}
    \end{minipage} \\
    \begin{tabular}{lcl } Break number &=&3 \\ WAIC &=& 14739 \\ -2*log marginal &=&14564 \\ -2*log likelihood &=& 14458 \\ Average Loss &=& 0.08 \end{tabular} 
  &
	\begin{minipage}{.18\textwidth}
      \includegraphics[width=\linewidth]{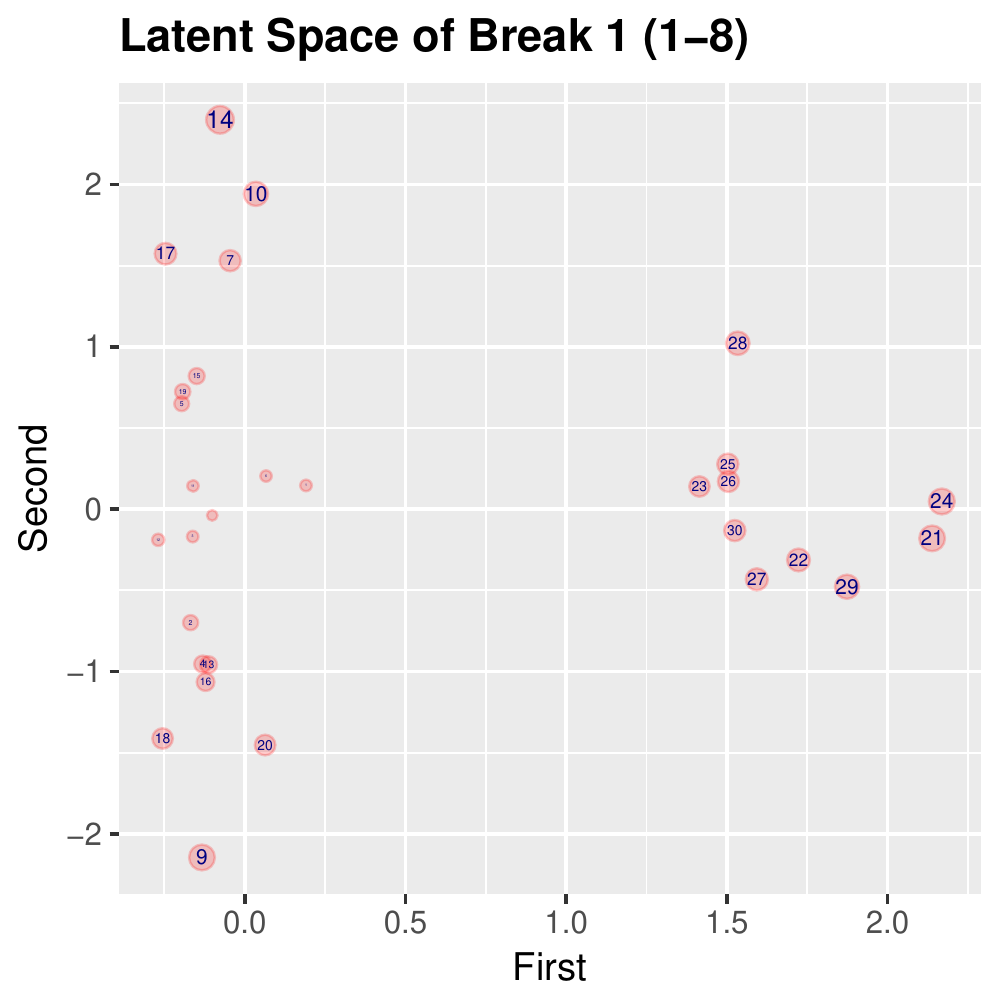}
    \end{minipage}
    &
        \begin{minipage}{.18\textwidth}
      \includegraphics[width=\linewidth]{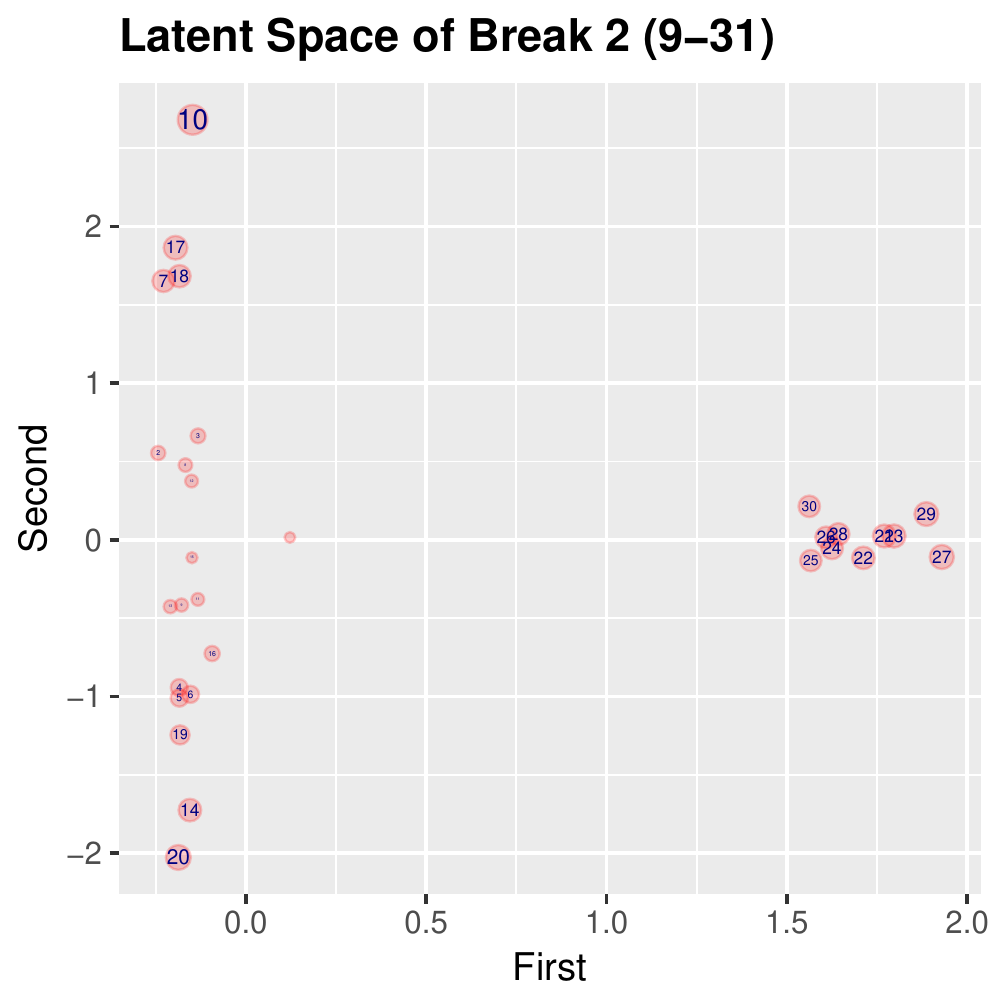}
    \end{minipage}
    &
            \begin{minipage}{.18\textwidth}
      \includegraphics[width=\linewidth]{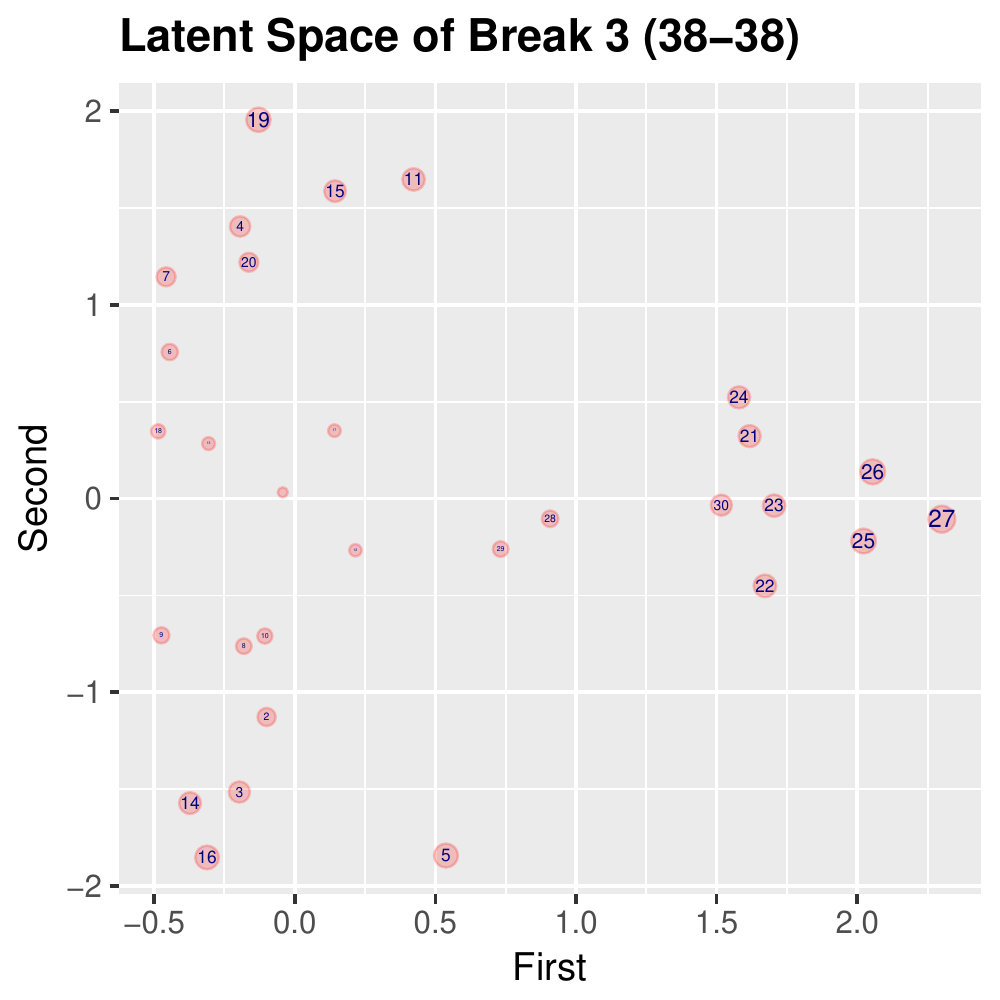}
    \end{minipage}
	&
	           \begin{minipage}{.18\textwidth}
      \includegraphics[width=\linewidth]{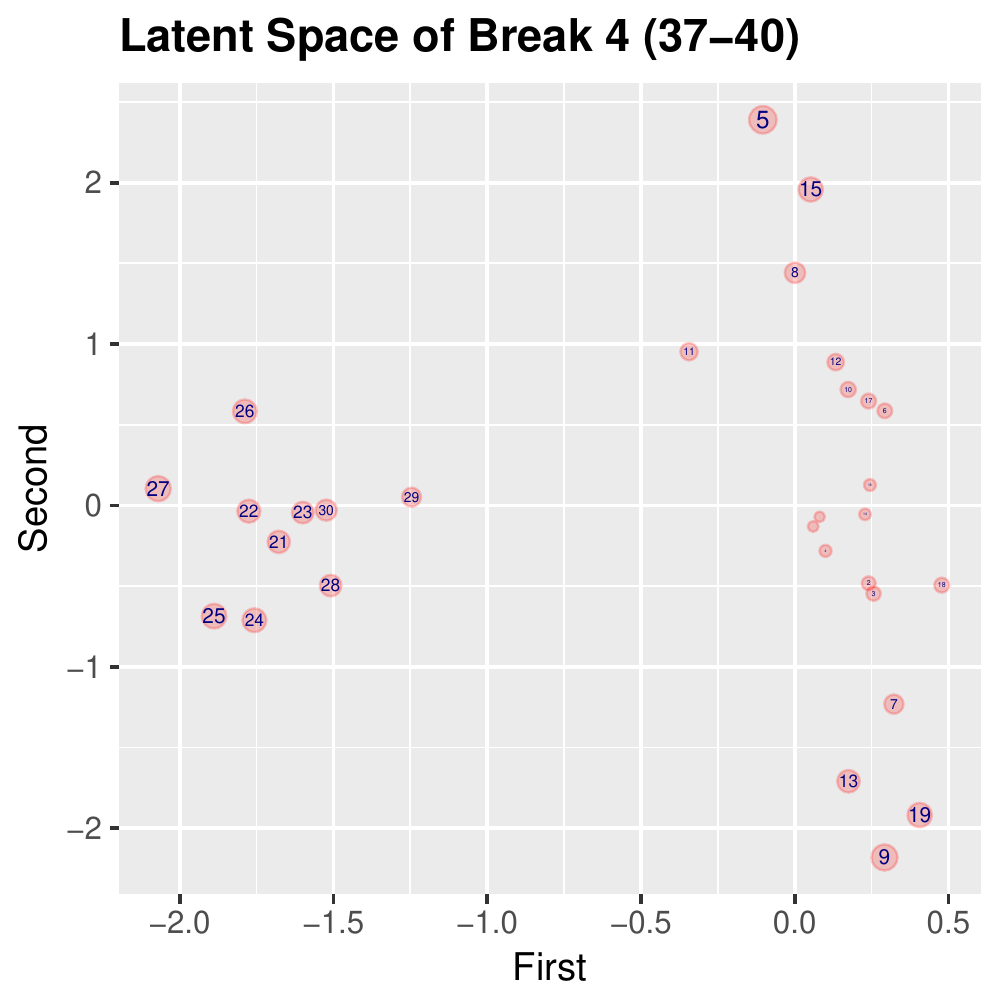}
    \end{minipage}
&
    \begin{minipage}{.18\textwidth}
      \includegraphics[width=\linewidth]{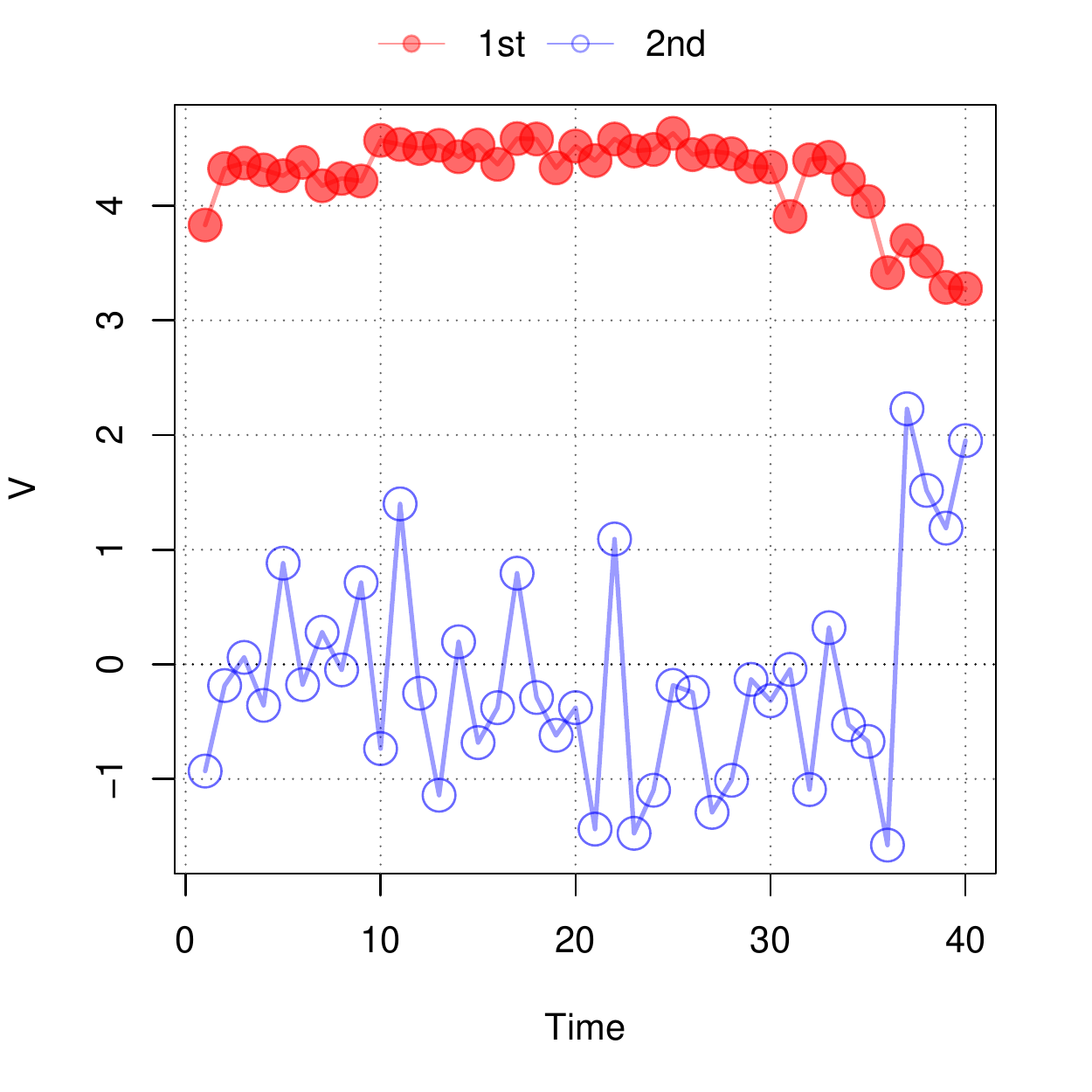}
    \end{minipage} \\
     \bottomrule
  \end{tabular}}
  \caption{Simulation Results of Block-structured Networks with No Break. The ground truth is no break and the underlying group structure is a two-group network. }\label{sim.constant}
\end{sidewaystable}


\begin{sidewaystable}\footnotesize 
  \centering
\resizebox{\textwidth}{!}{  \begin{tabular}{p{4cm} cccc c}
    \toprule
    Model Fit& \multicolumn{4}{c}{Latent Space ($\mathbf{U}_m$) Changes}&  \multicolumn{1}{c}{Generation Rule ($\mathbf{v}_t$)} \\ 
           \cline{2-5} 
    & Regime 1& Regime 2&Regime 3& Regime 4& \\\midrule
     \begin{tabular}{lcl} Break number &=& 0 \\ WAIC &=& 13744  \\ -2*log marginal &=& 13648 \\ -2*log likelihood &=& 13592 \end{tabular} 
  &
    \begin{minipage}{.18\textwidth}
      \includegraphics[width=\linewidth]{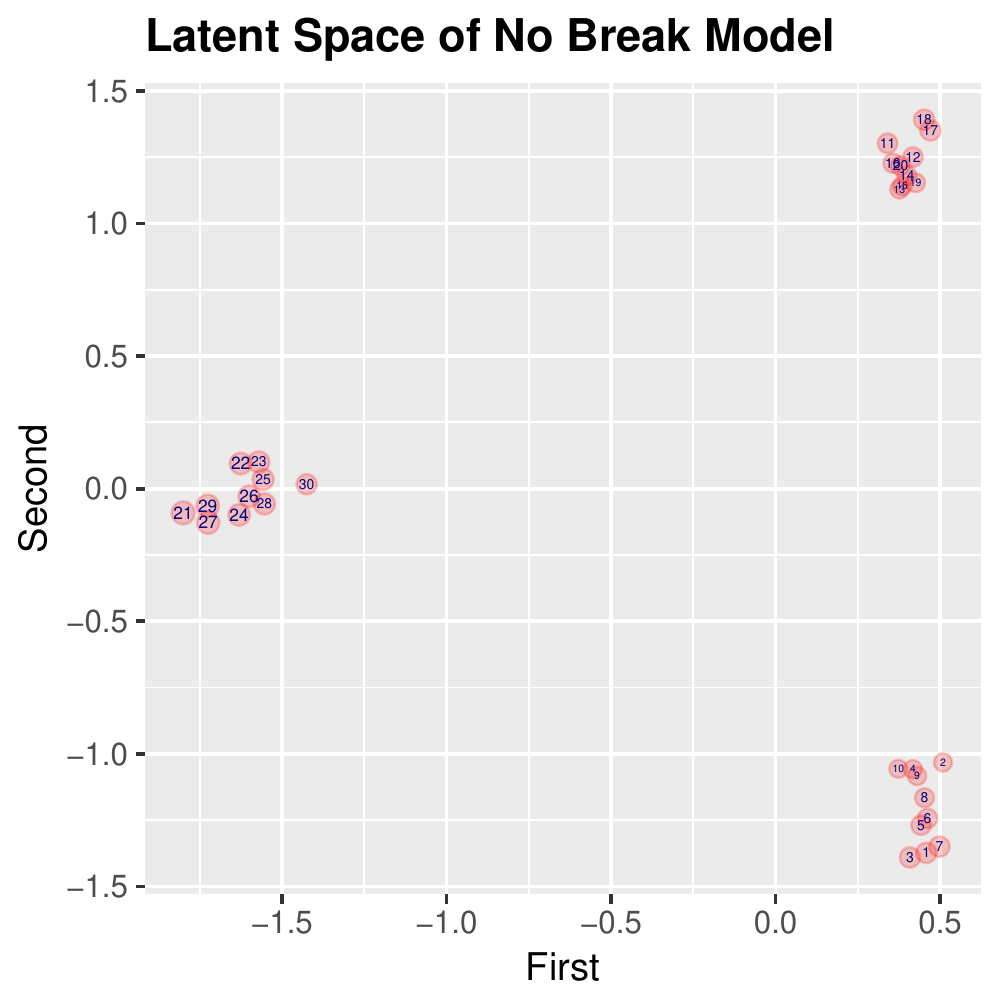}
    \end{minipage}
    &&&&
    \begin{minipage}{.18\textwidth}
      \includegraphics[width=\linewidth]{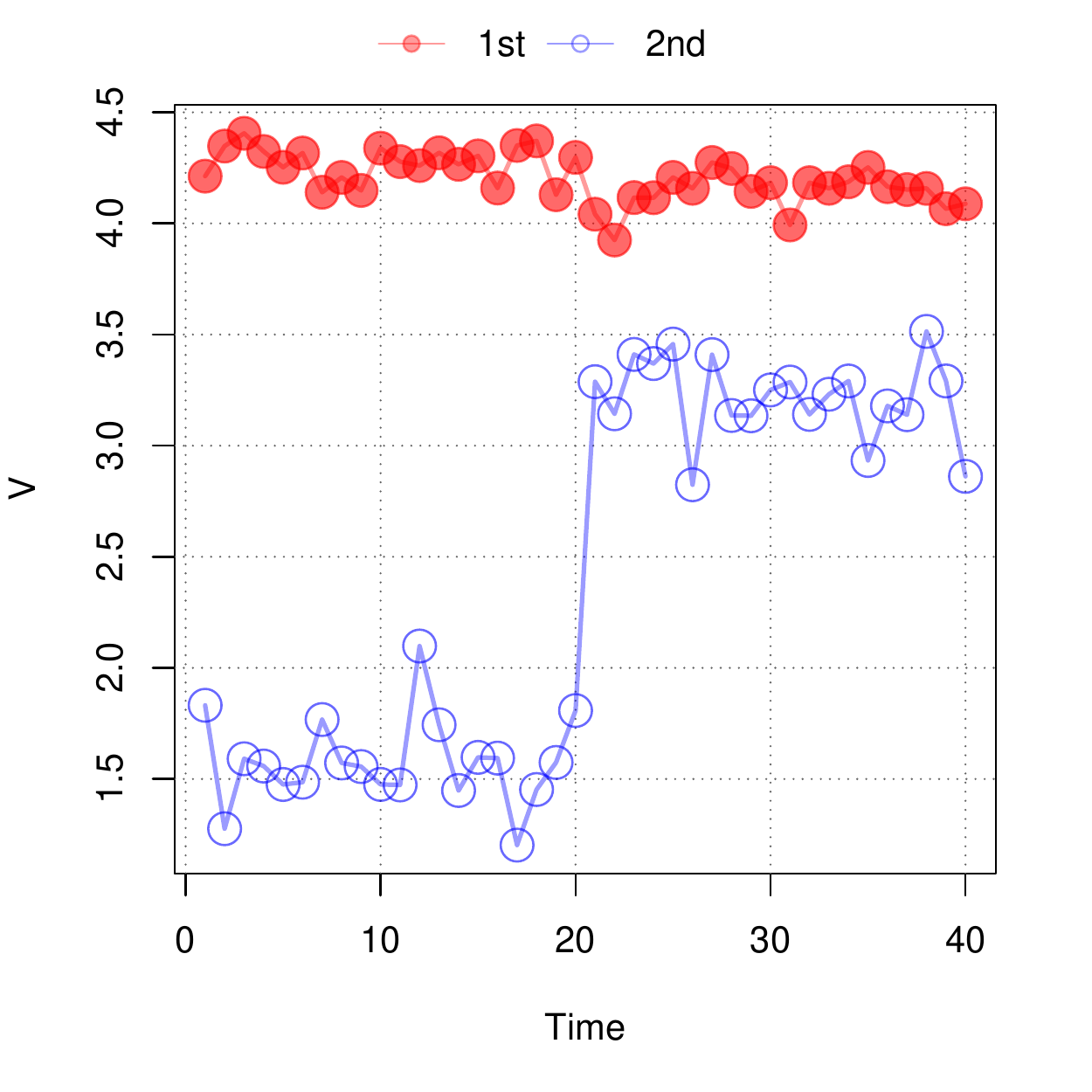}
    \end{minipage} \\
     \begin{tabular}{lcl} Break number &=& 1\\ WAIC &=& 13016  \\ -2*log marginal &=& 12922 \\ -2*log likelihood &=& 12810 \\ Average Loss &=& 0.00\end{tabular} 
&
    \begin{minipage}{.18\textwidth}
      \includegraphics[width=\linewidth]{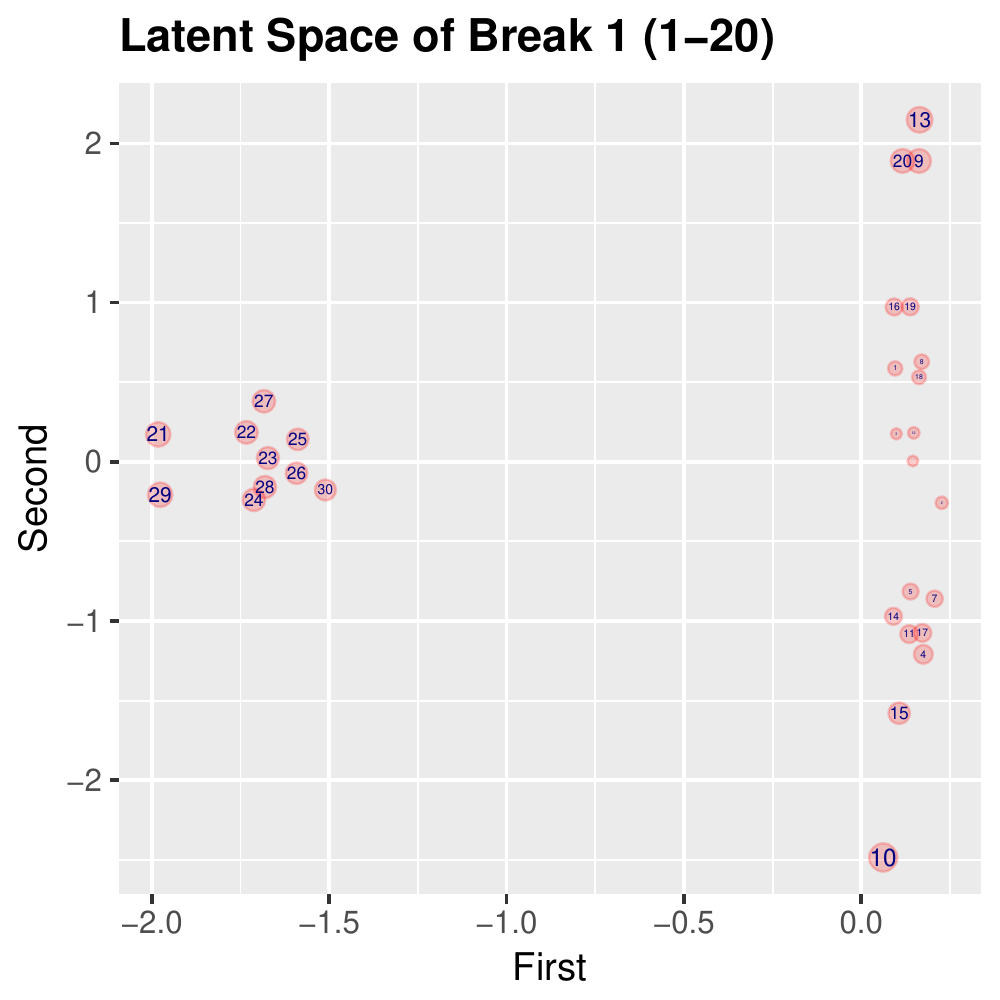}
    \end{minipage}
    &
        \begin{minipage}{.18\textwidth}
      \includegraphics[width=\linewidth]{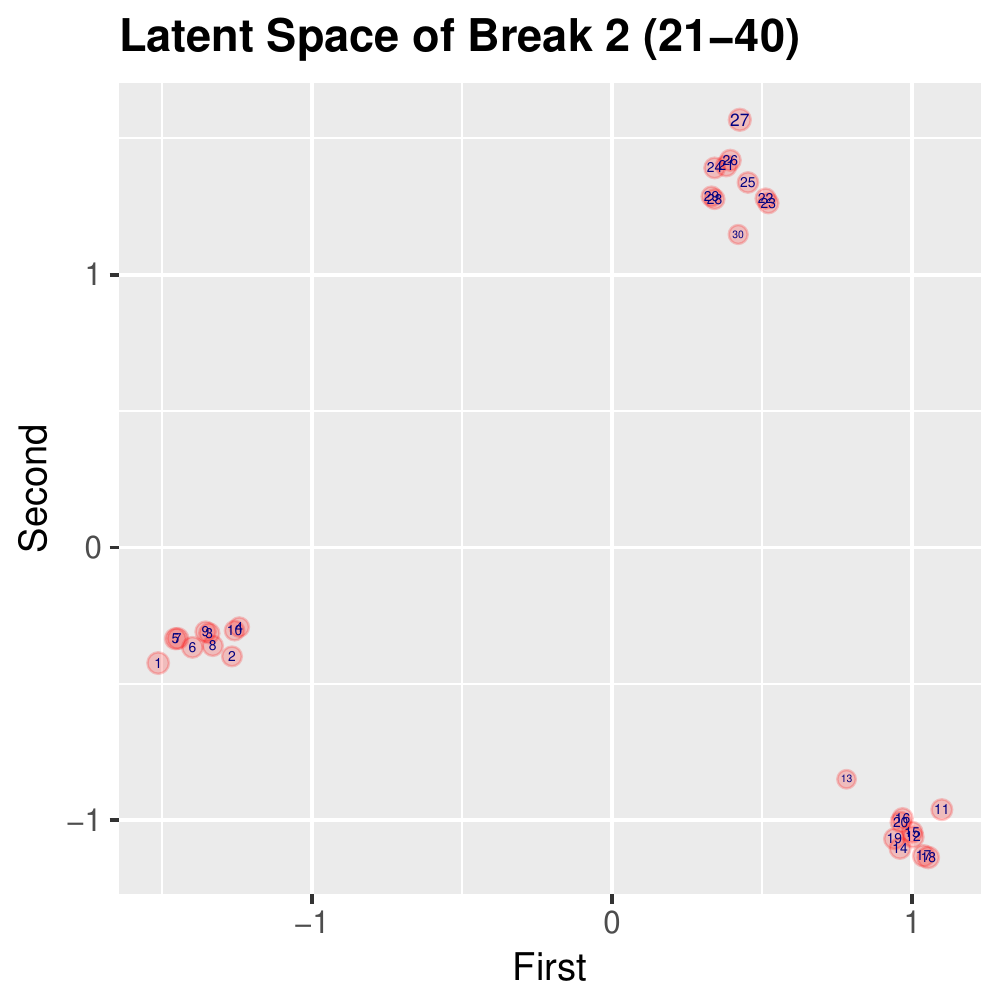}
    \end{minipage}
    &&&
    \begin{minipage}{.18\textwidth}
      \includegraphics[width=\linewidth]{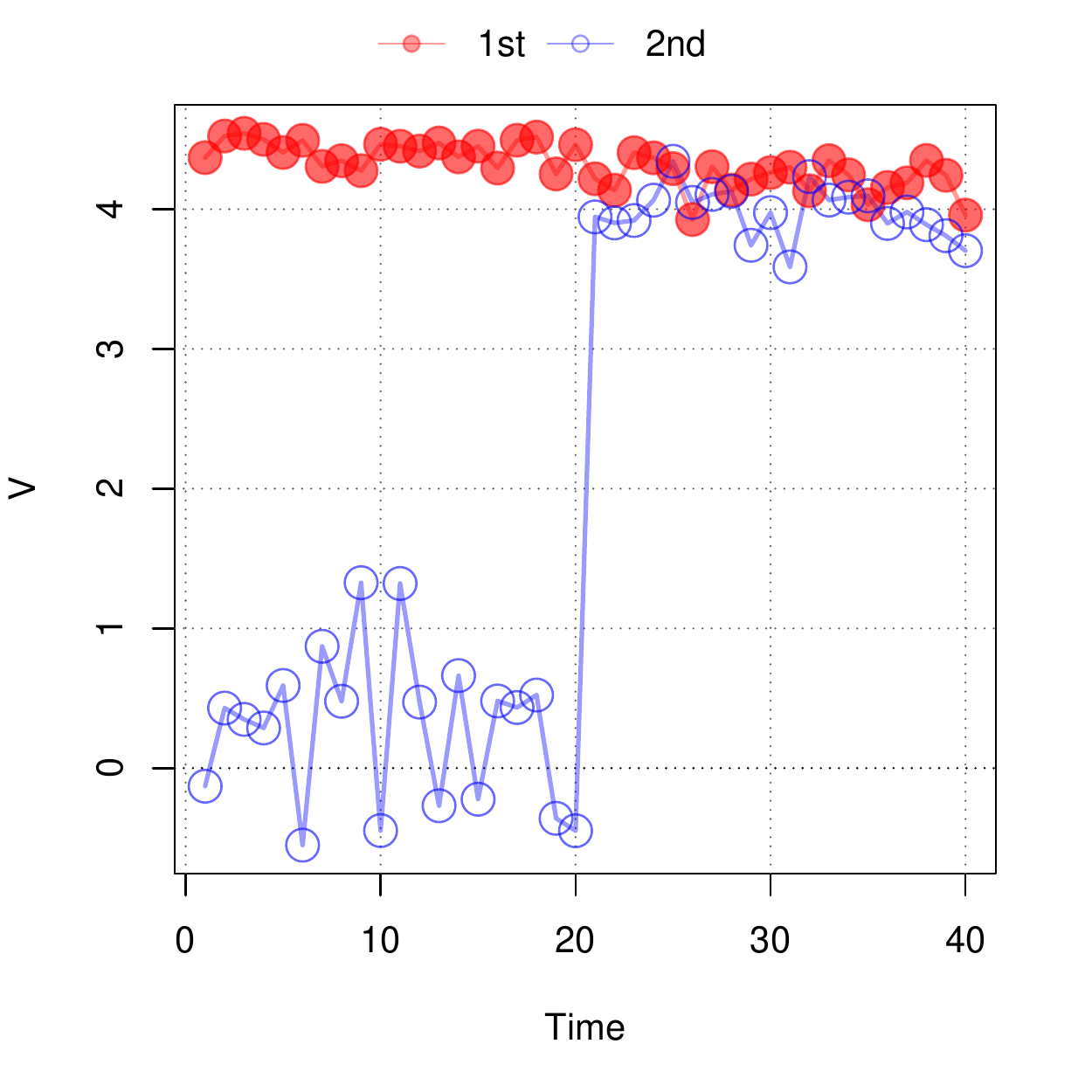}
    \end{minipage} \\
    \begin{tabular}{lcl} Break number &=& 2 \\ WAIC &=& 13067\\ -2*log marginal &=&12929  \\ -2*log likelihood &=& 12786 \\ Average Loss &=& 0.18 \end{tabular} 
	&
	 \begin{minipage}{.18\textwidth}
      \includegraphics[width=\linewidth]{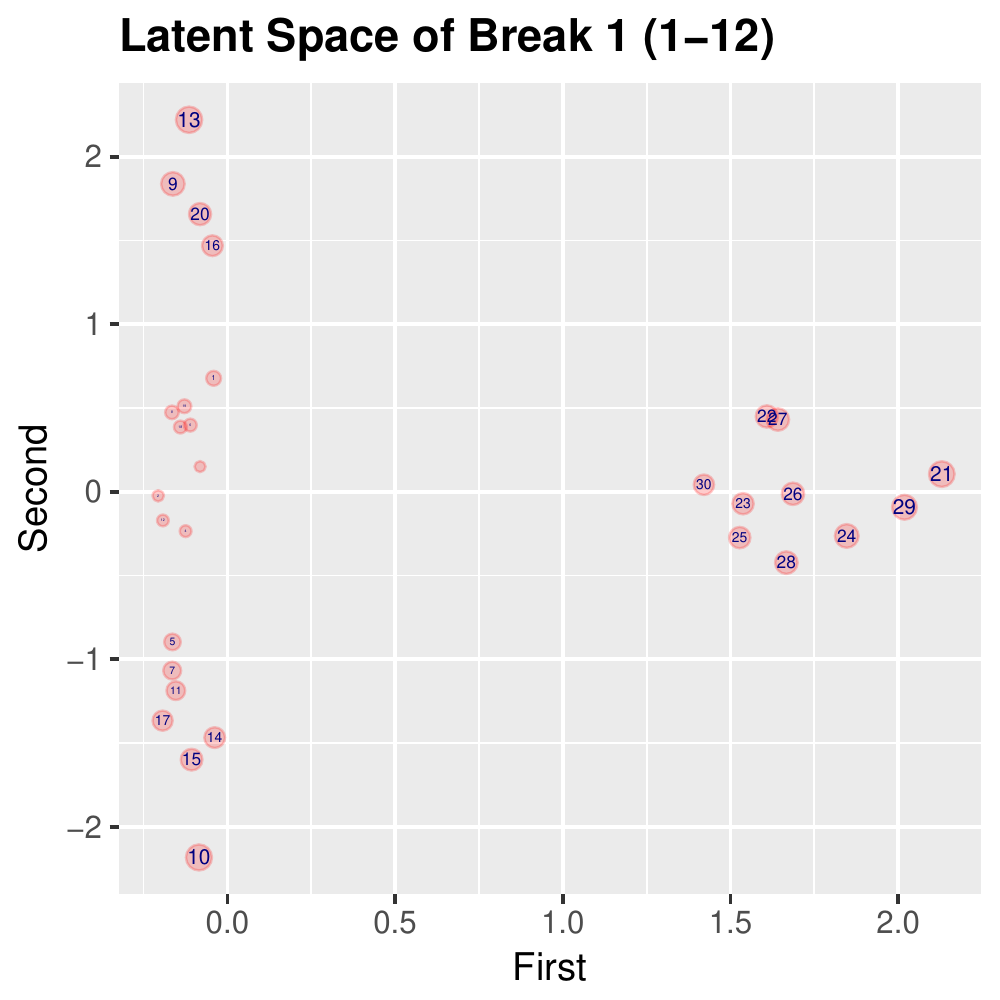}
    \end{minipage}
    &
        \begin{minipage}{.18\textwidth}
      \includegraphics[width=\linewidth]{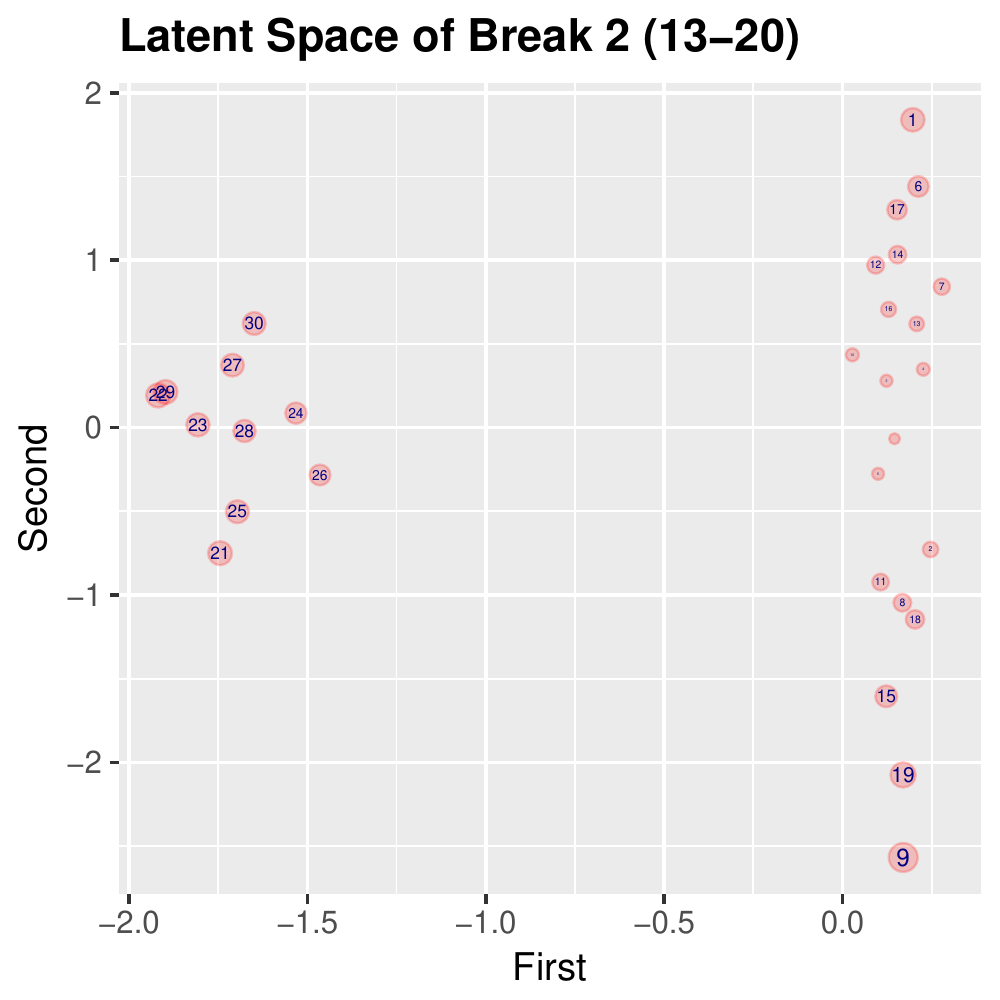}
    \end{minipage}
    &
            \begin{minipage}{.18\textwidth}
      \includegraphics[width=\linewidth]{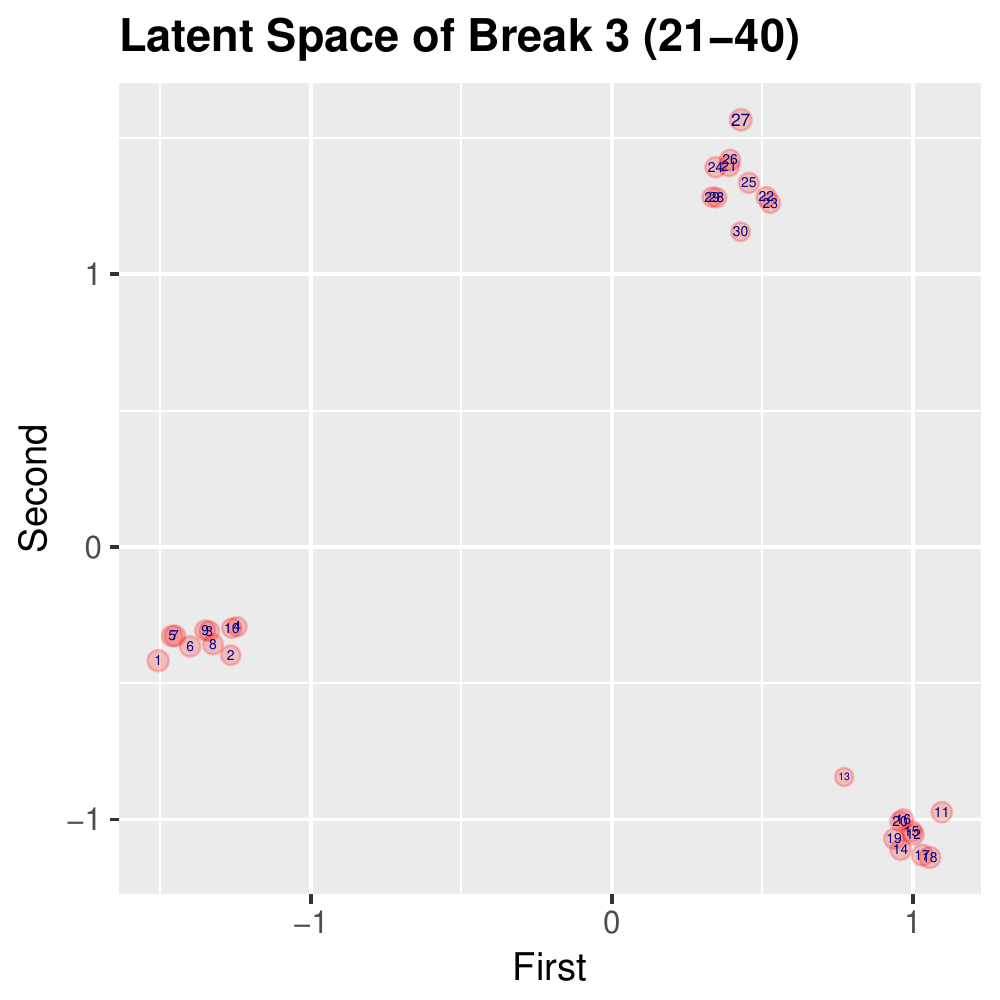}
    \end{minipage}
	&&
    \begin{minipage}{.18\textwidth}
      \includegraphics[width=\linewidth]{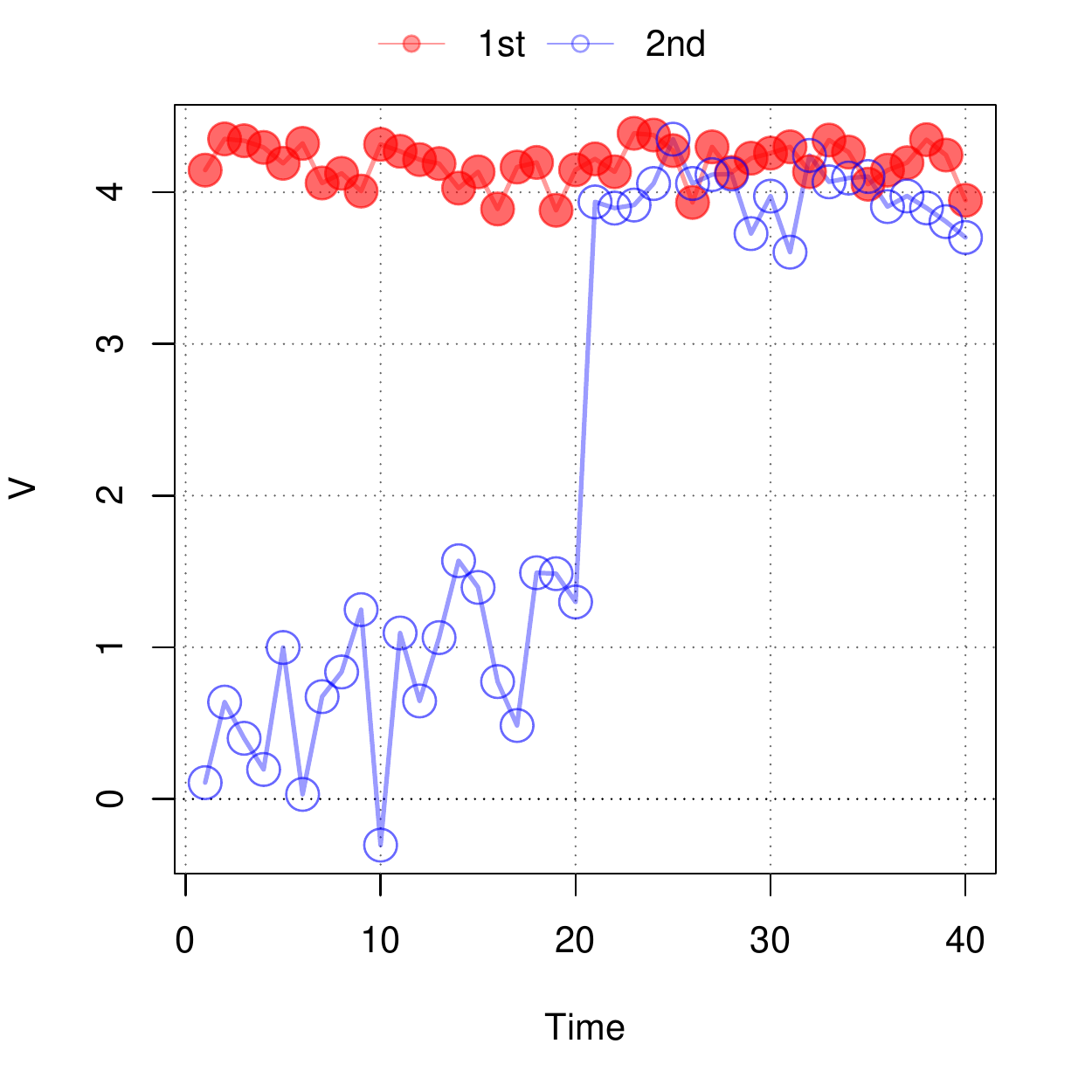}
    \end{minipage} \\
    \begin{tabular}{lcl } Break number &=& 3 \\ WAIC &=& 13064  \\ -2*log marginal &=&12877 \\ -2*log likelihood &=& 12738  \\ Average Loss &=& 0.49\end{tabular} 
  &
	\begin{minipage}{.18\textwidth}
      \includegraphics[width=\linewidth]{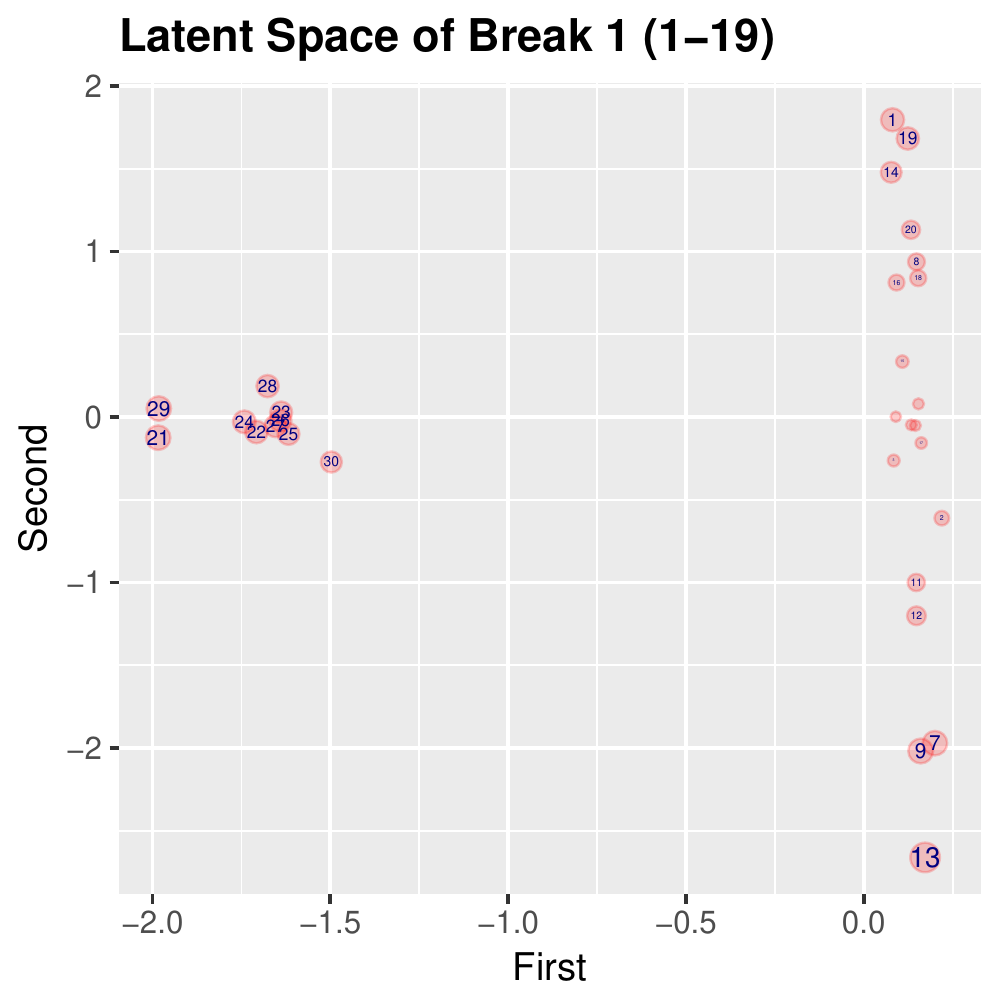}
    \end{minipage}
    &
        \begin{minipage}{.18\textwidth}
      \includegraphics[width=\linewidth]{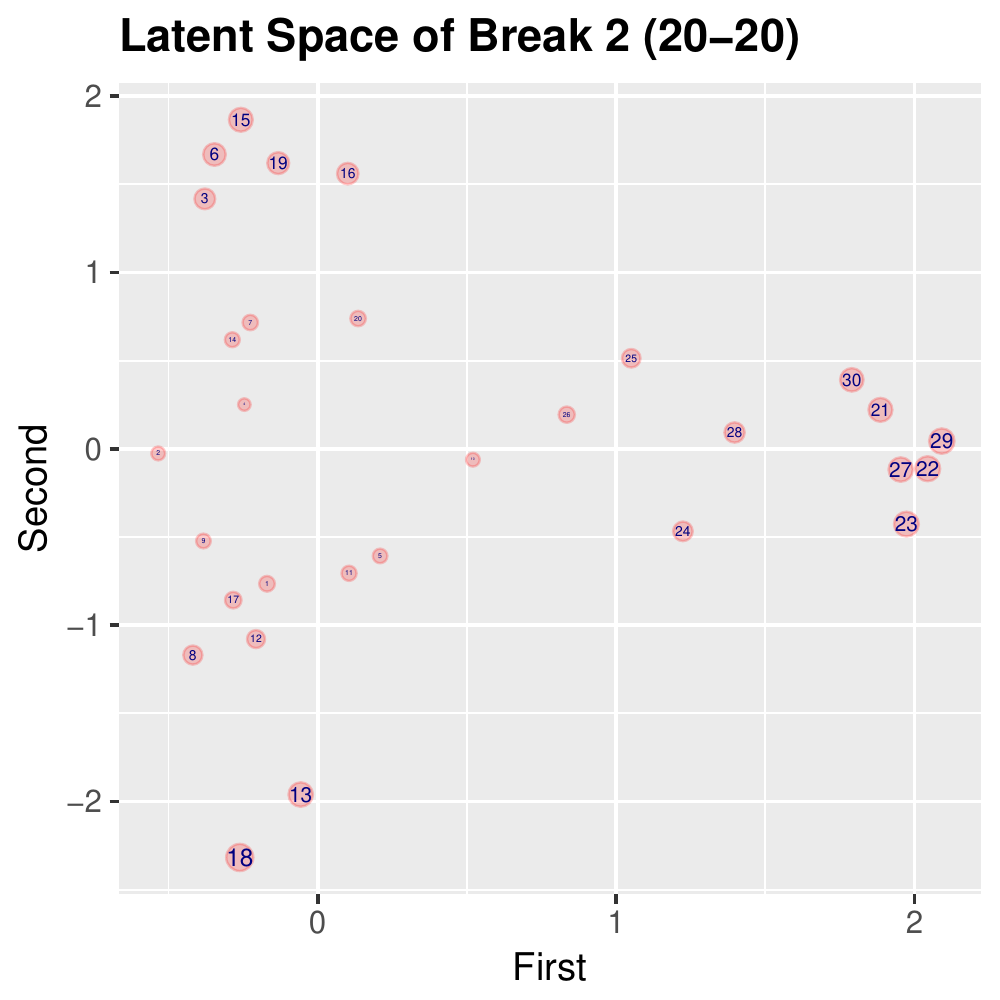}
    \end{minipage}
    &
            \begin{minipage}{.18\textwidth}
      \includegraphics[width=\linewidth]{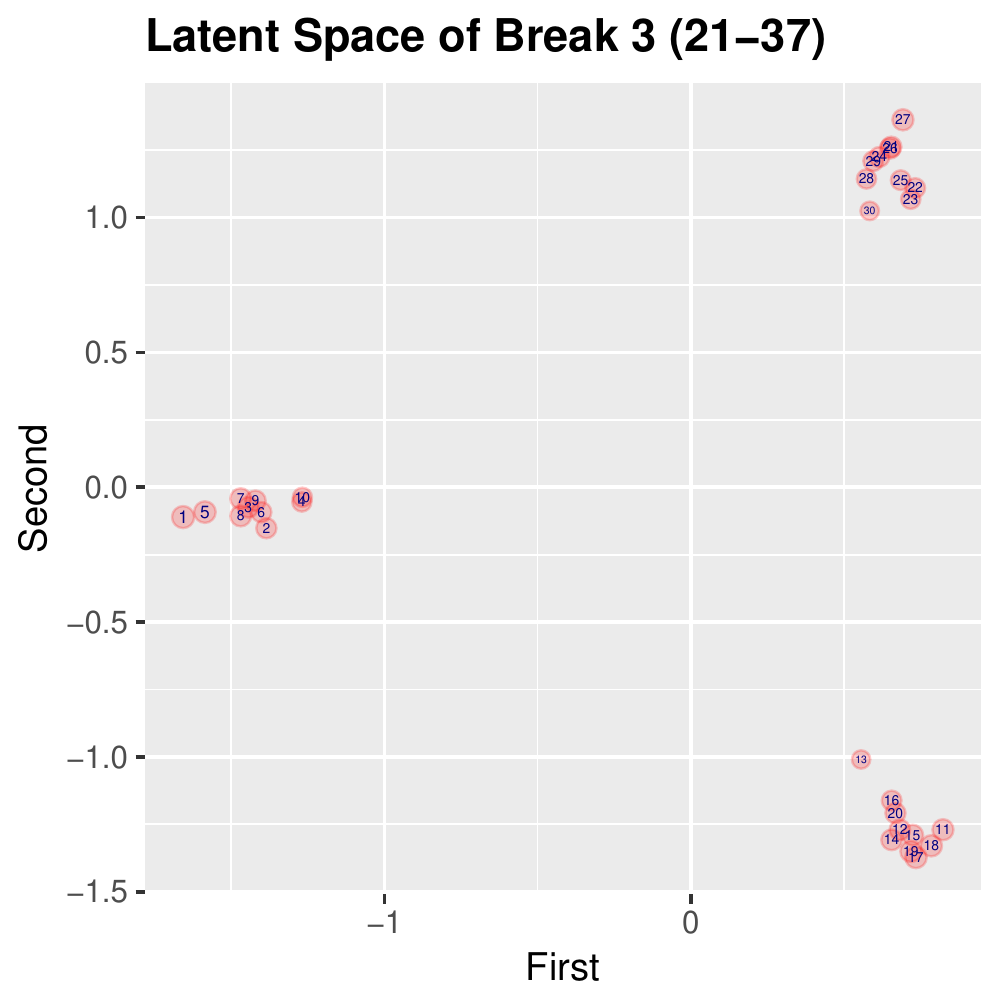}
    \end{minipage}
	&
	           \begin{minipage}{.18\textwidth}
      \includegraphics[width=\linewidth]{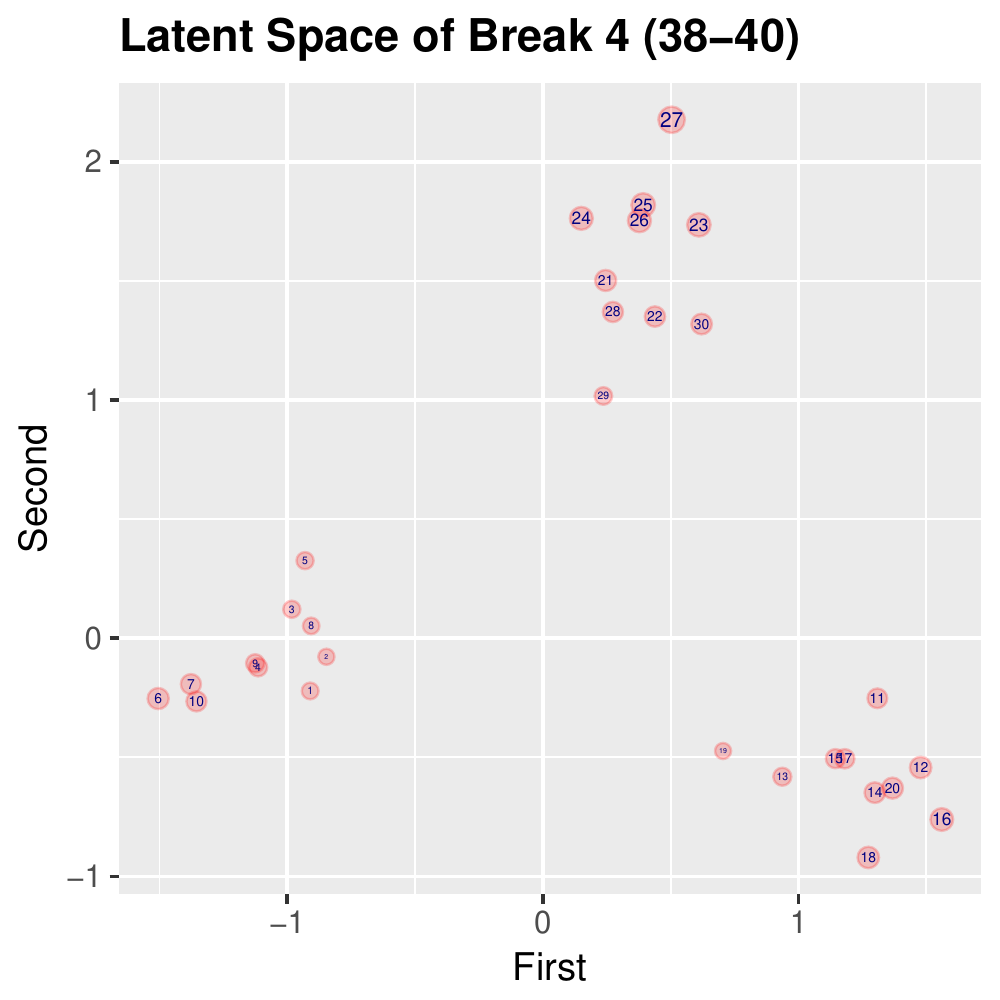}
    \end{minipage}
&
    \begin{minipage}{.18\textwidth}
      \includegraphics[width=\linewidth]{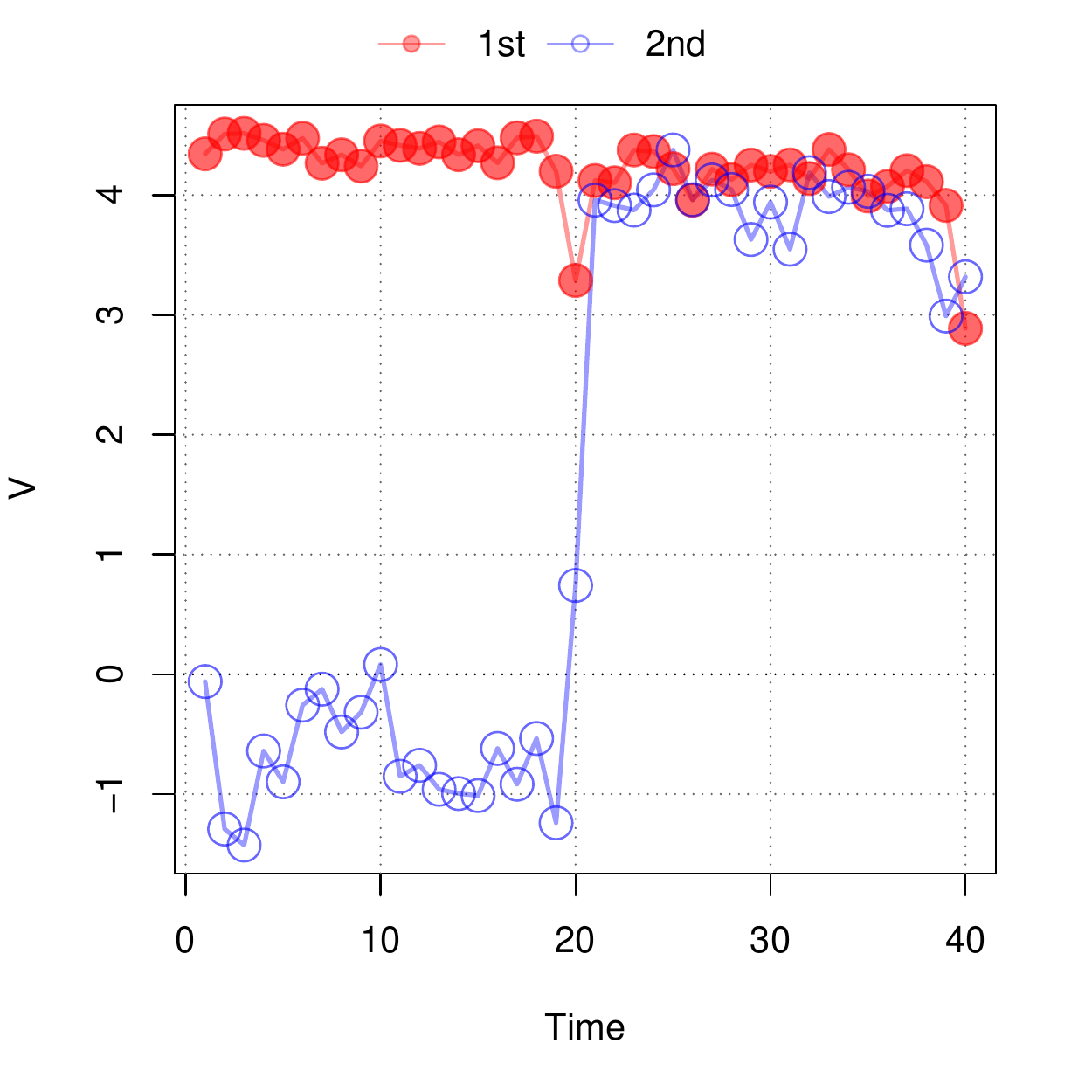}
    \end{minipage} \\
     \bottomrule
  \end{tabular}}
  \caption{Simulation Results of Block-structured Networks with a Block-splitting Break. The ground truth is one break and the underlying group structure changes from a two group structure to a three group structure in the middle. }\label{sim.split}
\end{sidewaystable}

\subsection{Block-structured Networks with No Break}
We first check how sensitive our proposed method is to the false positive bias (i.e. detecting breaks when there is no break). We generate block-structured networks with no break.\footnote{More specifically, the longitudinal network was generated from a two group structure, consisting of a 10 node group and a 20 node group respectively, and the underlying group structure remains constant.} Table \ref{sim.constant} summarizes the results from the no break case. 

The reading of the results starts from model diagnostics in the first column. While the approximate log marginal likelihood incorrectly favors the one break model ($\mathcal{M}_1$), WAIC correctly shows that the no break model ($\mathcal{M}_0$) fits the data best. Also, the average loss of break points shows somewhat unstable movements. On average, simulated break points of the one break model swing $\pm 0.57$ around the estimated break point and simulated break points of the two break model swing $\pm 1.48$ around the estimated break points.

If we look at the estimated latent space of the no break model, it correctly recovers the latent two-group structure while the one break model identifies another latent state almost identical to the first one. The approximate  log marginal likelihood fails to penalize the recovery of the redundant latent state. However, as we add more breaks, the approximate  log marginal likelihood successfully penalizes the model for the existence of redundant states in the two break model and in the three break model. 

\subsection{Block-structured Networks with a Block-splitting Break}
Now, we move the case of dynamic network data with a single group-splitting break. That is, the ground truth is that the number of latent groups changes from 2 to 3 in the middle. Table \ref{sim.split} shows the results of the simulation. Again, we start to read from the model diagnostic results in the first column. 

WAIC correctly identifies the single break model as the best-fitting model while the approximate log marginal likelihood favors the three break HMTM.  As we have seen in the previous example, the approximate log marginal likelihood fails to penalize the model with redundant breaks. The source of the problem is the singleton state (a latent state consisting of a single observation). A similar problem has been noticed in finite mixture models with singular components \citep{Hartigan1985, Bishop2006}.  In the three break model, the second state has only one observation, which increases the log likelihood dramatically. Since a singleton state is highly unlikely in reality, researchers can ignore false diagnostic results simply by checking the existence of singleton states.\footnote{\cite{Chib1995}'s algorithm is based on the summation of log likelihoods evaluated at posterior means and hence sensitive to the presence of singleton states in high dimensional time series data. In contrast, WAIC relies on  the log pointwise predictive density as a measure of the goodness of the fit and its variance as a penalty. Since the log pointwise predictive density is averaged over the entire MCMC scan ($\frac{1}{G} \sum_{g=1}^{G}p(\B_t^{\text{upper}}|\mathbf{\Theta}^{(g)}, \P^{(g)}, \mathcal{M}_M)$), it is less sensitive to singular components in high dimensional mixture models like HMTM. This is why WAIC outperforms in the break number detection in the context of HMTM.} Interestingly, the network generation rule in the last column shows almost identical patterns, regardless of the number of imposed breaks. Note that the second dimensional network generation rule ($v_2$) jumps to a large positive number in the middle as the number of groups increases from 2 to 3. 

The average loss of break points clearly favors the one break model. Adding redundant breaks increases the average loss of break points significantly. For example, simulated break points of the three break model swing $\pm 1.4$ around the estimated break points on average while simulated break points of the one break model stay constant.


\begin{sidewaystable}\footnotesize 
  \centering
\resizebox{\textwidth}{!}{  \begin{tabular}{p{4cm} cccc c}
    \toprule
    Model Fit& \multicolumn{4}{c}{Latent Space ($\mathbf{U}_m$) Changes}&  \multicolumn{1}{c}{Generation Rule ($\mathbf{v}_t$)} \\ 
           \cline{2-5} 
    & Regime 1& Regime 2&Regime 3& Regime 4& \\\midrule
     \begin{tabular}{lcl } Break number &=&0 \\ WAIC &=& 13727  \\ -2*log marginal &=& 13630 \\ -2*log likelihood &=& 13574 \end{tabular} 
  &
    \begin{minipage}{.18\textwidth}
      \includegraphics[width=\linewidth]{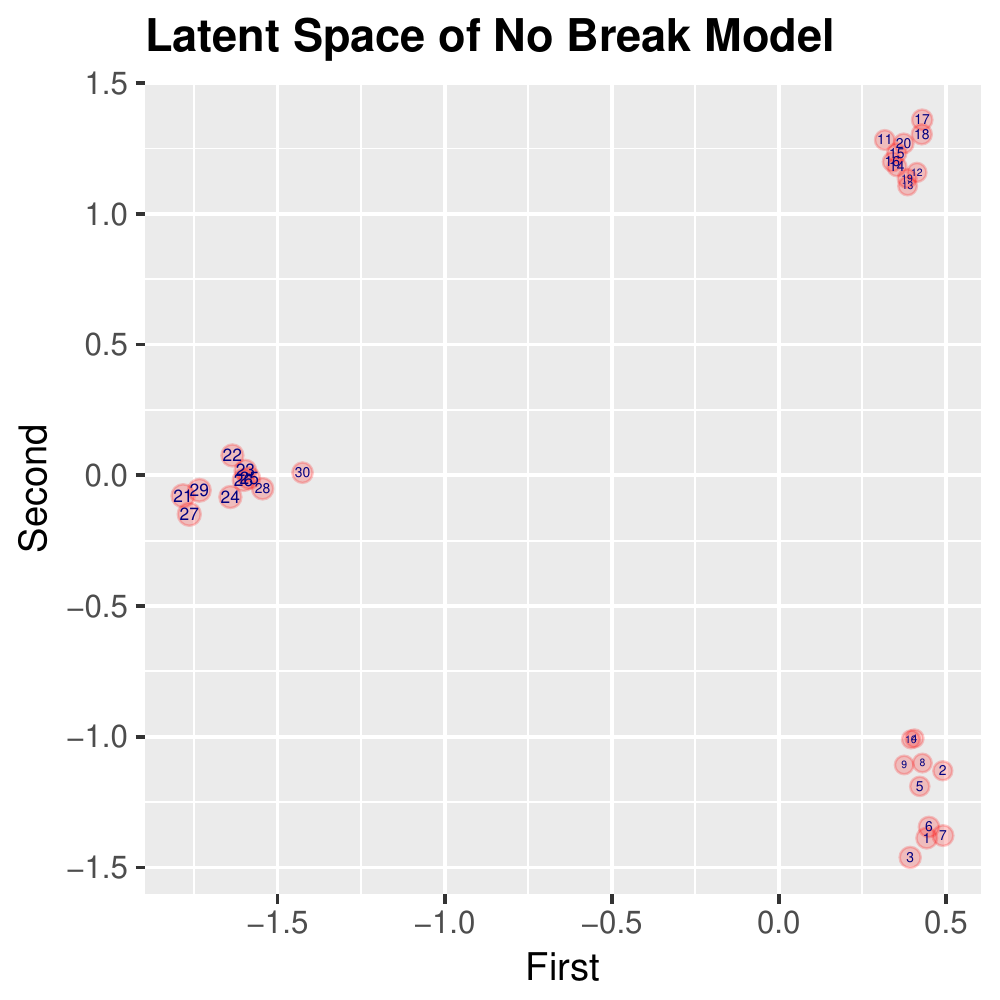}
    \end{minipage}
    &&&&
    \begin{minipage}{.18\textwidth}
      \includegraphics[width=\linewidth]{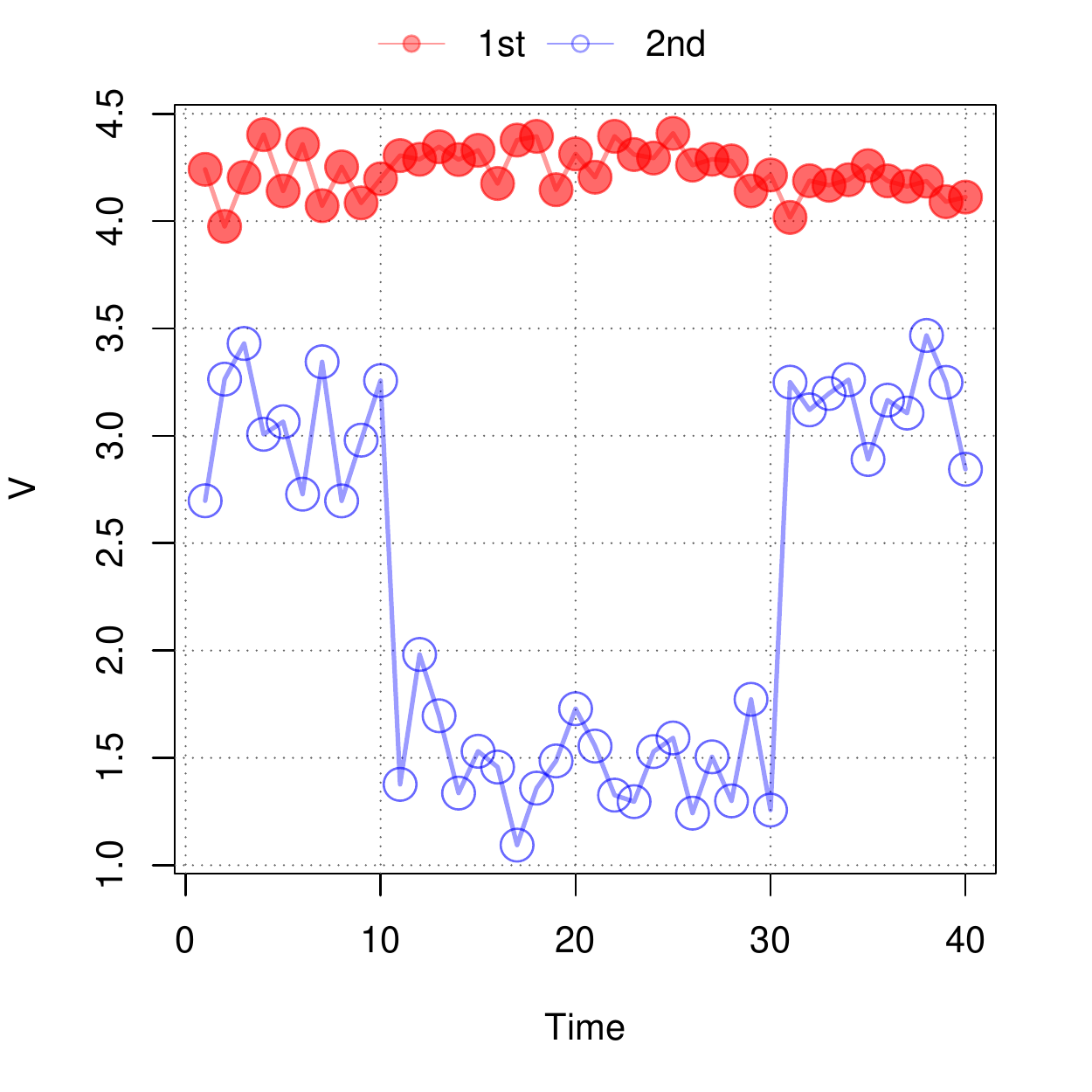}
    \end{minipage} \\
     \begin{tabular}{lcl } Break number &=&1\\ WAIC &=& 13583  \\ -2*log marginal &=& 13431\\ -2*log likelihood &=& 13324  \\ Average Loss &=& 0.27 \end{tabular} 
&
    \begin{minipage}{.18\textwidth}
      \includegraphics[width=\linewidth]{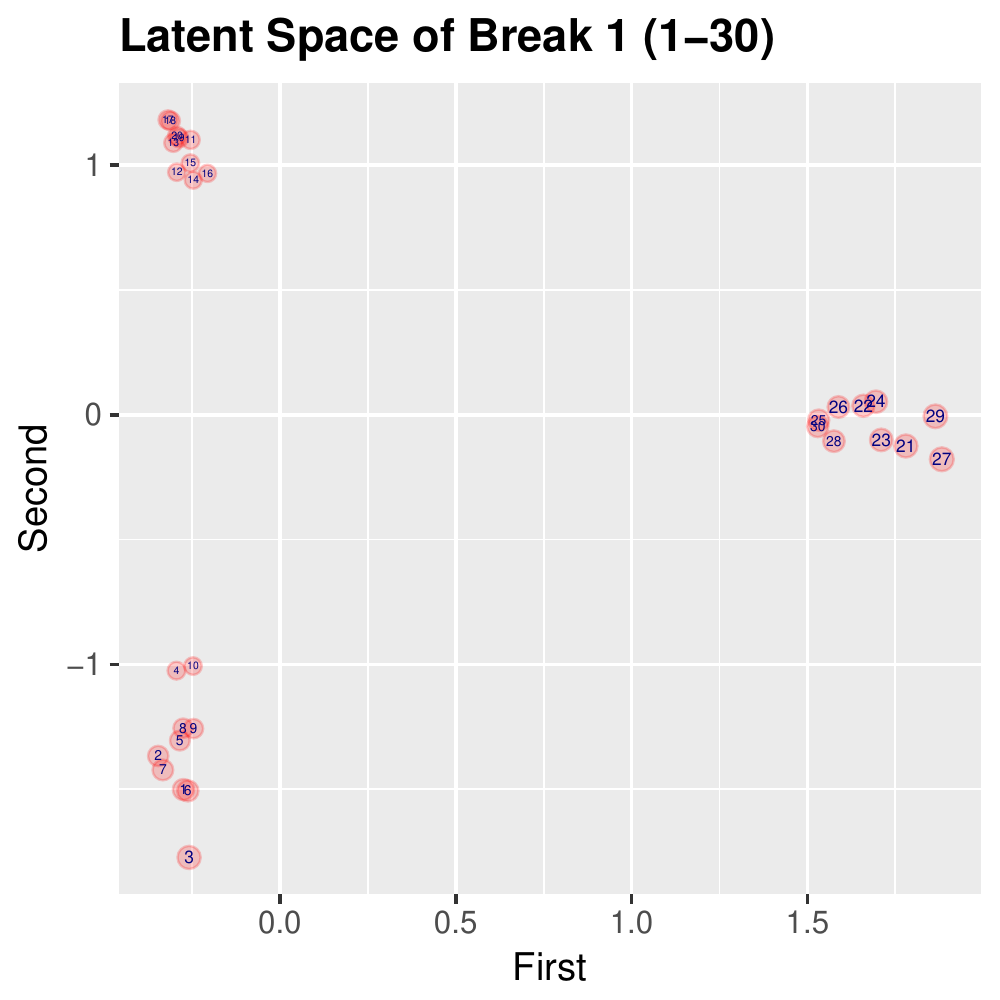}
    \end{minipage}
    &
        \begin{minipage}{.18\textwidth}
      \includegraphics[width=\linewidth]{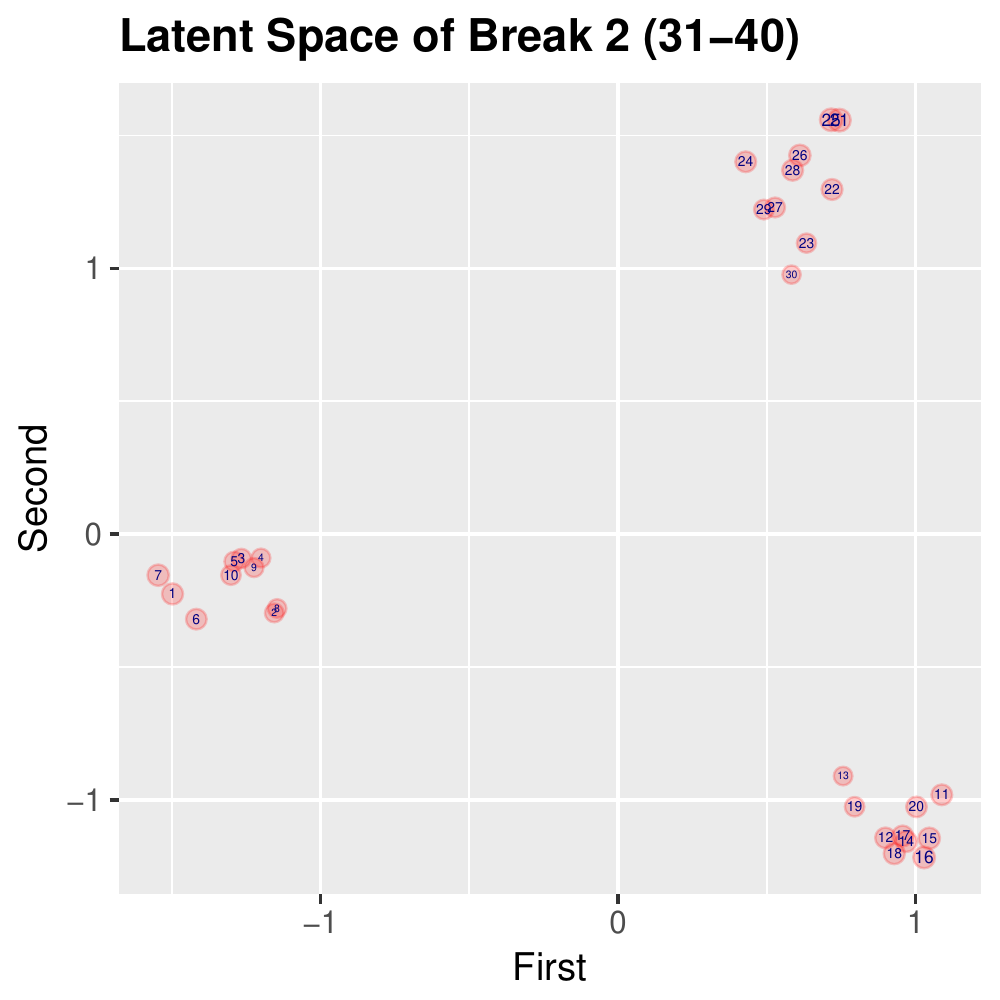}
    \end{minipage}
    &&&
    \begin{minipage}{.18\textwidth}
      \includegraphics[width=\linewidth]{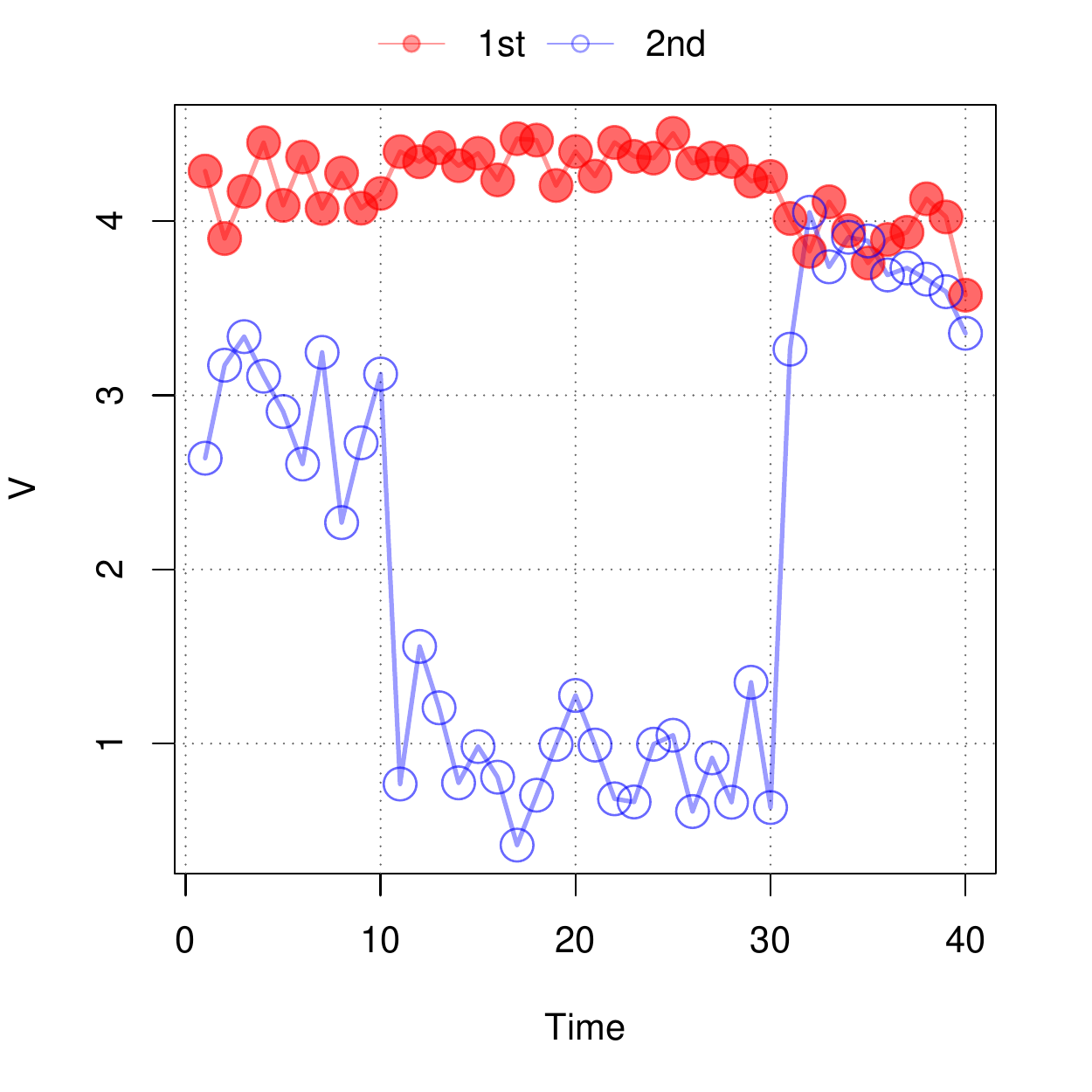}
    \end{minipage} \\
    \begin{tabular}{lcl } Break number &=&2 \\ WAIC &=& 13202\\ -2*log marginal &=&13043  \\ -2*log likelihood &=& 12892  \\ Average Loss &=& 0.00\end{tabular} 
	&
	 \begin{minipage}{.18\textwidth}
      \includegraphics[width=\linewidth]{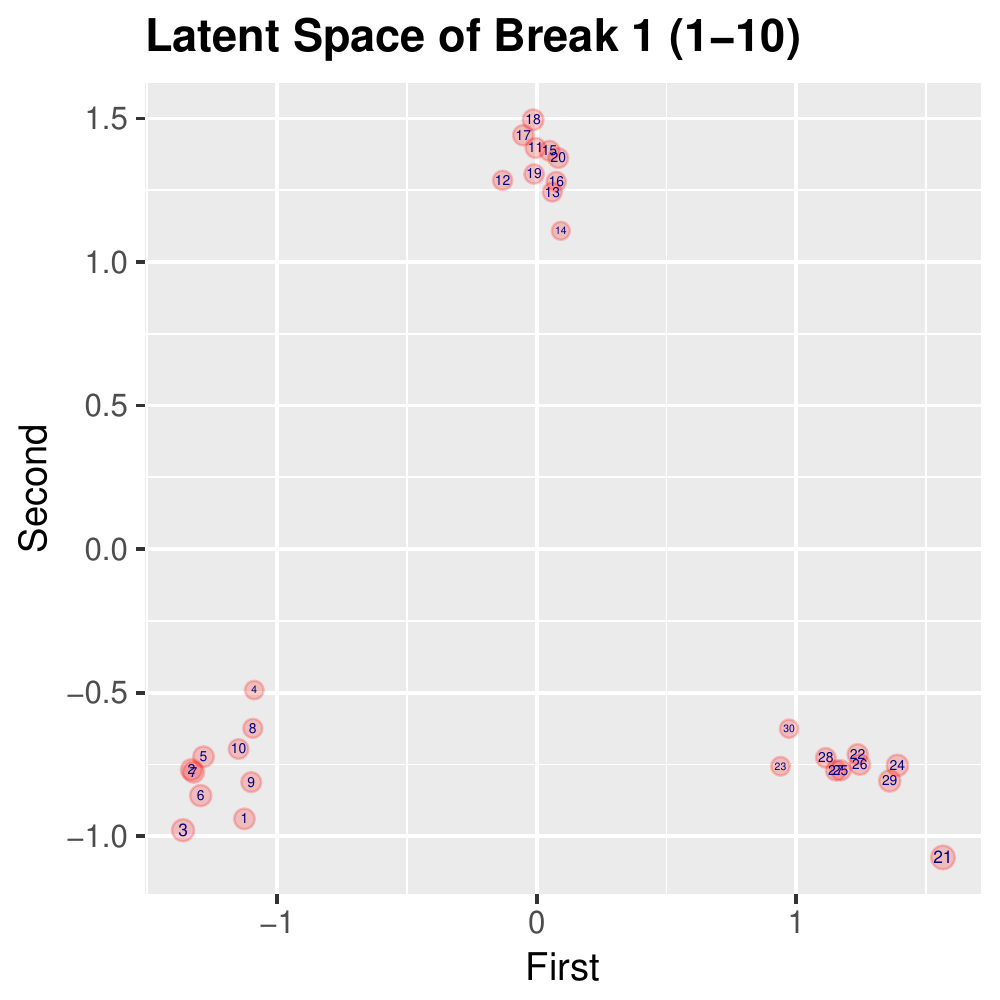}
    \end{minipage}
    &
        \begin{minipage}{.18\textwidth}
      \includegraphics[width=\linewidth]{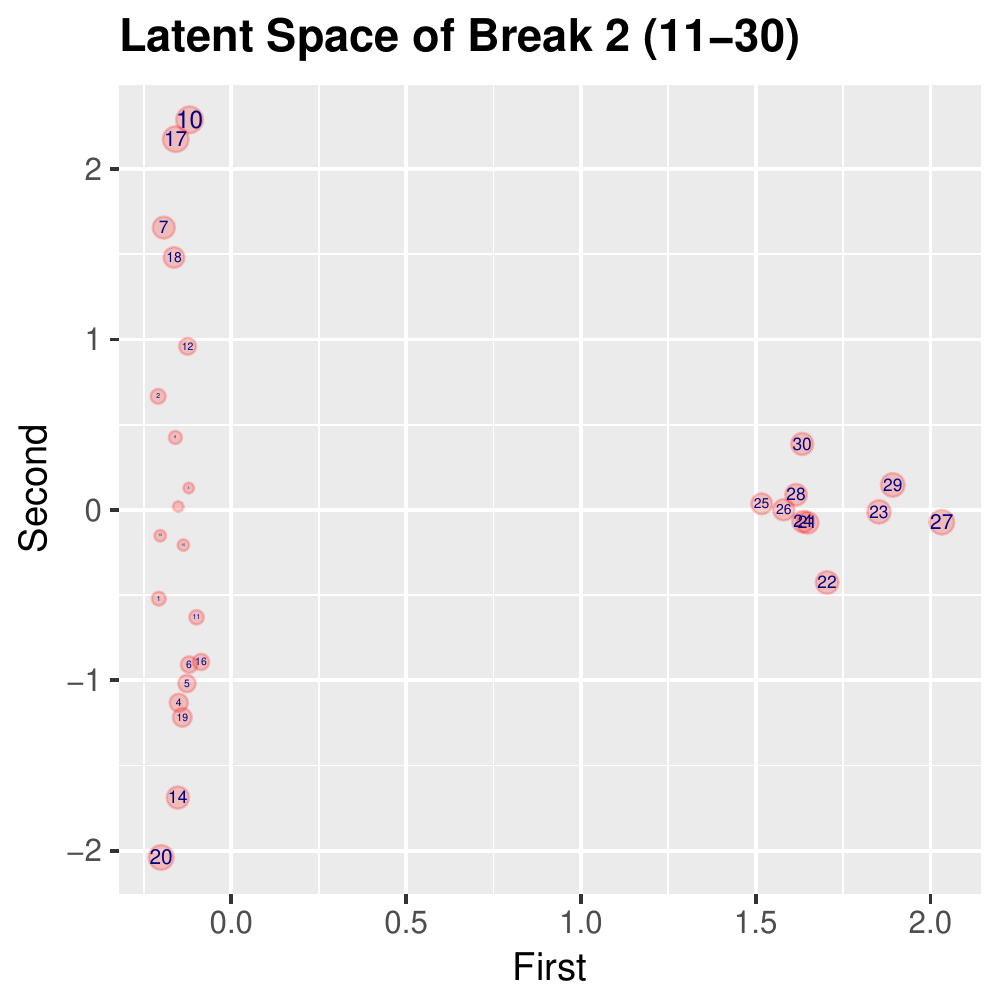}
    \end{minipage}
    &
            \begin{minipage}{.18\textwidth}
      \includegraphics[width=\linewidth]{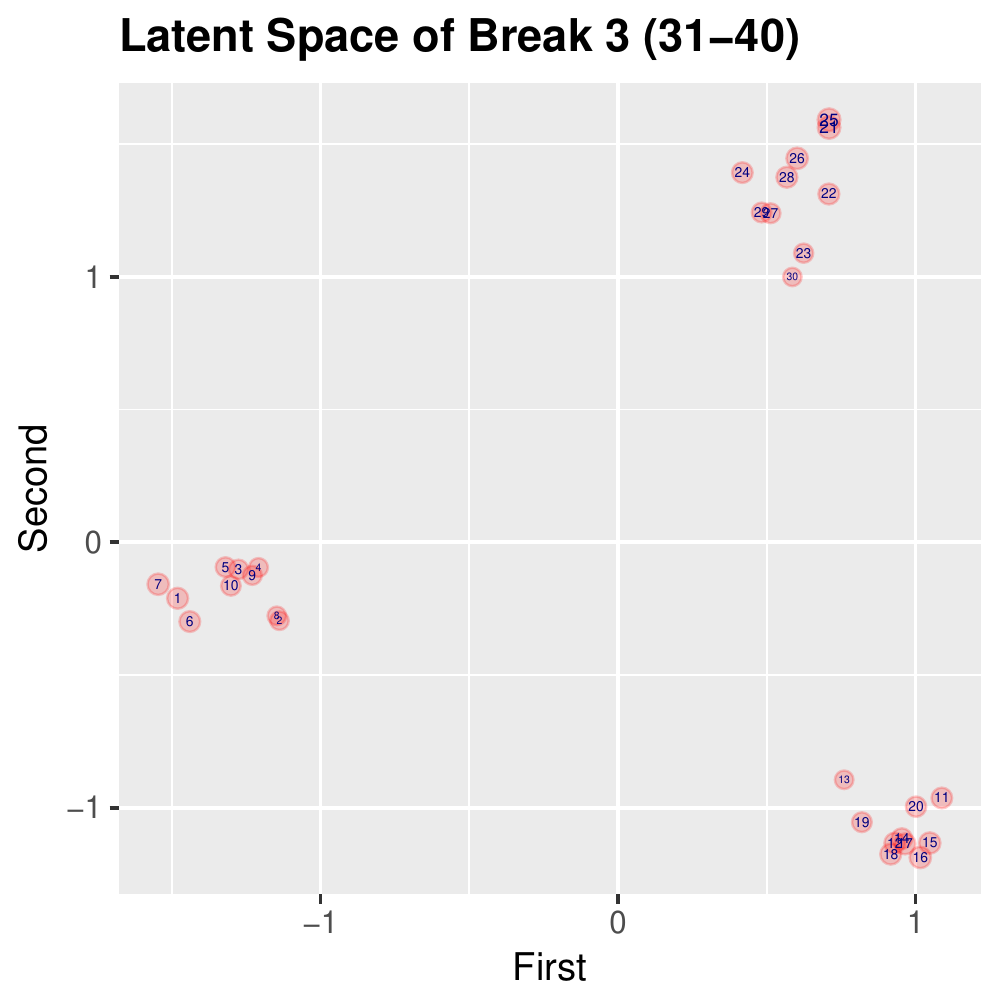}
    \end{minipage}
	&&
    \begin{minipage}{.18\textwidth}
      \includegraphics[width=\linewidth]{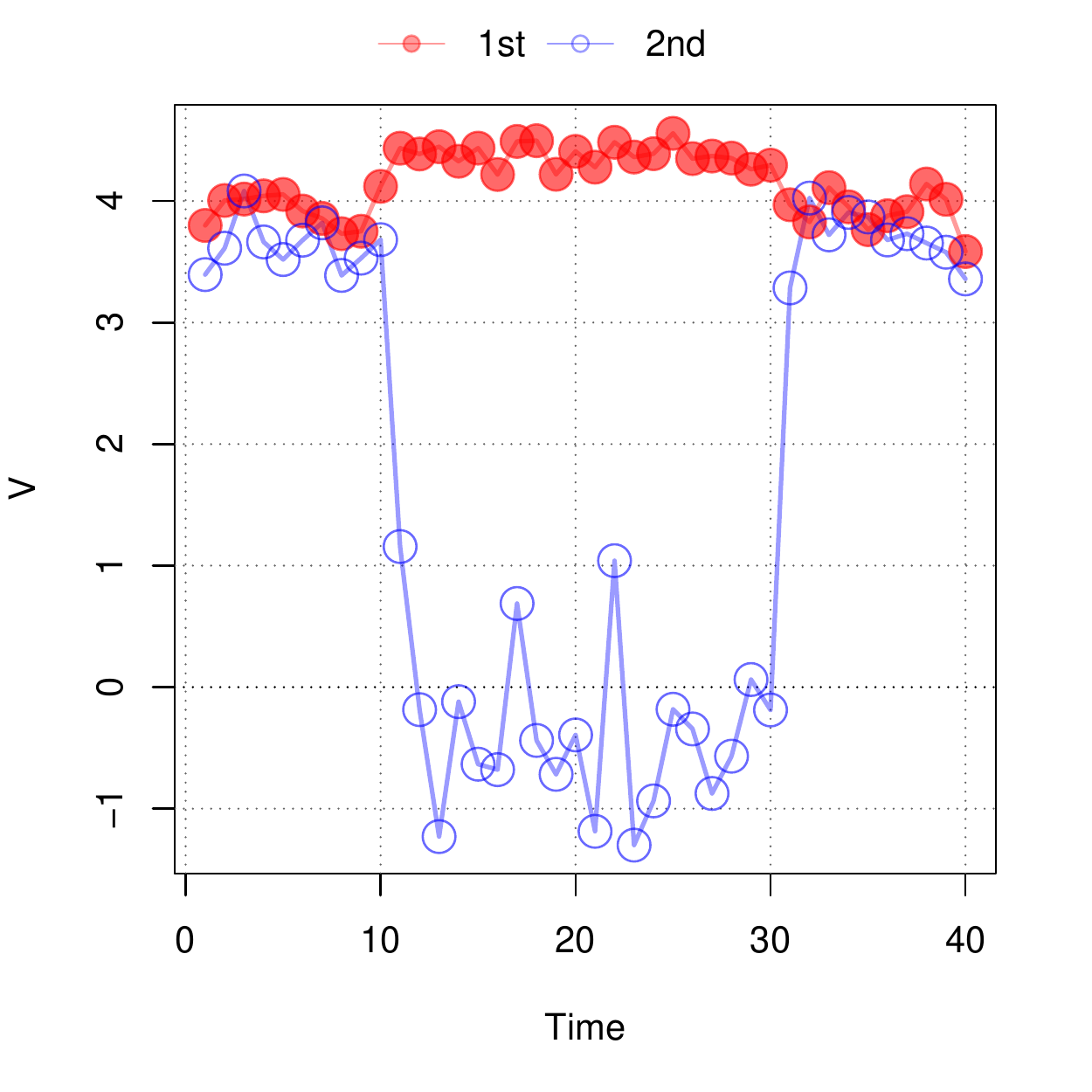}
    \end{minipage} \\
    \begin{tabular}{lcl } Break number &=&3 \\ WAIC &=& 13203 \\ -2*log marginal &=&13028 \\ -2*log likelihood &=& 12856  \\ Average Loss &=& 1.59 \end{tabular} 
  &
	\begin{minipage}{.18\textwidth}
      \includegraphics[width=\linewidth]{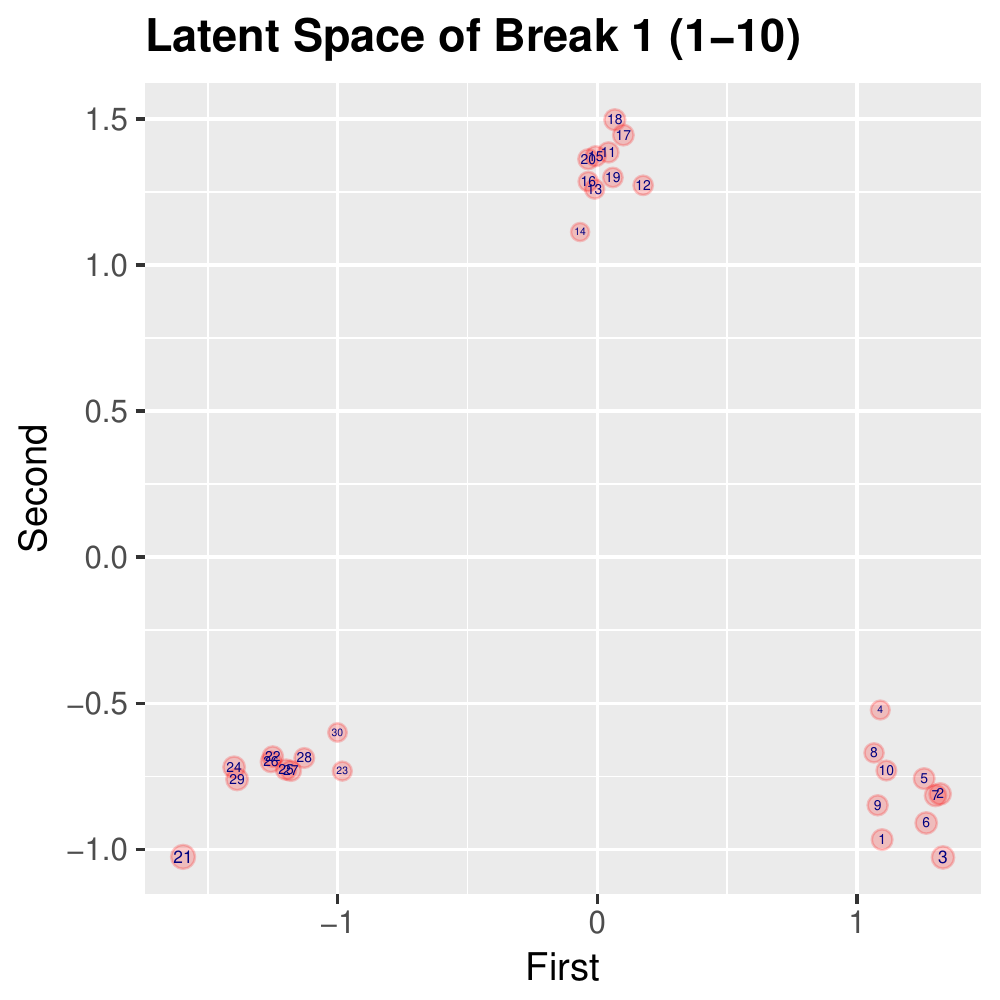}
    \end{minipage}
    &
        \begin{minipage}{.18\textwidth}
      \includegraphics[width=\linewidth]{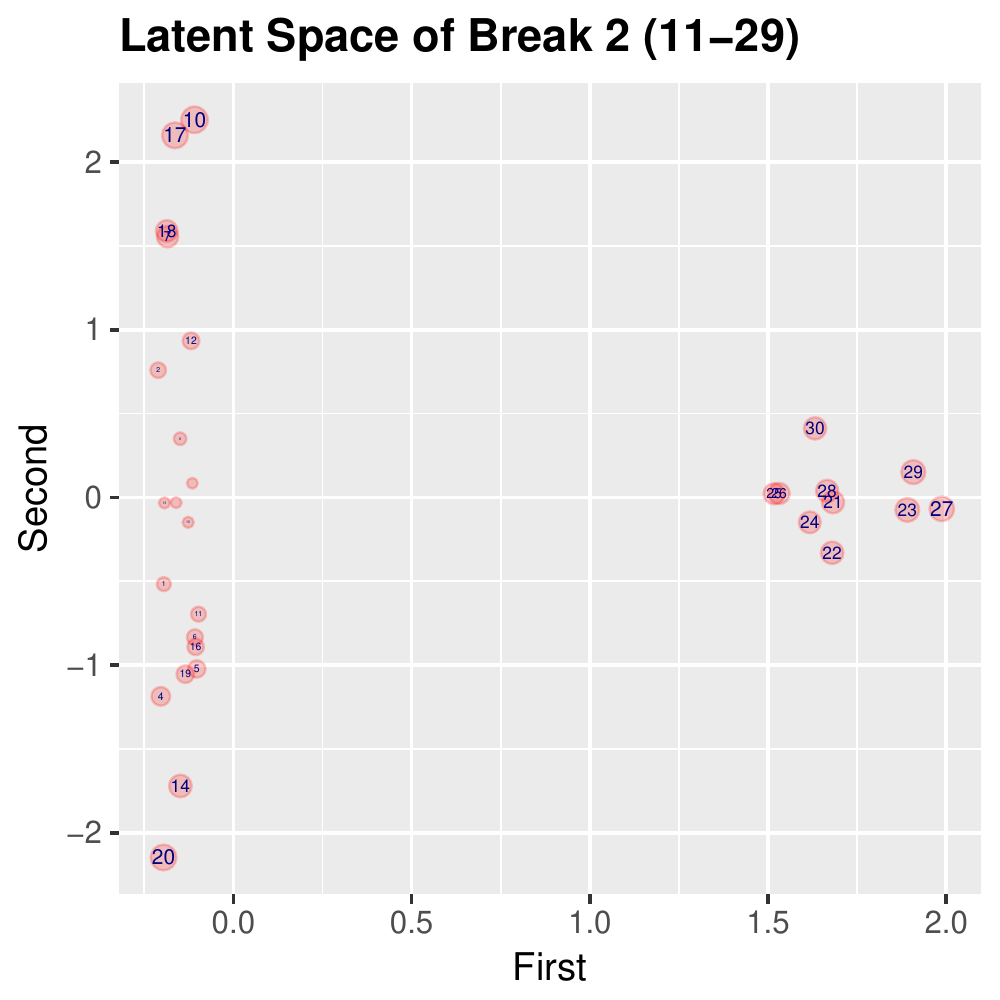}
    \end{minipage}
    &
            \begin{minipage}{.18\textwidth}
      \includegraphics[width=\linewidth]{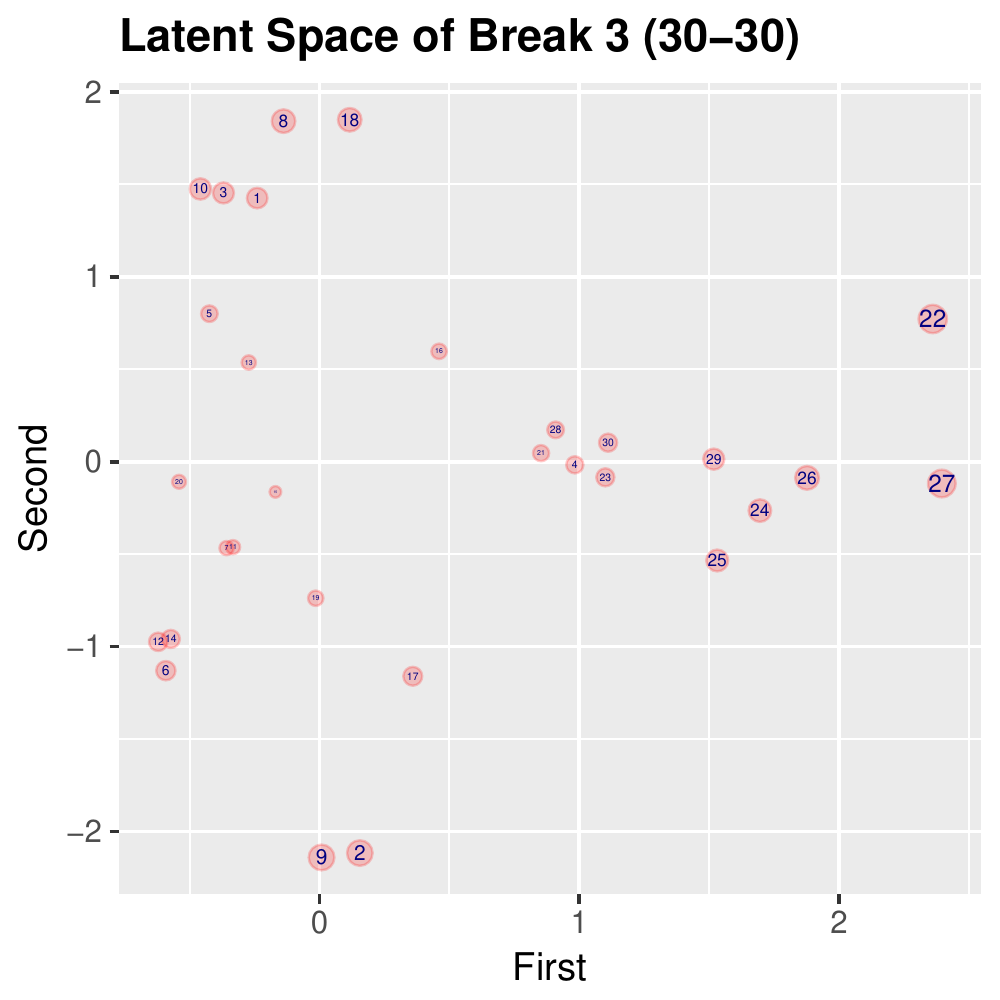}
    \end{minipage}
	&
	           \begin{minipage}{.18\textwidth}
      \includegraphics[width=\linewidth]{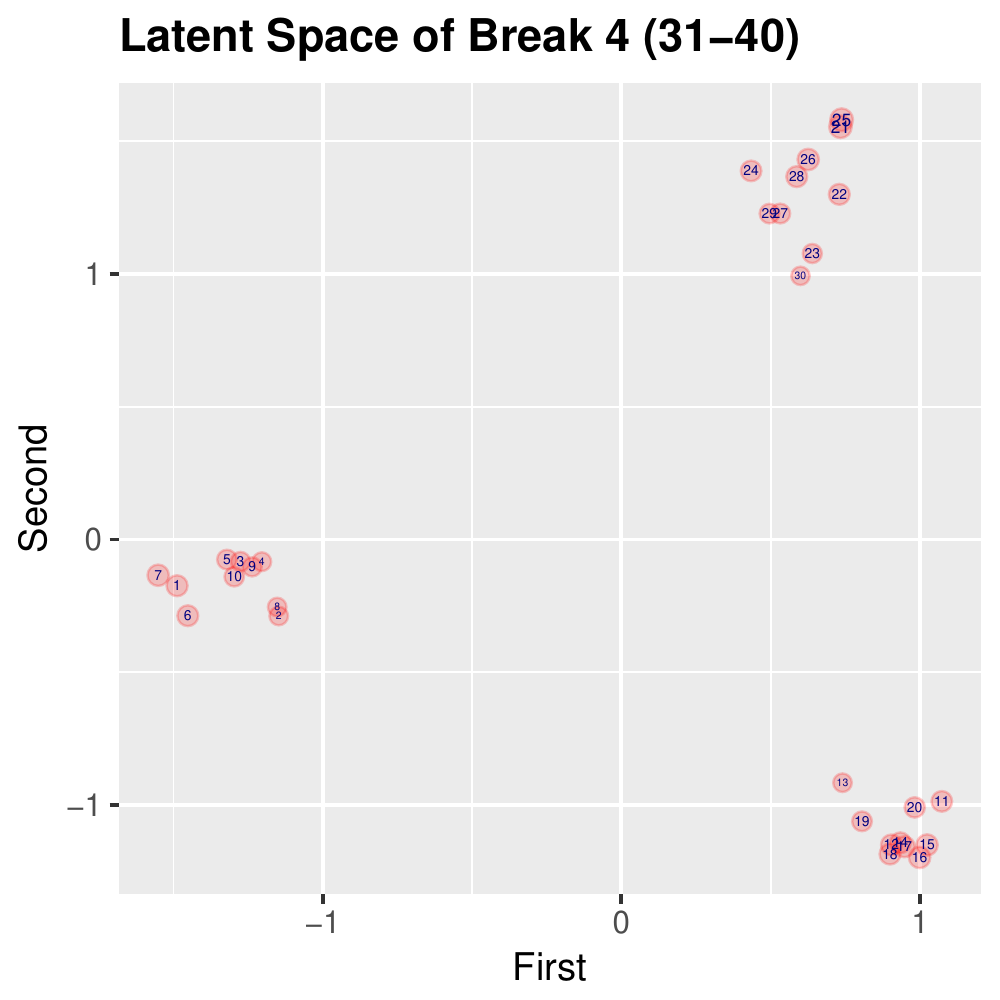}
    \end{minipage}
&
    \begin{minipage}{.18\textwidth}
      \includegraphics[width=\linewidth]{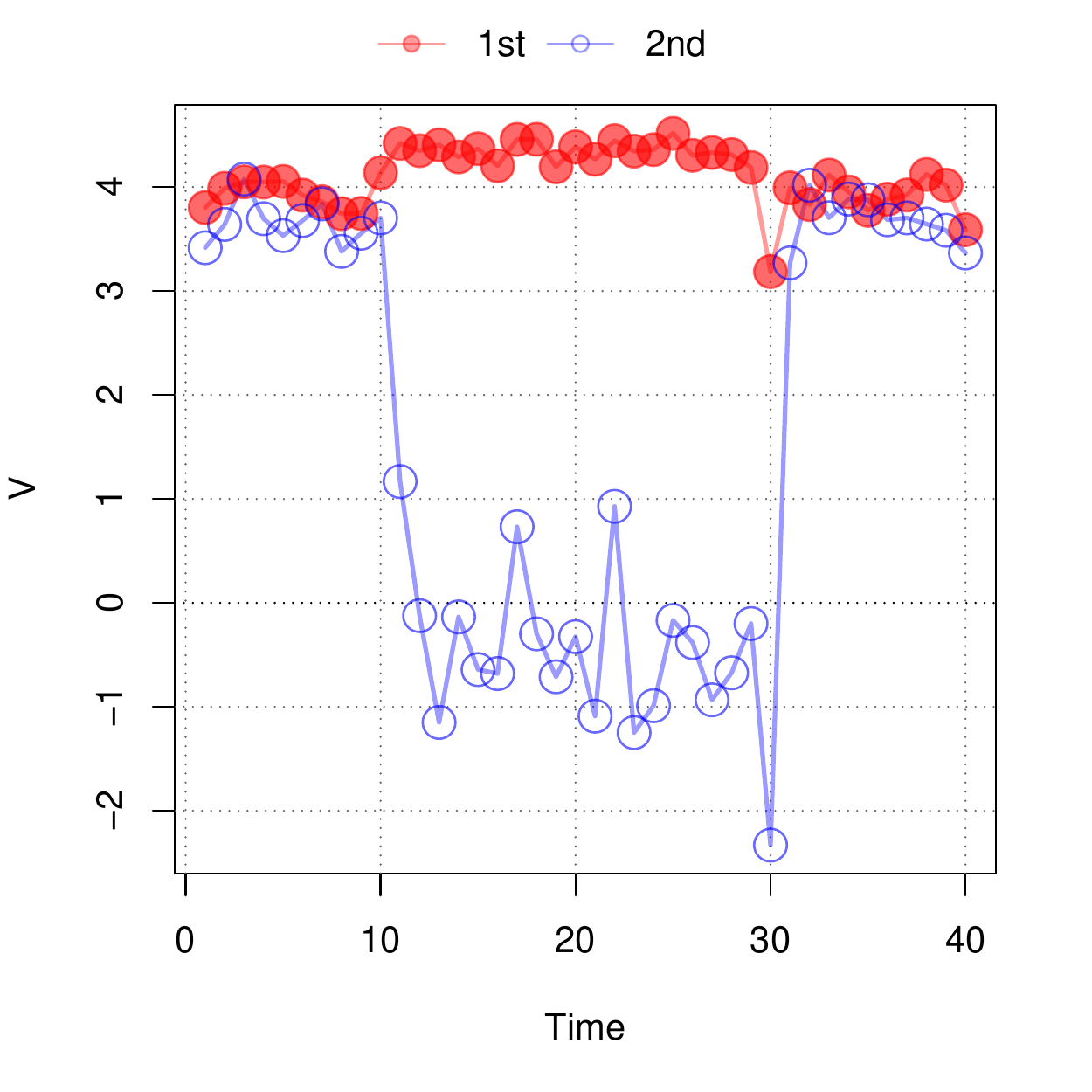}
    \end{minipage} \\
     \bottomrule
  \end{tabular}}
  \caption{Simulation Results of Block-structured Networks with Two Breaks. The ground truth is two breaks and the underlying group structure changes from a two group structure to a three group structure right after $t = 10$ and from a three group structure to a two group structure right after $t = 30$. The ground truth is two breaks. }\label{sim.merge-split}
\end{sidewaystable}

\subsection{Block-structured Networks with Two Breaks}
Last, we check whether our proposed method correctly recovers more complicated network changes. The first break is planted at $t = 10$ and it corresponds to a group-merging change. Another break is planted at $t = 30$, which corresponds to a group-splitting change. Thus, the number of latent groups changes 3, 2, and 3. 
Table \ref{sim.merge-split} reports the results of the two break test. 

WAIC correctly detects $\mathcal{M}_2$ as the best-fitting model while the approximate log likelihood favors $\mathcal{M}_3$, the pattern of which is constant in our simulation. Again, the presence of a singleton state in the three break model is the source of the problem for the approximate log likelihood. The average loss of break points correctly favors the two break model. Fitting the one break model  increases the average loss of break points because the latent state sampler falls into either of the breaks. In contrast, adding more than two breaks increases the average loss of break points significantly because the existence of a redundant break makes it difficult to pin down break points in simulations. 

If we look the recovered latent states, the two break HMTM correctly recovers the two underlying changes between $t = 10$ and $t = 11$ (group merging) and between $t = 30$ and $t = 31$ (group splitting).  Changes in the relative size of network generation rules (the last column) inform us the types of changes underlying network structures go through. For example, when the number of groups changes from 2 to 3 in the transition to Regime 3, $v_2$ returns to its previous level at Regime 1.

Overall, our simulation results clearly show that our proposed model and multiple metrics for model diagnostics work very well in (1) correctly identifying the number of breaks, (2) recovering latent group structure changes, and (3) identifying state-specific latent node positions. The approximate log marginal likelihood performs well where there is no singleton state while WAIC performs steadily regardless of the existence of singleton states. The average loss also shows steady performance, signaling the existence of redundant states in a tested model.

\section{Applications}
In this section, we apply our proposed method to the analysis of structural changes in military alliance networks. The structure of military alliance networks reflects the distribution of military power among coalitions of states, which changes over time in response to exogenous shocks to the international system or endogenous network dynamics. However, there has been no study that investigates changes in coalition structures of military alliance networks over time. A main reason is the lack of statistical methods that model unknown structural changes in longitudinal network data. 

To illustrate our method, we start from a simple example using a small data set, consisting of seven ``major powers" (Austria-Hungary, France, Germany, Italy, Japan, Russia, and the United Kingdom) from  1816 to 1938 \citep{Gibler2009}. We aggregated every 2 year network to increase the density of each layer.  Changing the granularity of aggregation does not change the substantive findings.  These seven major powers are main players of the balance of power system in Europe during the 19th century and two world wars in the 20th century. Also, the period from 1816 to 1938 corresponds to the era of shifting alliances among major powers. Thus, the structure of alliance among major powers will clearly display how the distribution of military power in the international system changes over time. Then, we apply our method to a larger data set of 104 postwar states after removing isolated nodes.

\subsection{Major Power Alliance Network Changes, 1816 - 1938}
\begin{figure}
	\includegraphics[width=1 \textwidth]{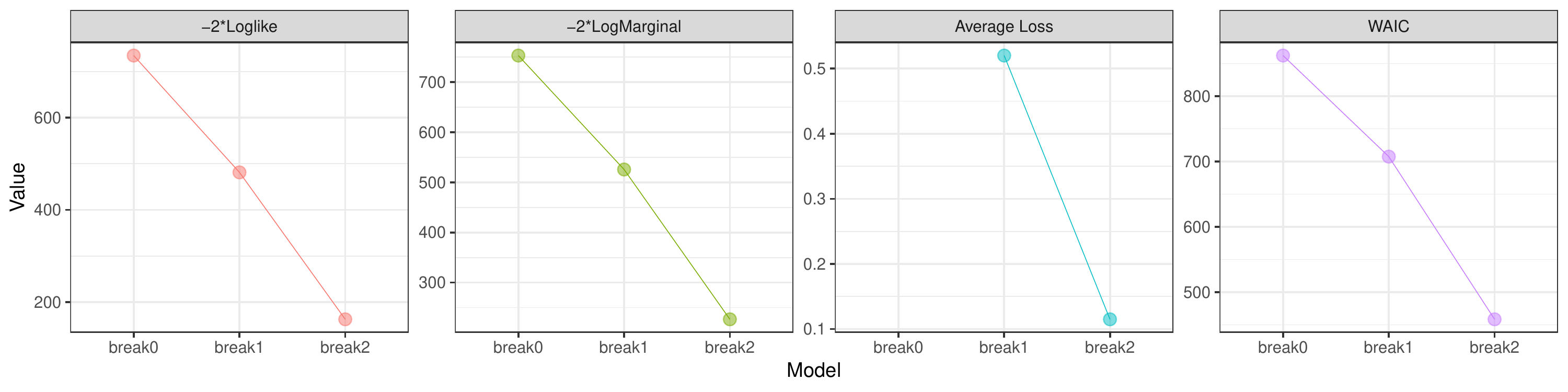}		
	\caption{\emph{Model Diagnostics of Major Power Alliance Network Changes, 1816 - 1938}}\label{majordiag}
\end{figure}

Figure \ref{majordiag} shows the model diagnostic results for the major power alliance data set. We dropped the results for the models with more than 3 breaks as they show strong signs of non-convergence, which indicate the existence of redundant states. All metrics of model diagnostics point to the two break model as the best-fitting model. In particular, the average loss of break points significantly drops in the two break model. 

\begin{figure}
	\includegraphics[width=1 \textwidth]{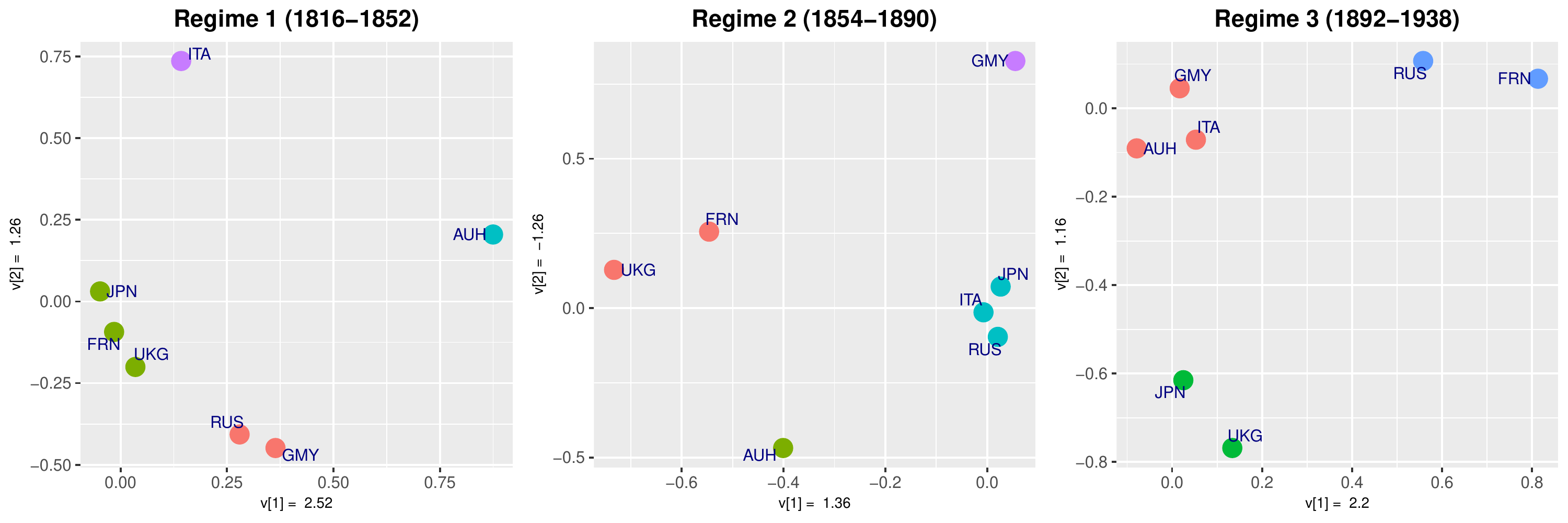}		
	\includegraphics[width=.98 \textwidth]{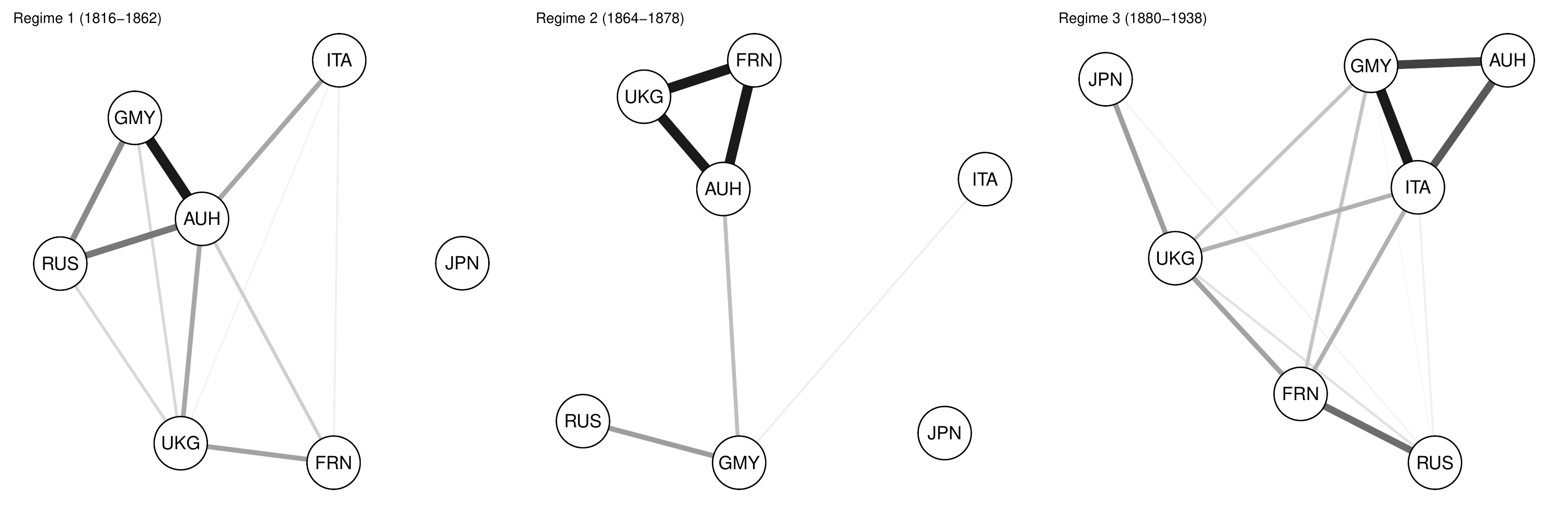}		
	\caption{\emph{Changing Node Positions and Network Topology of Military Alliance Networks Among Major Powers, 1816 - 1938: Node colors (online) indicate clusters. Regime averages of $v_t$ values for each dimension are reported in the axis (top panel).  Line widths (bottom panel) are proportional to the duration of alliance links. Included states are Austria-Hungary (AUT), France (FRA), Germany (GMY), Italy (ITA), Japan (JAP), Russia (RUS), and the United Kingdom (GBR). }}\label{majorraw}
\end{figure}

Figure \ref{majorraw} visualizes changes in latent node positions of major powers  (top) and changing patterns of the major-power network topology (bottom) from the two break model.\footnote{All network diagrams are drawn using a Fruchterman-Reingold layout, which locates nodes with more connections and short topological distance in proximal locations, for the better visibility of the state labels.} Node colors (online) indicate clusters of each node using the $k$ means clustering method. Regime-specific network generation rule parameters $v_{rt}$ are reported in axis labels. Several substantive findings are noteworthy.

The first notable finding is the centrality of Austria-Hungary, connecting groups of major powers, between 1816 and 1890. This period includes what historians call  ``the age of  Metternich'' (1815-1848) \citep{Rothenberg1968}. After the end of Napoleonic Wars, Chancellor of Austria-Hungary played an important role in maintaining the European balance of power system.  The first dimension of Regime 1 clearly distinguishes Austria-Hungary from the other major powers. In Regime 2 (1854-1890), Germany challenged the position of Austria-Hungary. However, throughout Regime 1 and Regime 2, the network position of Austria-Hungary remained highly critical in the sense that the removal of Austria-Hungary would have made the major power alliance network completely disconnected. In the language of social network analysis, Austria-Hungary filled a ``structural hole" in the major power alliance network at the time, playing the role of broker \citep{Burt2005, Burt2009, Stovel2012}. 


The second notable finding is the timing of the first break between 1852 and 1854. This break point coincides with the outbreak of the Crimean War. In this war, Russia was defeated by the united powers of Britain, France, Austria-Hungary, and Prussia (Germany). The rise of Germany led by Otto von Bismarck and the defeat of Russia marked the first break in the balance of power system. 

The third notable finding is the timing of the second break between 1890 and 1892. Scholars of international relations and historians consider the formation of the Dual Alliance between Germany and Austria-Hungary in 1879 and a sequence of alliances that followed it as a structural change in the  balance of power system \citep{Snyder1997,  Vermeiren2016}.\footnote{First, Russia formed alliances with Germany and Austria-Hungary (Three Emperors' Alliance) in 1881. Then, Italy joined Germany and Austria-Hungary (Triple Alliance) in 1882. France, a long-time rival of Germany, formed an alliance with Russia in 1894 to check Germany and Austria-Hungary. In this process, an important cleavage in the alliance networks emerged. } These series of events transformed a web of shifting alliances into a clearly diverged group structure, consisting of two clusters: Austria-Hungary, Germany, and Italy on the one hand and France, Russia, and the United Kingdom on the other. The network diagram of the third regime (bottom-right) shows members of the two clusters, which formed each side of belligerents in World War I. 

\subsection{Postwar Alliance Network Changes, 1946 - 2012}
\begin{figure}
	\includegraphics[width=1 \textwidth]{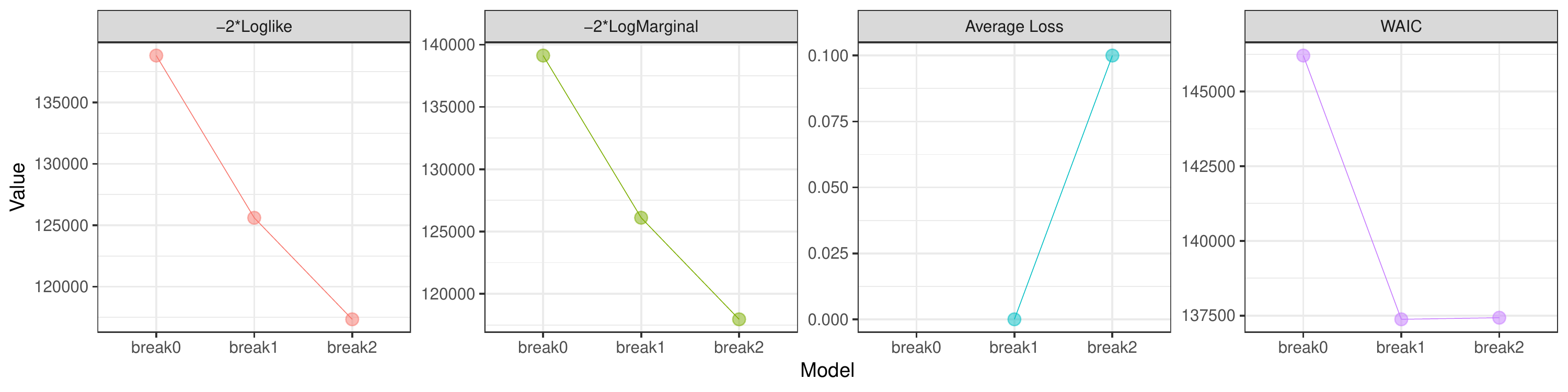}		
	\caption{\emph{Model Diagnostics of Postwar Alliance Network Changes, 1946 - 2012}}\label{alldiag}
\end{figure}

Now, we focus on the postwar period in which the number of states and alliance links among them exploded. After removing isolated nodes, our sample contained 104 states. Figure \ref{alldiag} shows the results of the model diagnostics. Although the two break model is preferred by the approximate log marginal likelihood, the average loss of break points and WAIC favor the one break model. The estimated break point is between 1978 and 1980. 

\begin{figure}
	\includegraphics[width=1 \textwidth]{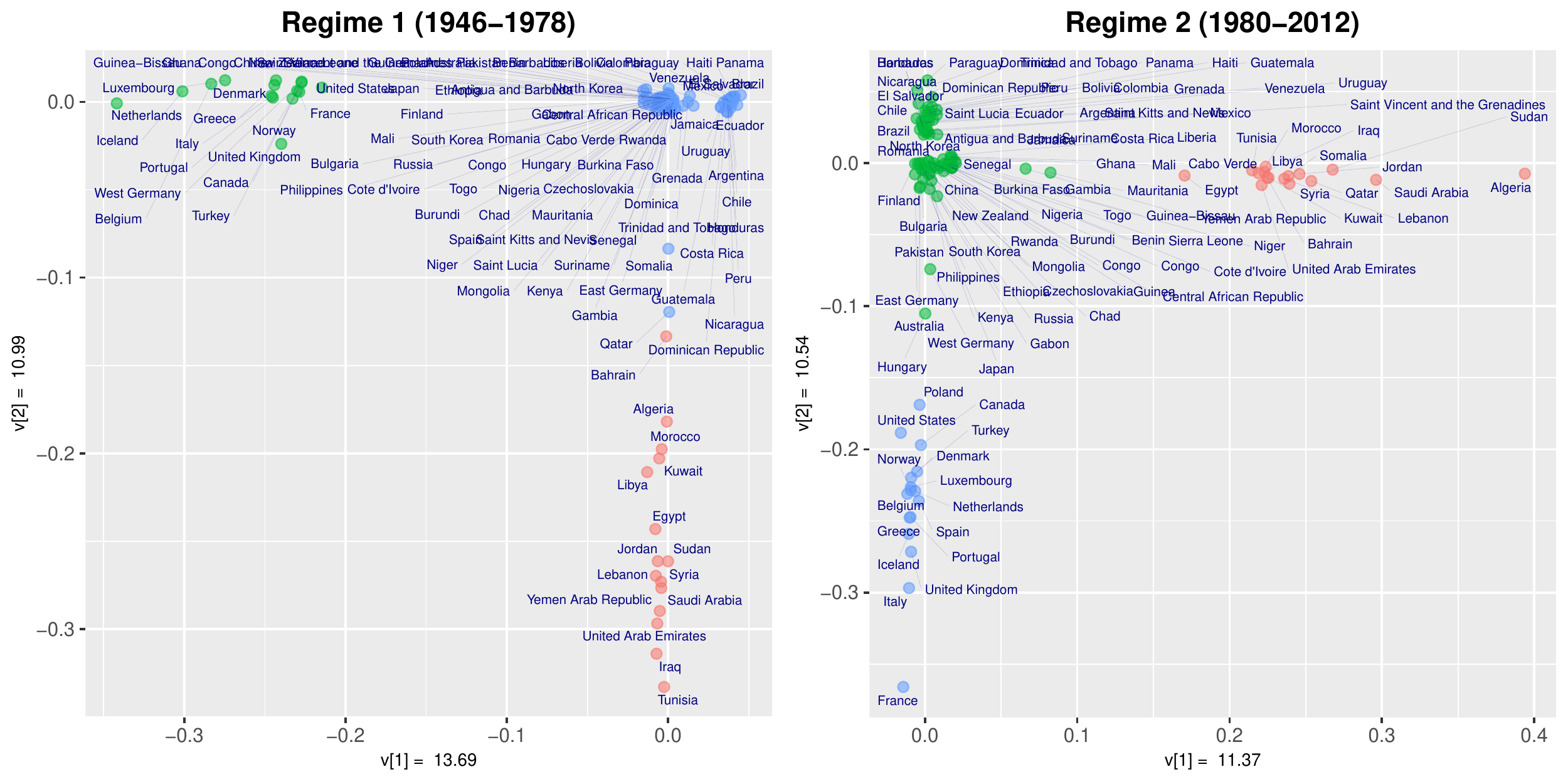}		
	\includegraphics[width=.98 \textwidth]{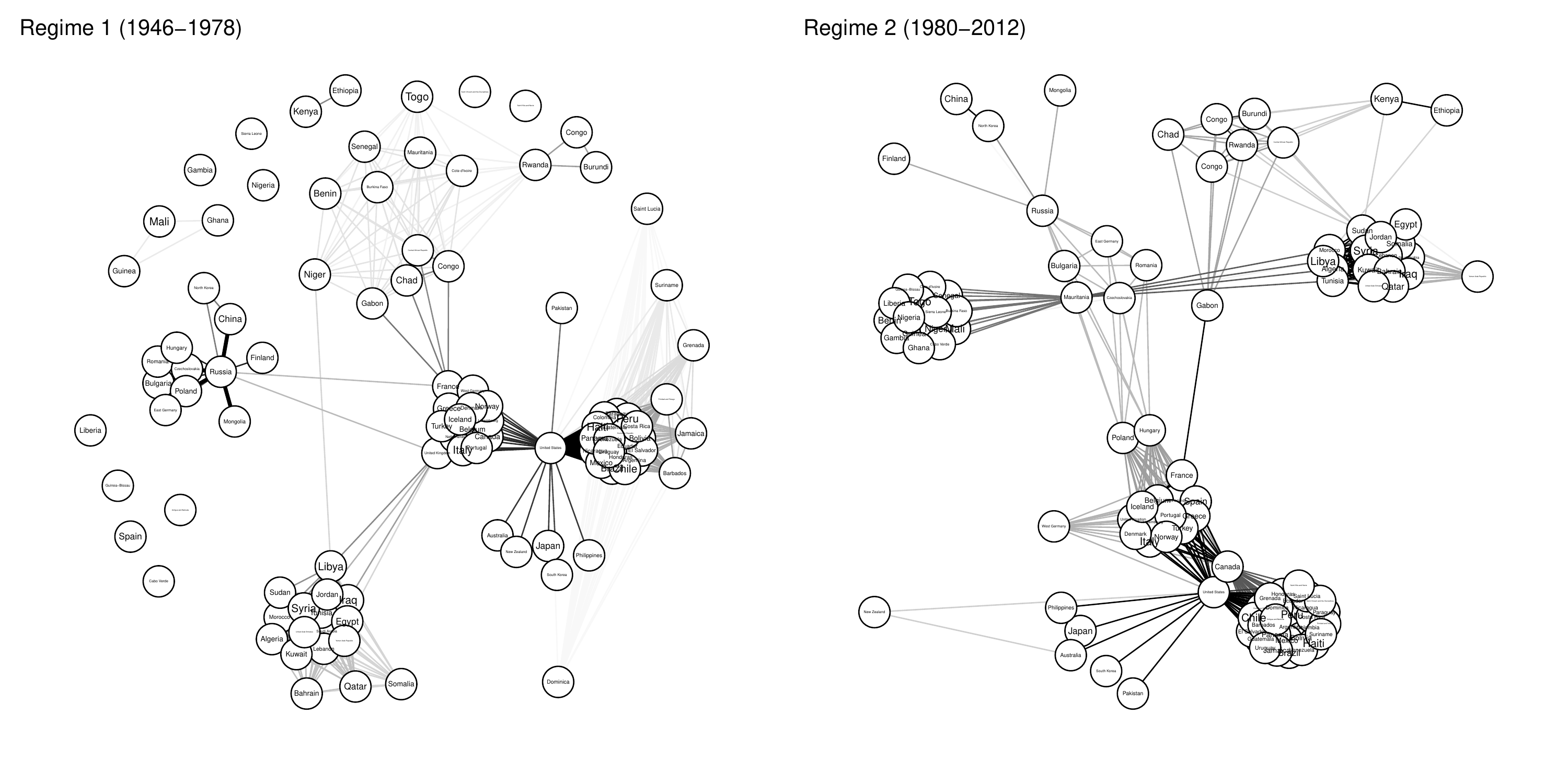}		
	\caption{\emph{Changing Node Positions and Network Topology of Postwar Military Alliance Networks, 1946 - 2012: Node colors (online) indicate clusters. Regime averages of $v_t$ values for each dimension are reported in the axis (top panel).  Line widths (bottom panel) are proportional to the duration of alliance links. }}\label{allnode}
\end{figure}

Figure \ref{allnode} shows changes in latent node positions (top) and changing patterns of the network topology (bottom). What we can see is that the number of latent groups does not change during the postwar period. What changed is the strength of the connection among groups of nodes, denoted by $v_{rt}$ in each axis. 
Both numbers in the $x$ and $y$ axes are large and positive, implying that the coalition structure of postwar military alliance networks is multidimensional with a sign of a strong homophily. However, after 1980, the strength of the within-group connection rule became weaker as more countries are connected with outside group members. This reflects the fact that the coalition structure of postwar military alliance networks was highly dense in the beginning of the Cold War. The Cold War division became less important over time due to the det\'tent and the rise of the Non-Aligned Movement. The results suggest that the structural change in the postwar military alliance network came earlier than the collapse of the Soviet Union. 

\section{Concluding Remarks}
In this article, we presented HMTM as a statistical method to detect and analyze changes in structural properties of longitudinal network data. The proposed method has several advantages over existing dynamic network models. 

First, the proposed method combines a highly flexible generative model of multilayer networks (MTRM) with a HMM, which has proved to be an effective tool to model irregular dynamics in temporal data. This formulation is flexible enough to accommodate a variety of network representations such as graph Laplacian \citep{rohe2011spectral} and motif Laplacian \citep{benson2016higher} as an input data format. Our simulation studies showed that our generative approach is a powerful tool to detect and analyze diverse types of network changes. 

Second, the Bayesian inference of HMTM enables researchers to identify the number of network changes in a principled way. Our simulation studies show that WAIC correctly identifies the number of breaks and the type of network changes in all tests while the approximate log marginal likelihood consistently favor overfitted models. 

Finally, HMTM provides an important tool to investigate changes in meso-scale structures of longitudinal network data. Meso-scale structural changes are important quantities that reflect fundamental changes in the network generating process, that are unable to be captured by local network properties or global summary indices.  


While we only consider undirected networks, our model can be extended to analyze other types of longitudinal network data consisting of directed networks or bipartite networks using a singular value decomposition-based framework \citep{de2000multilinear, hoff_2007_jasa} and the hierarchical multilinear framework \cite{Hoff2011} in general. Also, a hierarchical Dirichlet process prior can be used to endogenously detect the number of breaks \citep{Beal2002, Ko2015, Yee2006, Fox2011}. Another interesting extension of HMTM is the inclusion of nodal covariates \citep{volfovsky_hoff_2015} or covariates for network effects, where the latent space formulation may serve as an instrument to control for unobserved heterogeneous effects on tie formation.  One difficulty in adding covariates in hidden Markov models is to avoid the endogeneity between state transition and effects of covariates \citep{Kim2008}.  

  \clearpage\vfill

\section*{MCMC Algorithm for Hidden Markov Tensor Model}
\begin{description}
\item[] For each $t$ layer, generate $\mathbf{B}_t = \mathbf{Y}_t - \mathbf{\Omega}_t$ by choosing a null model ($\mathbf{\Omega}_t$).
\item[] Set the total number of changepoints $M$ and initialize ($\mathbf{U}, \boldsymbol{\mu}_u, \Psi_v, \V, \boldsymbol{\mu}_v, \Psi_v, \beta, \sigma^2, \S, \P)$.
\end{description}

\noindent \textbf{Part 1}

\begin{description}\footnotesize
\item[Step 1] The sampling of regime specific $\U, \boldsymbol{\mu}, \Psi_{u}$ consists of the following three steps for each regime $m$.  Let $ \Psi_{u} = \left( \begin{array}{ccc}
\psi_{1, u, m} & \ldots & 0 \\
0 & \psi_{r, u, m} & 0 \\
0 &  \ldots & \psi_{R, u, m} \end{array} \right) $.
	\begin{enumerate}
	\item $p(\psi_{r, u, m} |  \B,  \P, \S, \mathbf{\Theta} ^{-\Psi_{u, m}}) \propto \mathcal{IG}\left (\frac{u_0 + N}{2}, \frac{\U_{r, m}^T\U_{r, m} + u_{1}}{2}\right)$.
	
	\item $p(\boldsymbol{\mu}_{u, m}|  \B,  \P, \S, \mathbf{\Theta} ^{-\boldsymbol{\mu}_{u, m}}) \propto  \textrm{multivariate normal}(\U_m^T\mathbf{1}/(N + 1), \Psi_{u, m}/(N + 1))$.
	\item $p(\U_m|  \B,  \P, \S, \mathbf{\Theta} ^{-\U_m}) \propto \text{matrix normal}_{N \times R}(\tilde{\M}_{u, m}, \mathbf{I}_{N}, \tilde{\Psi}_{u, m})$ 
	where 
	\begin{eqnarray*}
	\tilde{\Psi}_{u, m} &=& (\mathbf{Q}_{u, m}/\sigma_m^2 + \Psi_{u, m}^{-1})^{-1} \\
	\tilde{\M}_{u, m} &=& (\mathbf{L}_{u, m}/\sigma_m^2 + \mathbf{1}\boldsymbol{\mu}_{u, m}^T \Psi_{u, m}^{-1}) \tilde{\Psi}_{u, m}\\
	\mathbf{Q}_{u, m} &=& (\U_m^T\U_m)\circ(\V_m^T\V_m) \\
	\mathbf{L}_{u, m} &=& \sum_{j, t:\; t \in S_t = m}b_{\cdot,j, t} \otimes (\U_{m, j, \cdot} \circ   \V_{m, t, \cdot} )
	\end{eqnarray*}
	\item Orthogonalize $\U_m$ using the Gram-Schmidt  algorithm. 
	\end{enumerate}

\item[Step 2] The sampling of $\V, \boldsymbol{\mu}_v, \Psi_v$ is done for each regime.  Let $ \Psi_{v} = \left( \begin{array}{ccc}
\psi_{1, v, m} & \ldots & 0 \\
0 & \psi_{r, v, m} & 0 \\
0 &  \ldots & \psi_{R, v, m} \end{array} \right) $.
	\begin{enumerate}
	\item $p(\psi_{r, v, m} |  \B,  \P, \S, \mathbf{\Theta} ^{-\Psi_{v, m}}) \propto\mathcal{IG}\left (\frac{v_0 + T}{2}, \frac{\V_{r, m}^T\V_{r, m} + v_{1}}{2} \right)$.
	\item  $p(\boldsymbol{\mu}_{v, m}|  \B,  \P, \S, \mathbf{\Theta} ^{-\boldsymbol{\mu}_{v, m}}) \propto\textrm{multivariate normal}(\V_m^T\mathbf{1}/(T_m + 1), \Psi_{v, m}/(T_m + 1))$.
	\item  $p(\V_m|  \B,  \P, \S, \mathbf{\Theta} ^{-\V_m}) \propto\text{matrix normal}_{T_m \times R}(\tilde{\M}_{v, m}, \mathbf{I}_{T_m}, \tilde{\Psi}_{v, m})$ 
	where 
	\begin{eqnarray*}
	\tilde{\Psi}_{v, m} &=& (\mathbf{Q}_{v, m}/\sigma_m^2 + \Psi_{v, m}^{-1})^{-1} \\
	\tilde{\M}_{v, m} &=& (\mathbf{L}_{v, m}/\sigma_m^2 + \mathbf{1}\boldsymbol{\mu}_{v, m}^{T_m}\Psi_{v, m}^{-1}) \tilde{\Psi}_{v, m}\\
	\mathbf{Q}_{v, m} &=& (\U_m^{T}\U_m)\circ(\U_m^{T}\U_m) \\
	\mathbf{L}_{v, m} &=& \sum_{i, j}b_{i, j, \cdot} \otimes (\U_{m, i, \cdot} \circ   \U_{m, j, \cdot} )
	\end{eqnarray*}
	\end{enumerate}
\item[Step 3] The sampling of $\beta$ from $\normdist{b_1}{B_1}$ where 
\begin{eqnarray*}
B_1&=& (B_0^{-1} + \sum_{m=1}^{M}\sigma^{-2}_{m} N^2 \mathbf{1}(\S = m))^{-1} \\
b_1&=& B_1 \times \Big(B_0^{-1}b_0 + \sum_{i = 1}^N\sum_{j = 1}^N \sum_{t = 1}^T b_{i, j, t} - \mu_{i, j, t} \Big).
\end{eqnarray*}
$\mathbf{1}(\S = m)$ is the number of time units allocated to state $m$ and $\mu_{i, j, t}$ is an element of $\U_{S_t}\boldsymbol{\Lambda}_t\U_{S_t}^T$.

\item[Step 4] The sampling of $\sigma^2_m$ from $\mathcal{IG}\left (\frac{c_0 +  N_m \cdot N_m \cdot T_m}{2}, \frac{d_{0} + \sum_{i = 1}^N\sum_{j = 1}^N\sum_{t=1}^T b_{i, j, t} - \beta - \mu_{i, j, t}}{2} \right)$. \end{description}

\noindent \textbf{Part 2} 
\begin{description}\footnotesize
\item[Step 5] Sample $\S$ recursively using \cite{Chib1998}'s algorithm. The joint conditional distribution of the latent states $p(S_0, \ldots, S_T | \mathbf{\Theta}, \B, \P) $ can be written as the product of $T$ numbers of independent conditional distributions: 
\begin{equation*}
 p(S_0, \ldots, S_T |\mathbf{\Theta}, \B, \P) = p(S_T| \mathbf{\Theta}, \B, \P)\ldots p(S_t|\S^{t+1},  \mathbf{\Theta}, \B, \P) \ldots p(S_0|\S^{1},  \mathbf{\Theta}, \B, \P). 
\end{equation*}

Using Bayes' Theorem, \cite{Chib1998} shows that 
\begin{eqnarray*}
p(S_t|\S^{t+1}, \mathbf{\Theta}, \B, \P) &\propto& \underbrace{p(S_t|\mathbf{\Theta}, \mathbf{B}_{1:t}, \P)}_{\text{State probabilities given all data}} \underbrace{p(S_{t+1}|S_t, \P)}_{\text{Transition probability at $t$}}.
\end{eqnarray*}
The second part on the right hand side is a one-step ahead transition probability at $t$, which can be obtained from a sampled transition matrix ($\P$). The first part on the right hand side is state probabilities given all data, which can be simulated via a forward-filtering-backward-sampling algorithm as shown in \cite{Chib1998}.

\item[Step 5-1] During the burn-in iterations, if sampled $\S$ has a state with single observation, randomly sample  $\S$ with replacement using a pre-chosen perturbation weight ($\mathbf{w}_{\mathrm{perturb}} = (w_1, \ldots, w_{M})$).  
\end{description}

\noindent \textbf{Part 3: $p(\P| \B, \S, \mathbf{\Theta})$} 
\begin{description}\footnotesize
\item[Step 6] Sample each row of $\P$ from the following Beta distribution: \\
\begin{equation*}
p_{kk} \sim \betadist{a_0 + j_{k, k} - 1}{b_{0} + j_{k, k+1}}
\end{equation*}
where $p_{kk}$ is the probability of staying when the state is $k$, and $j_{k, k}$ is the number of jumps from state $k$ to $k$, and $j_{k, k+1}$ is the number of jumps from state $k$ to $k+1$.
\end{description}

\newpage
\section*{The Approximate Log Marginal Likelihood of a Hidden Markov Tensor Model}

The computation of the log posterior density of posterior means requires a careful blocking in a highly parameterized model. In our HMTM, the log posterior density of posterior means is decomposed into seven blocs:
\begin{eqnarray*}\nonumber
\log p(\boldsymbol{\mu}_u^*, \psi_{., u}^*, \boldsymbol{\mu}_v^*, \psi_{., v}^*, \beta^*, \sigma^{2*}, \P^*|\B) &=& 
 \log p(\boldsymbol{\mu}^*_u|\B) + \sum_{r=1}^R\log  p(\psi^*_{r, u}|\B, \boldsymbol{\mu}^*_u)\\\nonumber
&& + \log p(\boldsymbol{\mu}_v^*|\B, \boldsymbol{\mu}^*_u, \psi^*_{., u}) + \sum_{r=1}^R \log  p(\psi_{r, v}^*|\B, \boldsymbol{\mu}^*_u, \psi^*_{., u}, \boldsymbol{\mu}_v^*) \\ \label{marg}
 && + \log p(\beta^{*}|\B,\boldsymbol{\mu}^*_u, \psi^*_{., u}, \boldsymbol{\mu}_v^*, \psi_{., v}^*) \\ \nonumber 
 && + \log p(\sigma^{2*}|\B,\boldsymbol{\mu}^*_u, \psi^*_{., u}, \boldsymbol{\mu}_v^*, \psi_{., v}^*, \beta^*) \\\nonumber
 && + \log  p(\P^*|\B, \boldsymbol{\mu}^*_u, \psi^*_{., u}, \boldsymbol{\mu}_v^*, \psi_{., v}^*, \beta^*, \sigma^{2*}).
\end{eqnarray*}

Let $ \mathbf{\Theta}$ indicate a parameter vector beside hidden states ($\S$) and a transition matrix ($\P$): $\mathbf{\Theta} = \{\boldsymbol{\mu}_u, \psi_{., u}, \boldsymbol{\mu}_v, \psi_{., v}, \beta, \sigma^2\}$. Let $(\mathbf{\Theta}^*,  \P^*)$ be posterior means of $(\mathbf{\Theta},  \P)$. Using \cite{Chib1995}'s formula to compute the approximate log marginal likelihood,   
\begin{eqnarray*}
p(\mathbf{\Theta}^*,  \P^*|\B) &=& \frac{p(\B | \mathbf{\Theta}^*,  \P^*) p(\mathbf{\Theta}^*,  \P^*)}{m(\B)}\\
m(\B) &=& \frac{p(\B | \mathbf{\Theta}^*,  \P^*) p(\mathbf{\Theta}^*,  \P^*)}{p(\mathbf{\Theta}^*,  \P^*|\B)}\\
\log m(\B) &=& \log p(\B | \mathbf{\Theta}^*,  \P^*) + \log p(\mathbf{\Theta}^*,  \P^*)- 
\log p(\mathbf{\Theta}^*,  \P^*|\B).
\end{eqnarray*}

The quantities in the right hand side of Equation (\ref{marg}) can be computed by \cite{Chib1995}'s candidate formula: 
\begin{description}\footnotesize
\item[Step 1]\begin{eqnarray*}
p(\boldsymbol{\mu}^*_u|\B)  &\approx& \int p(\boldsymbol{\mu}^*_u|\B, \psi_{., u}, \boldsymbol{\mu}_v, \psi_{., v}, \beta, \sigma^{2}, \P, \S) d p(\psi_{., u}, \boldsymbol{\mu}_v, \psi_{., v}, \beta, \sigma^{2}, \P, \S| \B)
\end{eqnarray*}
\item[Step 2]\begin{eqnarray*}
p(\psi^*_{., u}|\B, \boldsymbol{\mu}^*_u)  &\approx& \int p(\psi^*_{., u}|\B, \boldsymbol{\mu}^*_u, \boldsymbol{\mu}_v, \psi_{., v}, \beta, \sigma^{2}, \P, \S) d p(\boldsymbol{\mu}_v, \psi_{., v}, \beta, \sigma^{2}, \P, \S| \B)
\end{eqnarray*}
\item[Step 3]\begin{eqnarray*}
p(\boldsymbol{\mu}_v^*|\B, \boldsymbol{\mu}^*_u, \psi^*_{., u})  &\approx& \int p(\boldsymbol{\mu}_v^*|\B, \boldsymbol{\mu}^*_u, \psi^*_{., u}, \psi_{., v}, \beta, \sigma^{2}, \P, \S) d p(\psi_{., v}, \beta, \sigma^{2}, \P, \S| \B)
\end{eqnarray*}
\item[Step 4]\begin{eqnarray*}
p(\psi_{., v}^*|\B, \boldsymbol{\mu}^*_u, \psi^*_{., u}, \boldsymbol{\mu}_v^*)  &\approx& \int p(\psi_{., v}^*|\B, \boldsymbol{\mu}^*_u, \psi^*_{., u}, \boldsymbol{\mu}_v^*, \beta, \sigma^{2}, \P, \S) d p(\beta, \sigma^{2}, \P, \S| \B)
\end{eqnarray*}
\item[Step 5]\begin{eqnarray*}
p(\beta^{*}|\B,\boldsymbol{\mu}^*_u, \psi^*_{., u}, \boldsymbol{\mu}_v^*, \psi_{., v}^*)  &\approx& \int p(\beta^{*}|\B,\boldsymbol{\mu}^*_u, \psi^*_{., u}, \boldsymbol{\mu}_v^*, \psi_{., v}^*, \sigma^{2}, \P, \S) d p(\sigma^{2}, \P, \S| \B)
\end{eqnarray*}
\item[Step 6]\begin{eqnarray*}
p(\sigma^{2*}|\B,\boldsymbol{\mu}^*_u, \psi^*_{., u}, \boldsymbol{\mu}_v^*, \psi_{., v}^*, \beta^*) &\approx& \int p(\sigma^{2*}|\B,\boldsymbol{\mu}^*_u, \psi^*_{., u}, \boldsymbol{\mu}_v^*, \psi_{., v}^*, \beta^*, \P, \S) d p(\P, \S| \B)
\end{eqnarray*}
\item[Step 7]\begin{eqnarray*}
p(\P^*|\B, \boldsymbol{\mu}^*_u, \psi^*_{., u}, \boldsymbol{\mu}_v^*, \psi_{., v}^*, \beta^*, \sigma^{2*}) &\approx& \int p(\P^*|\B, \boldsymbol{\mu}^*_u, \psi^*_{., u}, \boldsymbol{\mu}_v^*, \psi_{., v}^*, \beta^*, \sigma^{2*}, \S) d p(\S| \B)
\end{eqnarray*}
\end{description}

\newpage
\section*{Sampling of HMTM with a Student-$t$ Distributed Error}
We modify the sampling of $\U$ based on a Student-$t$ distributed error as follows:
	\begin{enumerate}
	\item $p(\U_m|  \B,  \P, \S, \mathbf{\Theta} ^{-\U_m}) \propto \text{matrix normal}_{N \times R}(\tilde{\M}_{u, m}, \mathbf{I}_{N}, \tilde{\Psi}_{u, m})$ 
	where 
	\begin{eqnarray*}
	\tilde{\Psi}_{u, m} &=& (\mathbf{Q}_{u, m}/\sigma_m^2 + \Psi_{u, m}^{-1})^{-1} \\
	\tilde{\M}_{u, m} &=& (\mathbf{L}_{u, m}/\sigma_m^2 + \mathbf{1}\boldsymbol{\mu}_{u, m}^T \Psi_{u, m}^{-1}) \tilde{\Psi}_{u, m}\\
	\mathbf{Q}_{u, m} &=& (\U_m^T\U_m)\circ(\V_m^T \boldsymbol{\Gamma}_m \V_m) \\
	\mathbf{L}_{u, m} &=& \sum_{j, t:\; t \in S_t = m}b_{\cdot,j, t} \otimes (\U_{m, j, \cdot} \circ   (\V_{m, t, \cdot} \gamma_t))
	\end{eqnarray*}
	where $\boldsymbol{\Gamma}_m$ is a $T_m$ by $T_m$ diagonal matrix with $\gamma_t$ corresponding to regime $m$.
	\end{enumerate}
	
Next, the sampling of $\sigma^2_m$ can be done using $\mathcal{IG}\left (\frac{c_0 +  N_m \cdot N_m \cdot T_m}{2}, \frac{d_{0} + \sum_{i = 1}^N\sum_{j = 1}^N\sum_{t=1}^T \gamma_t(b_{i, j, t} - \beta - \mu_{i, j, t})}{2} \right)$. 

Then, the draws of $\gamma_t$ are obtained from $\gamdist{\nu_2/2}{\nu_{3,t}/2}$ where 
\begin{eqnarray}\label{eq:cp3}
	\nu_2 &=& \nu_0 + 1\\
	\nu_{3,t} &=& \nu_1 + \sigma_{S_t}^{-2}\left (\sum_{i = 1}^N\sum_{j = 1}^N b_{i, j, t} - \beta - \mu_{i, j, t} \right).
\end{eqnarray}

\newpage
\section*{Block Structure Recovery from Latent Space Estimates}

\begin{figure}
\centering
\caption{Block Label Recovery from Block-Merging-Splitting Changes: $v[1]$ indicates the first element of regime-averaged network generation rule parameter and $v[2]$ is the second element of the same parameter. Node color and labels indicate the planted group membership.}\label{sim2}
\includegraphics[width=.8 \textwidth]{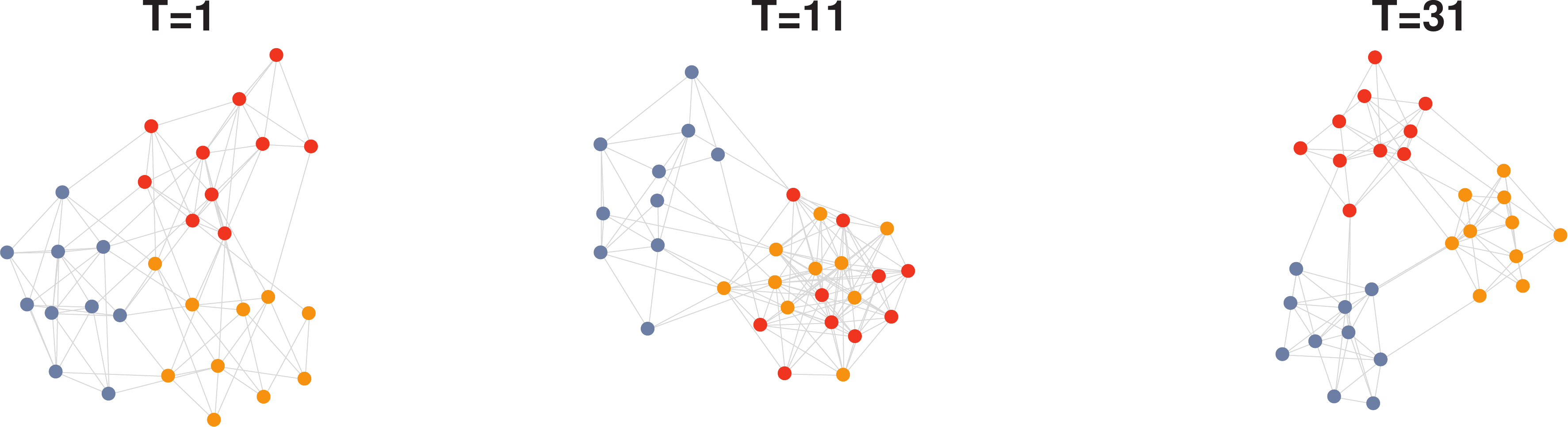}
\subcaption{\emph{Planted Network Change}}
\includegraphics[width=.9 \textwidth]{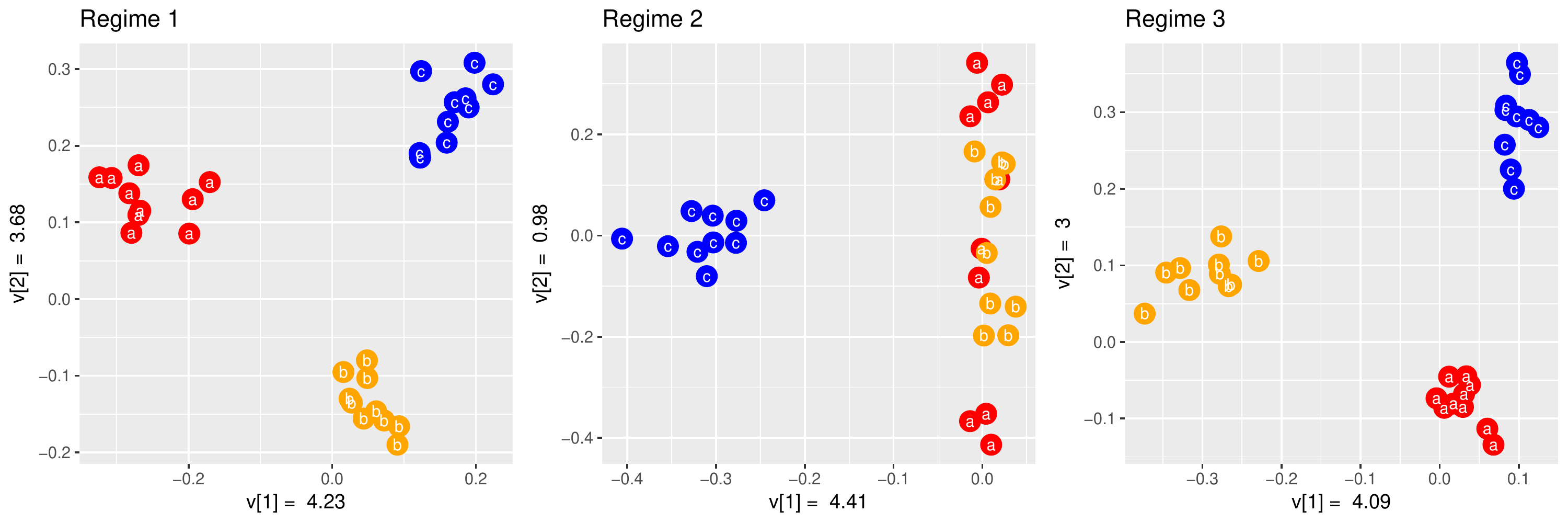}
\subcaption{\emph{Recovered Latent Node Positions and Block Structure}}
\end{figure}

Figure \ref{sim2} demonstrates the performance of discrete block label recovery using a block-merging-splitting example with $T = 40$. Panel (a) shows the planted block structures at the beginning of each regime and panel (b) shows recovered latent node positions and their block memberships that are identified by \cite{Hartigan1979}'s $k$-means clustering algorithm. The two break HMTM correctly recovers the two planted changes at $T = 11$ (block merging) and $T = 31$ (block splitting). When block $a$ and block $b$ are merged into one block in Regime 2, the second dimension becomes redundant and the second element of the regime-averaged network generation rule ($v_2$) becomes small (0.98). In Regime 3, the number of blocks increases back to 3 and $v_2$ returns to its previous level at Regime 1.

\newpage
\section*{Further Simulation Results}
In this section, we report additional simulation results that are not reported in the manuscript due to space limitation. First, we report simulation results from a block-merging network change. Then, we report simulation results from block-splitting-merging network changes. The structure of the reported tables is identical the ones discussed in the manuscript. 

\subsection*{Results of a Block-Merging Network Change}
Table \ref{sim.merge} summarizes the simulation results from the block-merging change example. The ground truth is the single break HMTM in the third row. WAIC correctly identifies the single break model as the best-fitting model while the approximate log marginal likelihood favors the three break HMTM.  The one break model ($\mathcal{M}_1$) in Table \ref{sim.merge} correctly recovers the block-merging latent change and the second dimensional network generation rule ($v_2$) drops from a positive number to 0 in the middle as the number of blocks decreases from 3 to 2. 

\begin{sidewaystable}\footnotesize
  \centering
\resizebox{\textwidth}{!}{  \begin{tabular}{p{2.5cm} cccc c}
    \toprule
    Model Fit& \multicolumn{4}{c}{Latent Space ($\mathbf{U}_m$) Changes}&  \multicolumn{1}{c}{Generation Rule ($\mathbf{v}_t$)} \\ 
    
    & Regime 1& Regime 2&Regime 3& Regime 4 \\\midrule
     {\tiny \begin{tabular}{lcl } Break number &=&0 \\ WAIC &=& 13728  \\ -2*log marginal &=& 13632 \\ -2*log likelihood &=& 12828  \end{tabular} }
  &
    \begin{minipage}{.2\textwidth}
      \includegraphics[width=\linewidth]{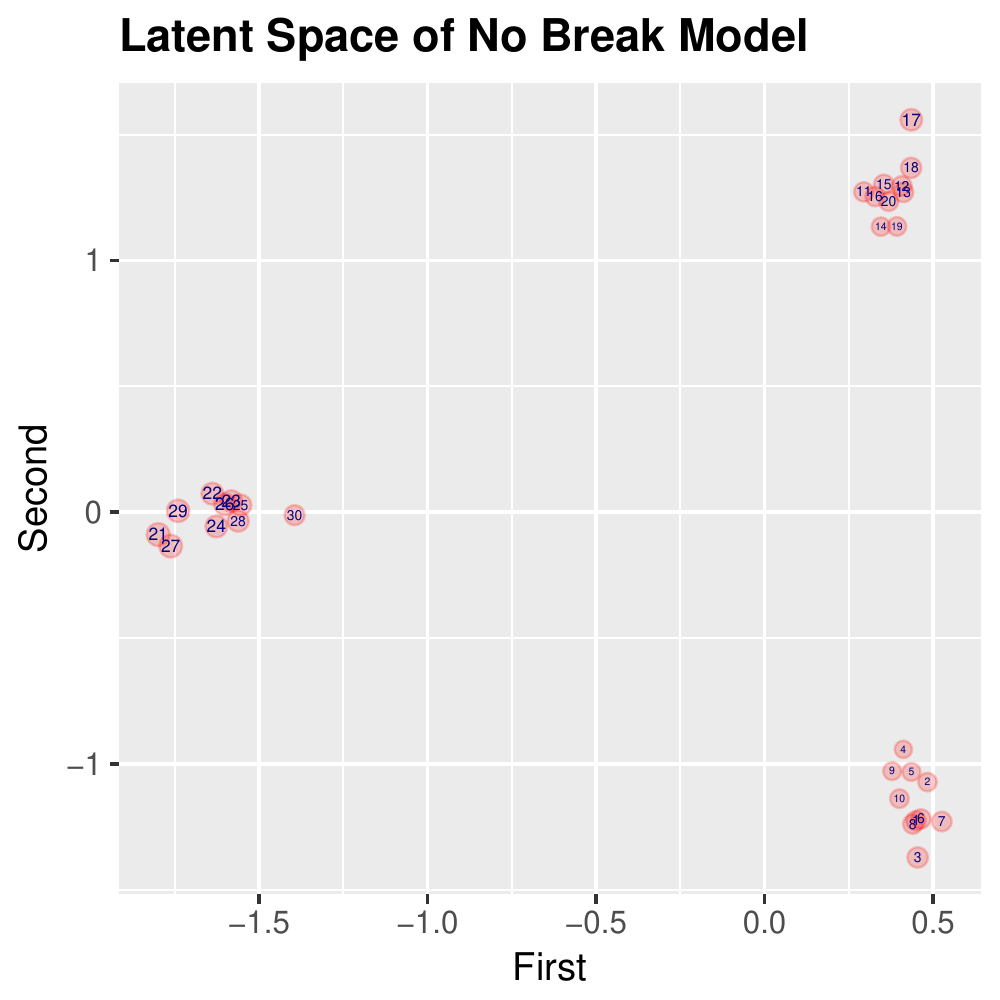}
    \end{minipage}
    &&&&
    \begin{minipage}{.18\textwidth}
      \includegraphics[width=\linewidth]{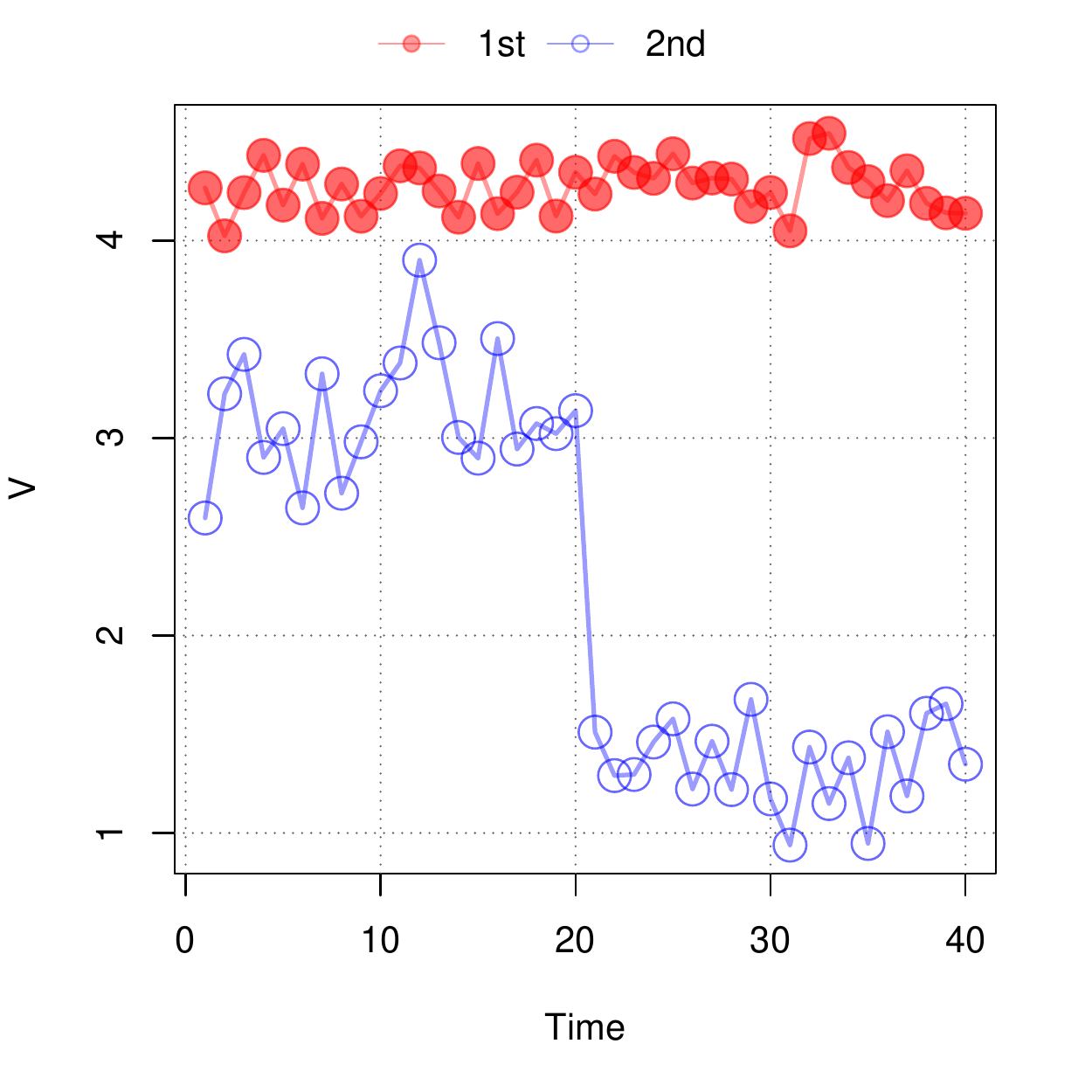}
    \end{minipage} \\
      {\tiny    \begin{tabular}{lcl } Break number &=&1\\ WAIC &=& 13080  \\ -2*log marginal &=& 12941 \\ -2*log likelihood &=& -6414 \\ Average Loss &=& 0.00 \end{tabular} }
&
    \begin{minipage}{.2\textwidth}
      \includegraphics[width=\linewidth]{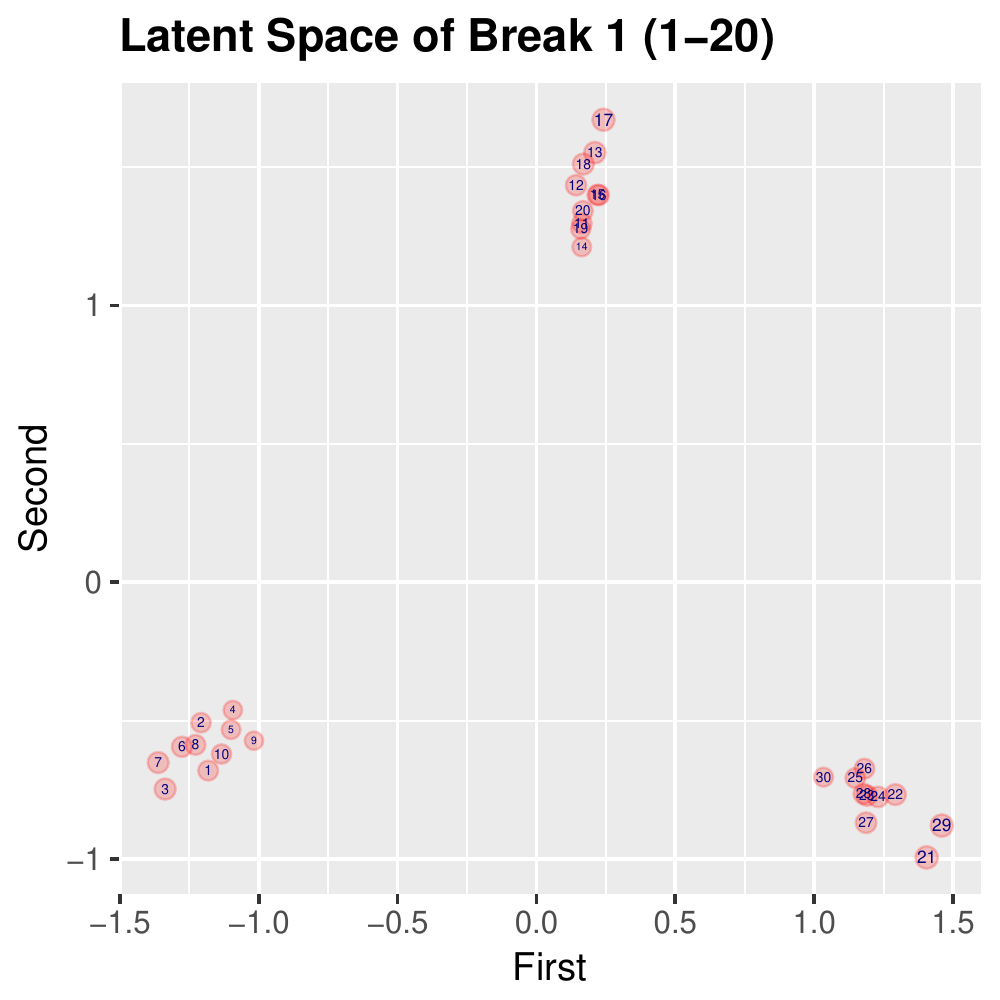}
    \end{minipage}
    &
        \begin{minipage}{.2\textwidth}
      \includegraphics[width=\linewidth]{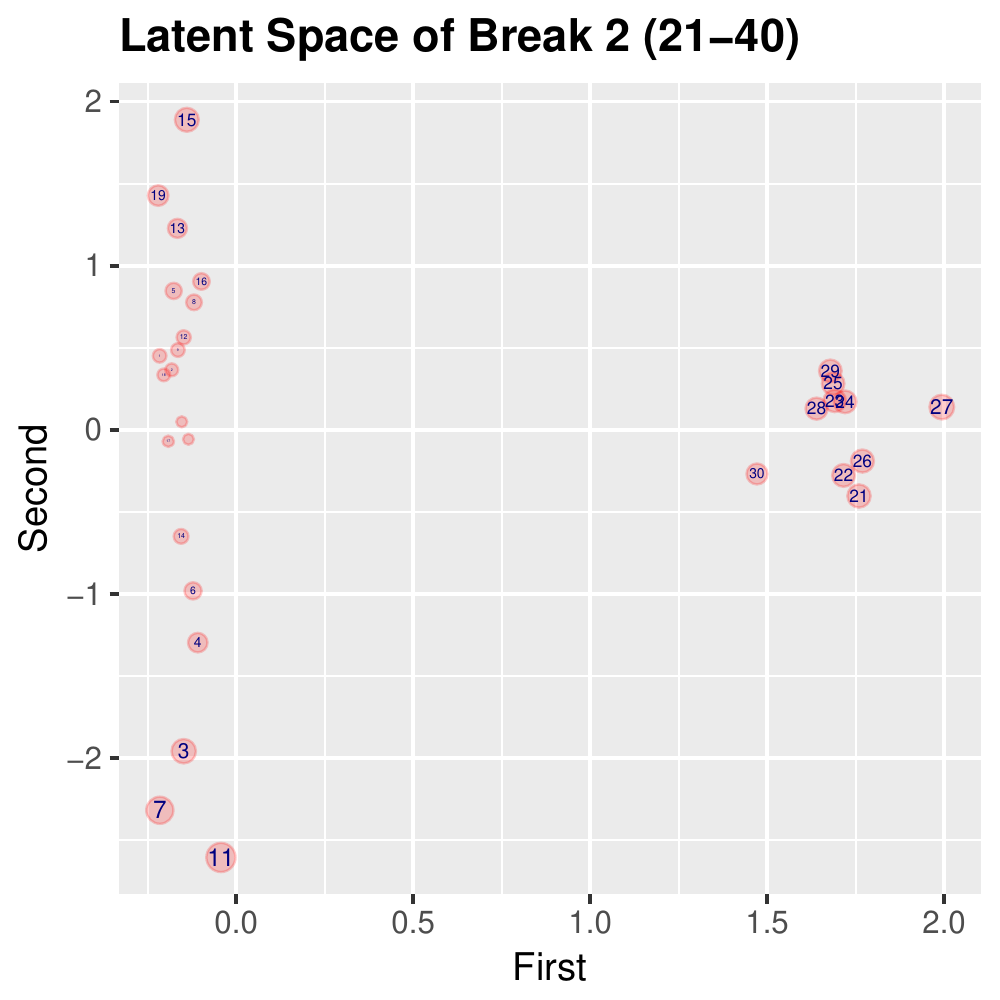}
    \end{minipage}
    &&&
    \begin{minipage}{.18\textwidth}
      \includegraphics[width=\linewidth]{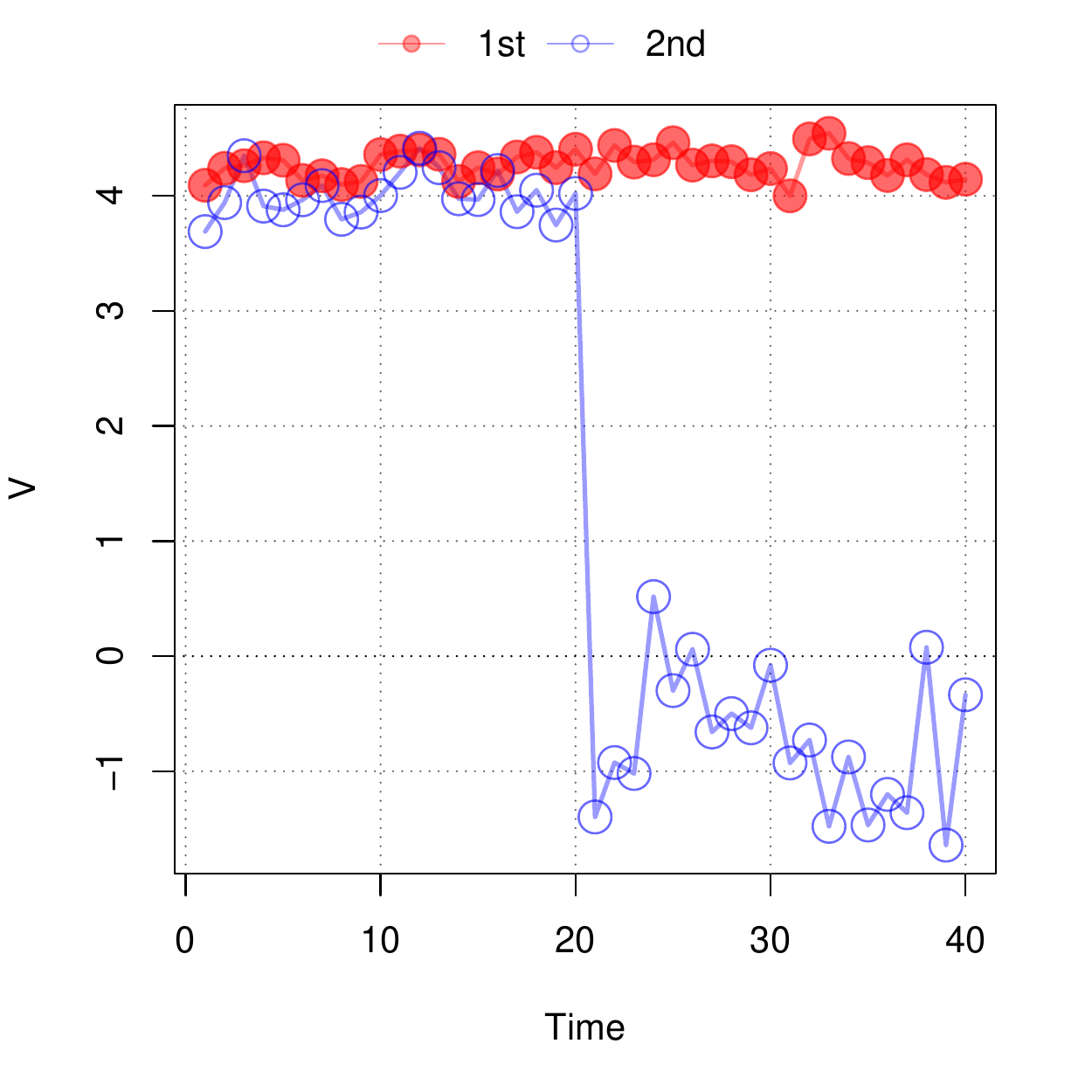}
    \end{minipage} \\
  {\tiny    \begin{tabular}{lcl } Break number &=&2 \\ WAIC &=& 13120\\ -2*log marginal &=&12962  \\ -2*log likelihood &=& 12840  \\ Average Loss &=& 0.01 \end{tabular} }	
&	 \begin{minipage}{.2\textwidth}
      \includegraphics[width=\linewidth]{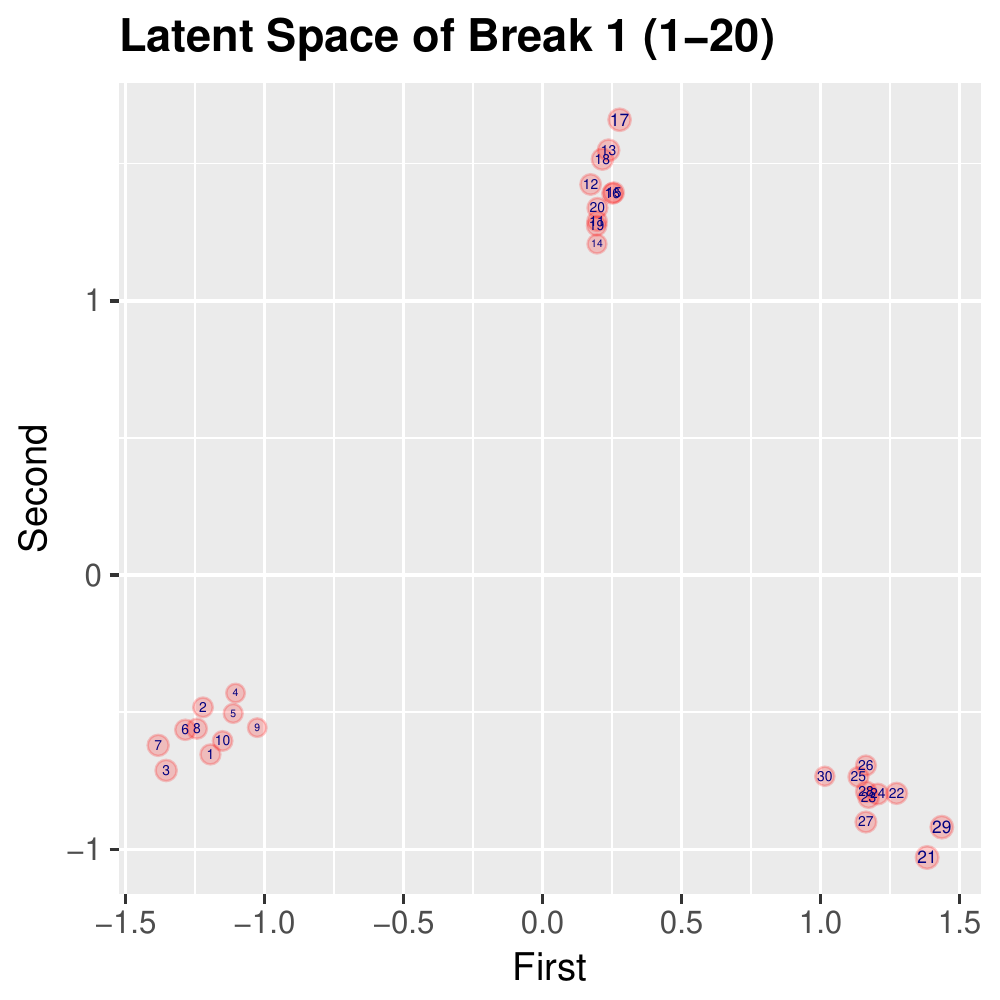}
    \end{minipage}
    &
        \begin{minipage}{.2\textwidth}
      \includegraphics[width=\linewidth]{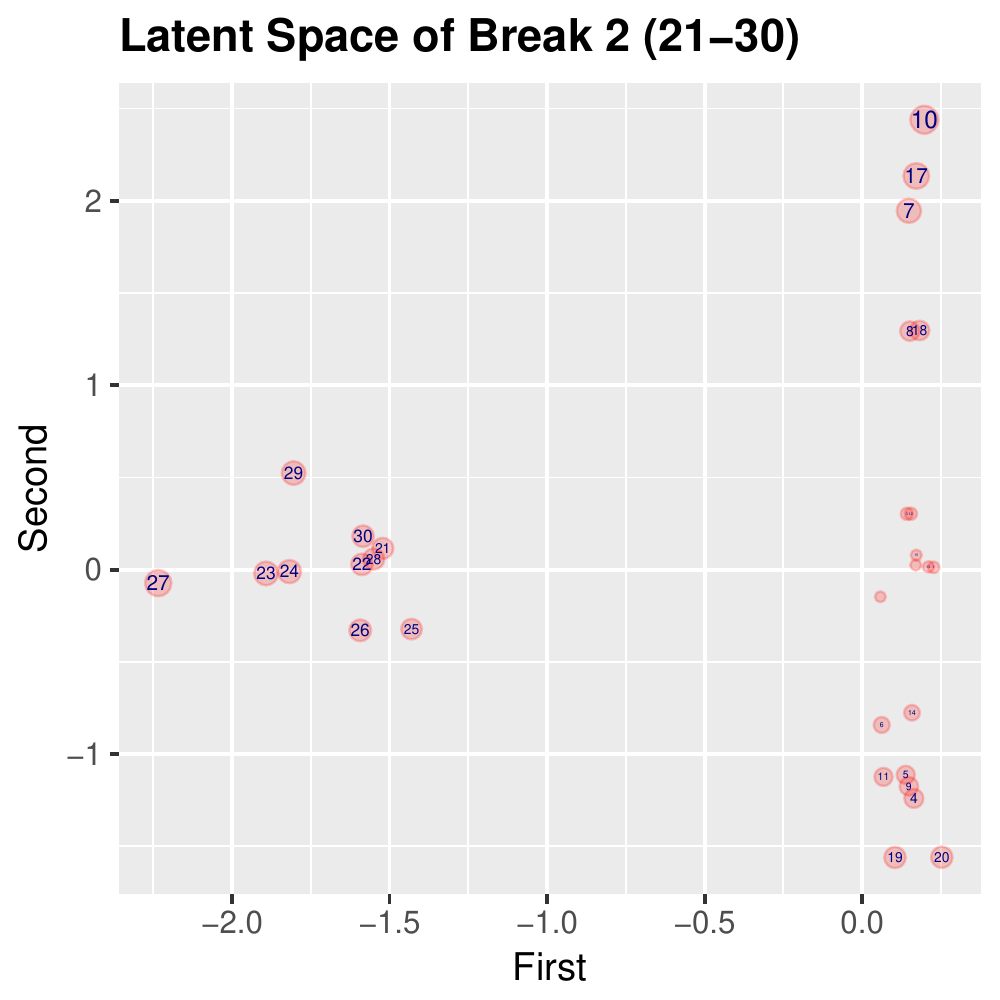}
    \end{minipage}
    &
            \begin{minipage}{.2\textwidth}
      \includegraphics[width=\linewidth]{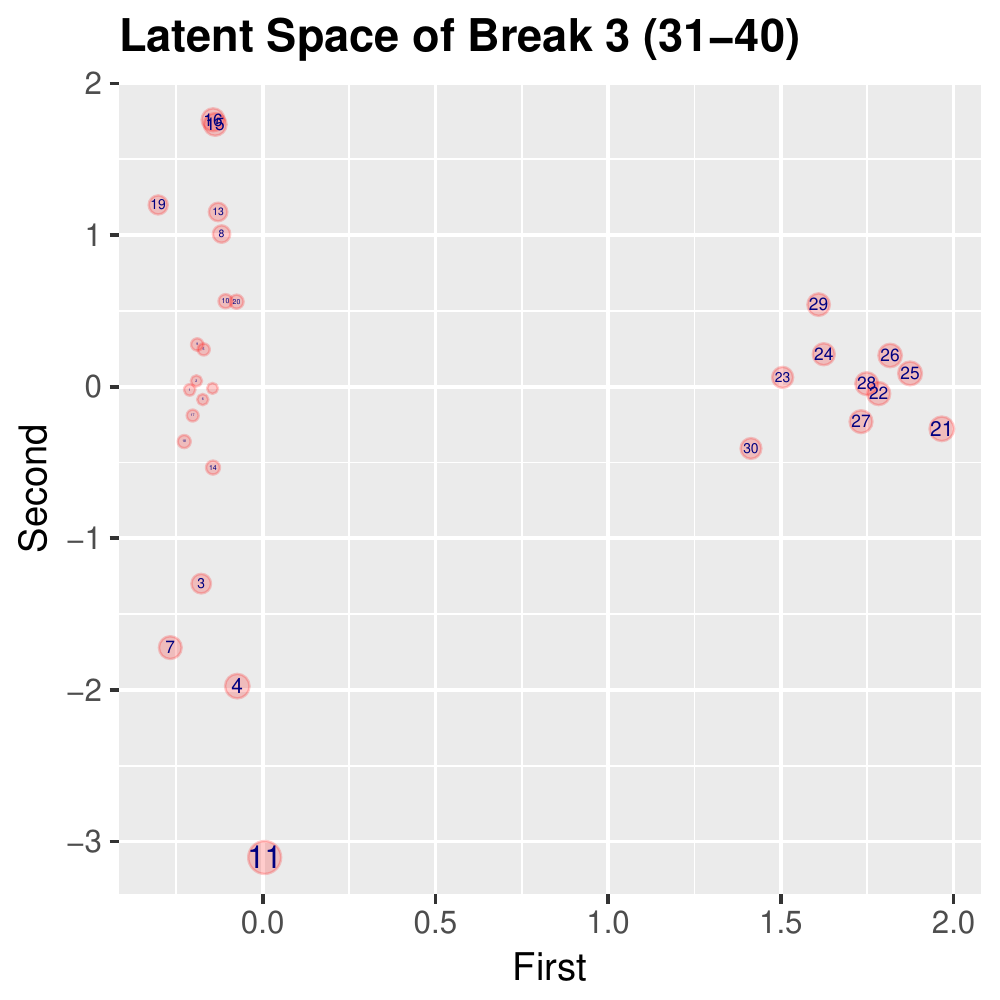}
    \end{minipage}
	    &&
    \begin{minipage}{.18\textwidth}
      \includegraphics[width=\linewidth]{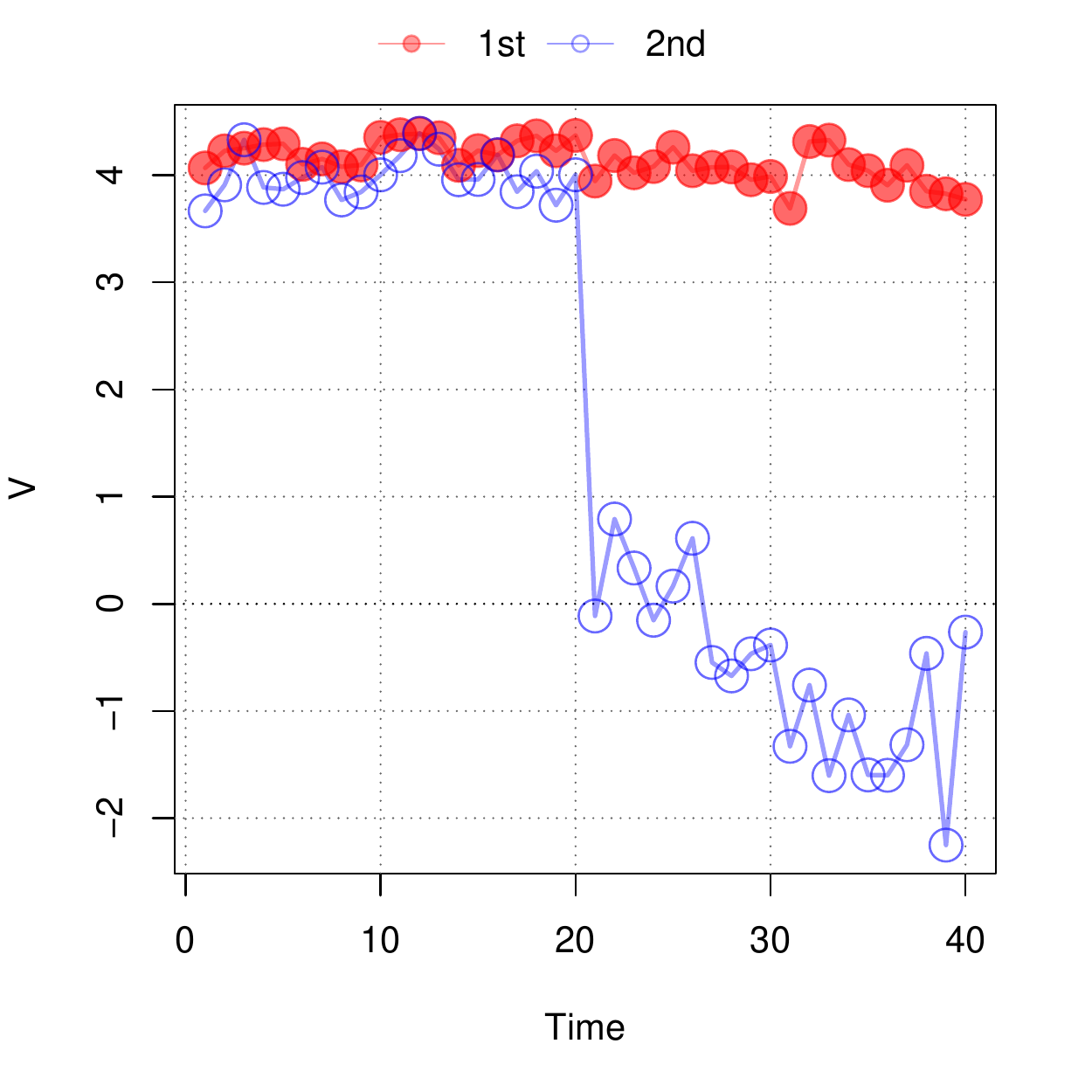}
    \end{minipage} \\
  {\tiny \begin{tabular}{lcl } Break number &=&3 \\ WAIC &=& 13092 \\ -2*log marginal &=&12885 \\ -2*log likelihood &=& 12698 \\ Average Loss &=& 0.01  \end{tabular} }
  &
	\begin{minipage}{.2\textwidth}
      \includegraphics[width=\linewidth]{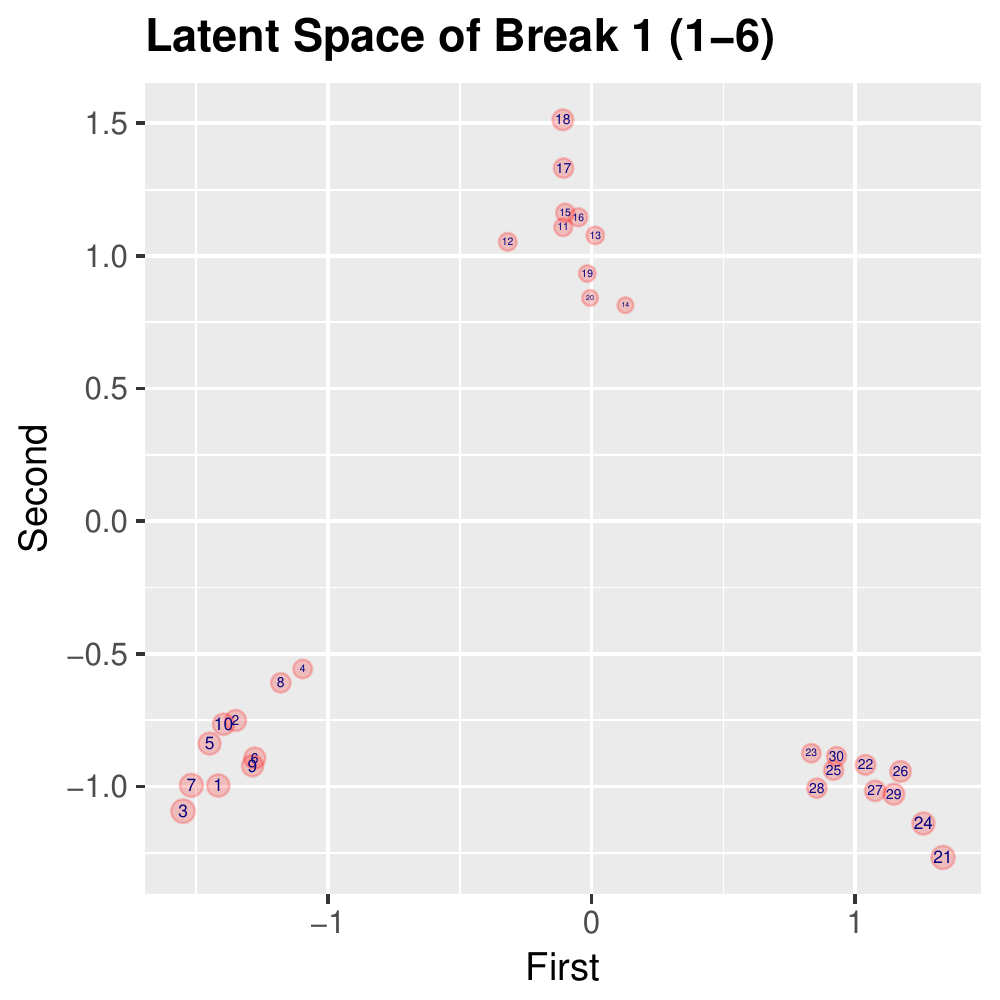}
    \end{minipage}
    &
        \begin{minipage}{.2\textwidth}
      \includegraphics[width=\linewidth]{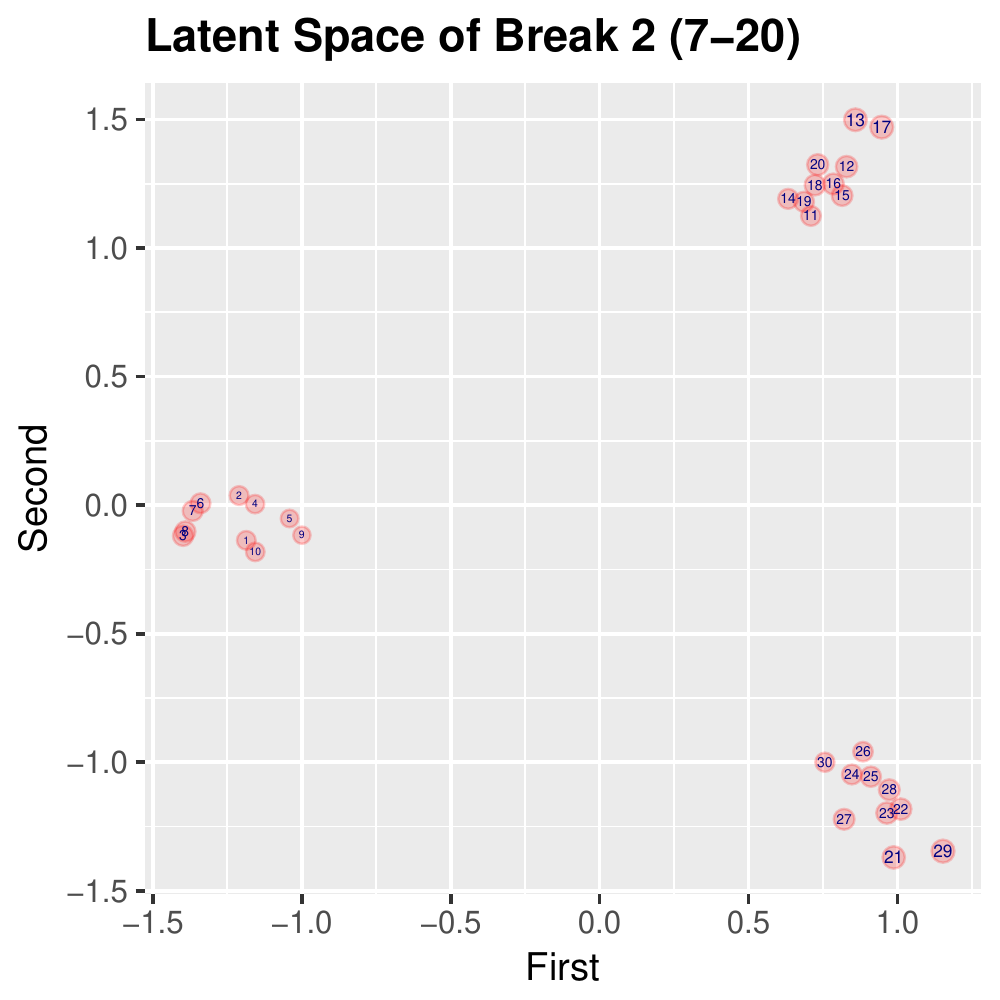}
    \end{minipage}
    &
            \begin{minipage}{.2\textwidth}
      \includegraphics[width=\linewidth]{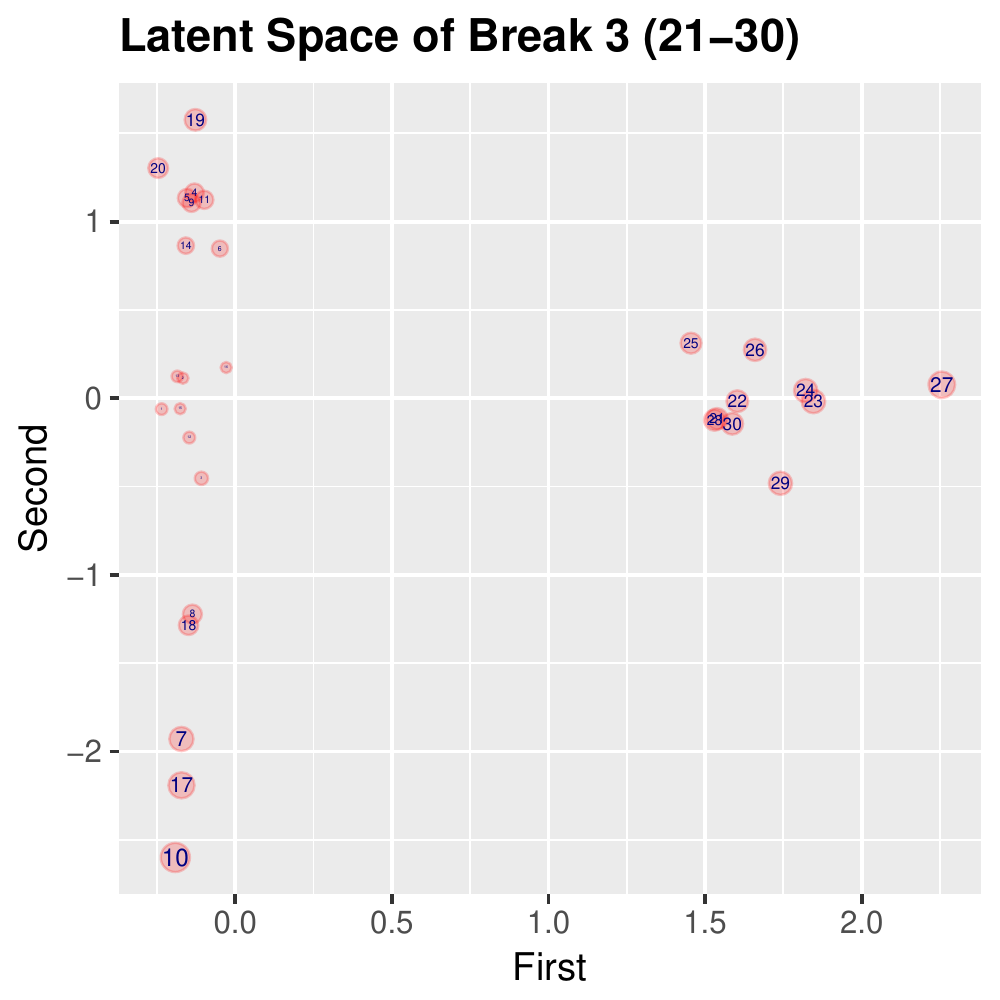}
    \end{minipage}
	&
	           \begin{minipage}{.18\textwidth}
      \includegraphics[width=\linewidth]{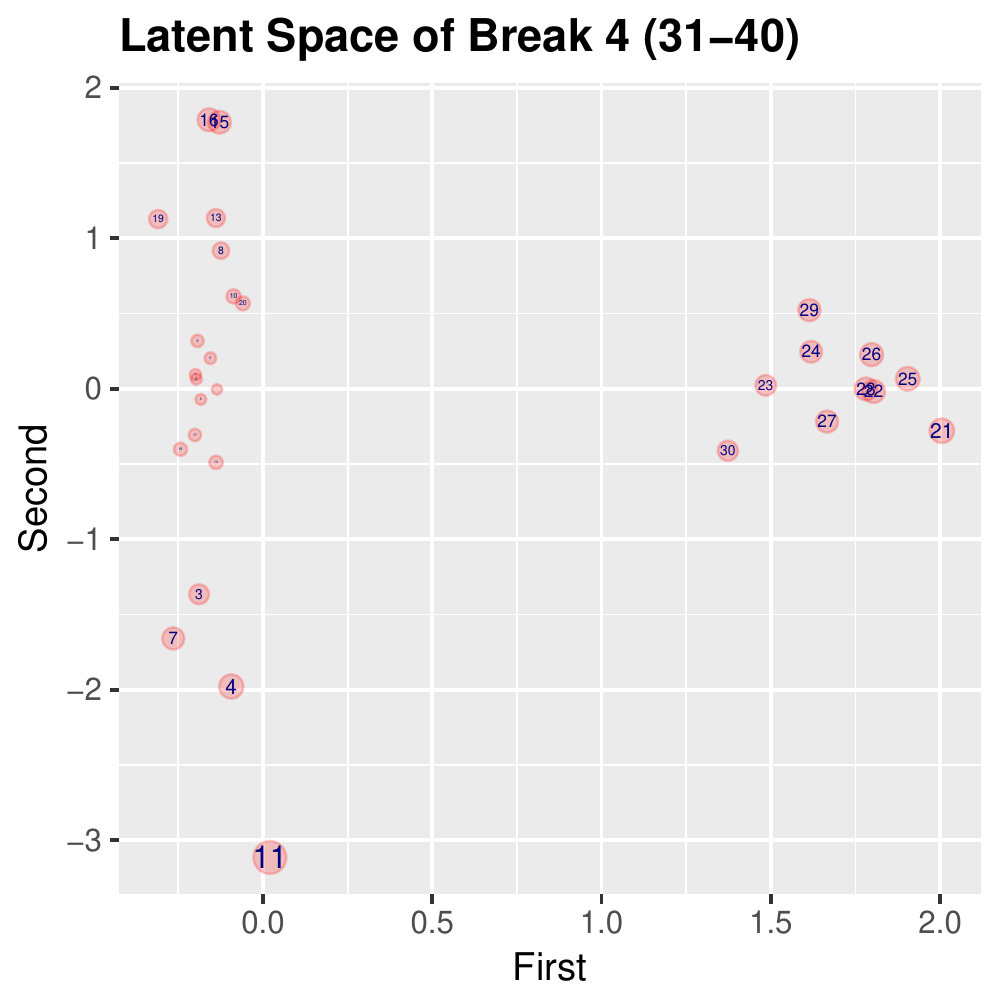}
    \end{minipage}
        &
    \begin{minipage}{.18\textwidth}
      \includegraphics[width=\linewidth]{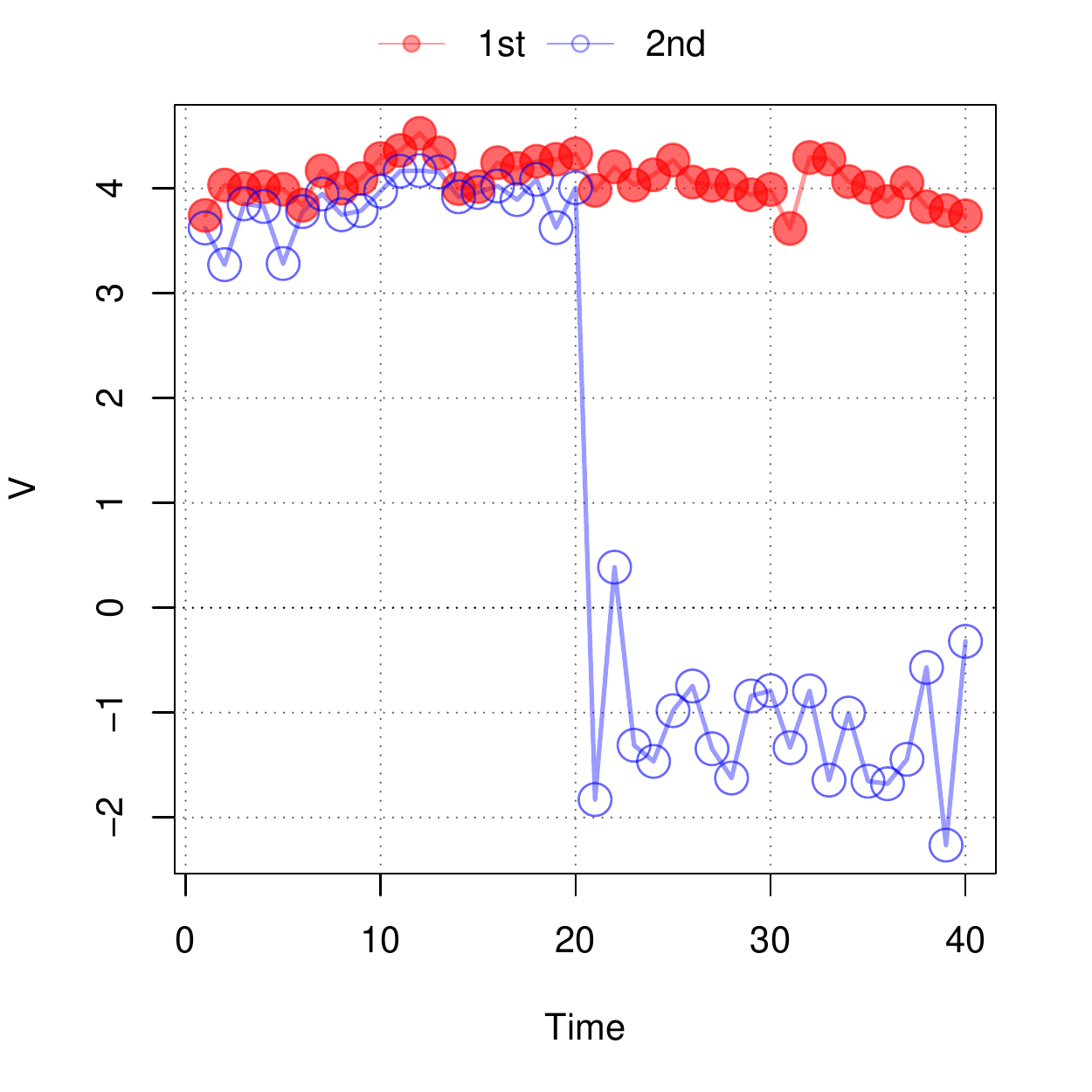}
    \end{minipage} \\
       \bottomrule
  \end{tabular}}
  \caption{Simulation Analysis of Block Merging MTRM. The ground truth is one break. All setup is analogous to the simulation examples in the main text. Two 10 node blocks were merged into a single block at $t=21$.}\label{sim.merge}
\end{sidewaystable}

\newpage
\subsection*{Results of Block-Splitting-Merging Network Changes}
Table \ref{sim.split-merge} summarizes the simulation results from the block-splitting-merging change example. The ground truth is the two break HMTM ($\mathcal{M}_2$). WAIC correctly identifies $\mathcal{M}_2$ as the best-fitting model while the approximate log marginal likelihood favors $\mathcal{M}_3$.  Note that Regime 3 of $\mathcal{M}_3$ has only two observations, which inflates the approximate log marginal likelihood. The results of $\mathcal{M}_2$ correctly recovers the block-splitting-merging change in the latent node positions and network generation rules. 

\begin{sidewaystable}\footnotesize  \centering
\resizebox{\textwidth}{!}{  \begin{tabular}{p{4cm} cccc c}
    \toprule
    Model Fit& \multicolumn{4}{c}{Latent Space ($\mathbf{U}_m$) Changes}&  \multicolumn{1}{c}{Generation Rule ($\mathbf{v}_t$)} \\ 
    
    & Regime 1& Regime 2&Regime 3& Regime 4& \\\midrule
     \begin{tabular}{lcl } Break number &=&0 \\ WAIC &=& 13744  \\ -2*log marginal &=& 13648 \\ -2*log likelihood &=& 13592 \end{tabular} 
  &
    \begin{minipage}{.18\textwidth}
      \includegraphics[width=\linewidth]{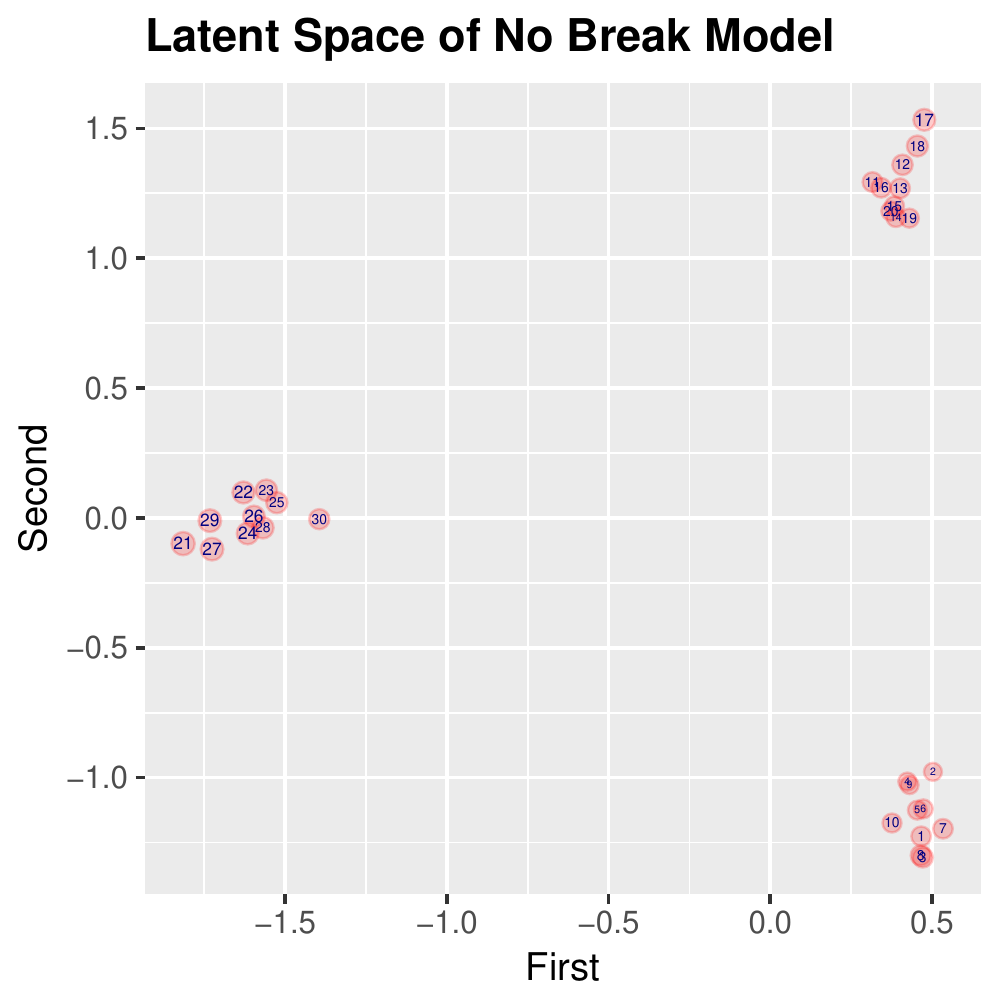}
    \end{minipage}
    &&&&
    \begin{minipage}{.18\textwidth}
      \includegraphics[width=\linewidth]{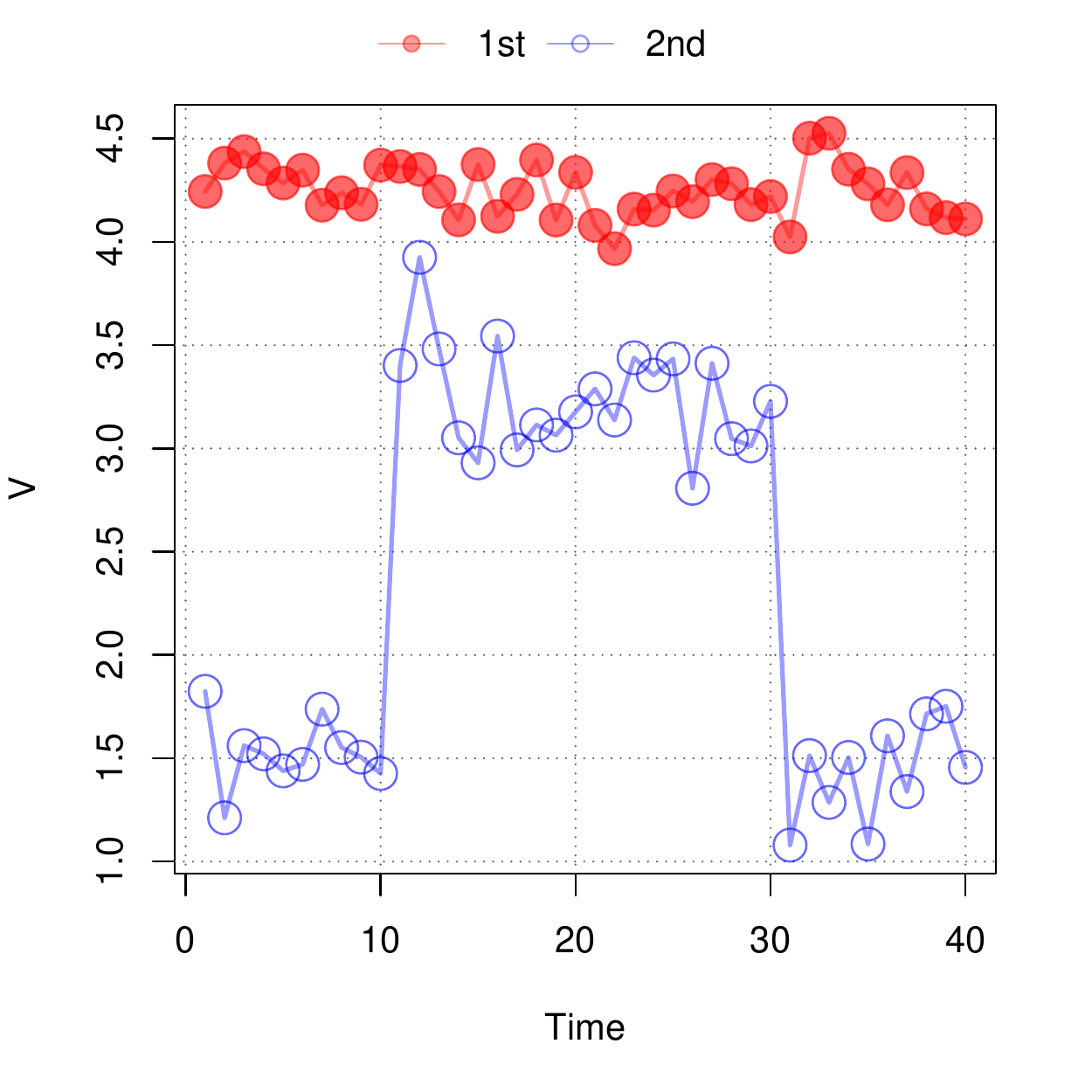}
    \end{minipage} \\
     \begin{tabular}{lcl } Break number &=&1\\ WAIC &=& 13608  \\ -2*log marginal &=& 13469 \\ -2*log likelihood &=& 13366  \\ Average Loss &=& 0.02\end{tabular} 
&
    \begin{minipage}{.18\textwidth}
      \includegraphics[width=\linewidth]{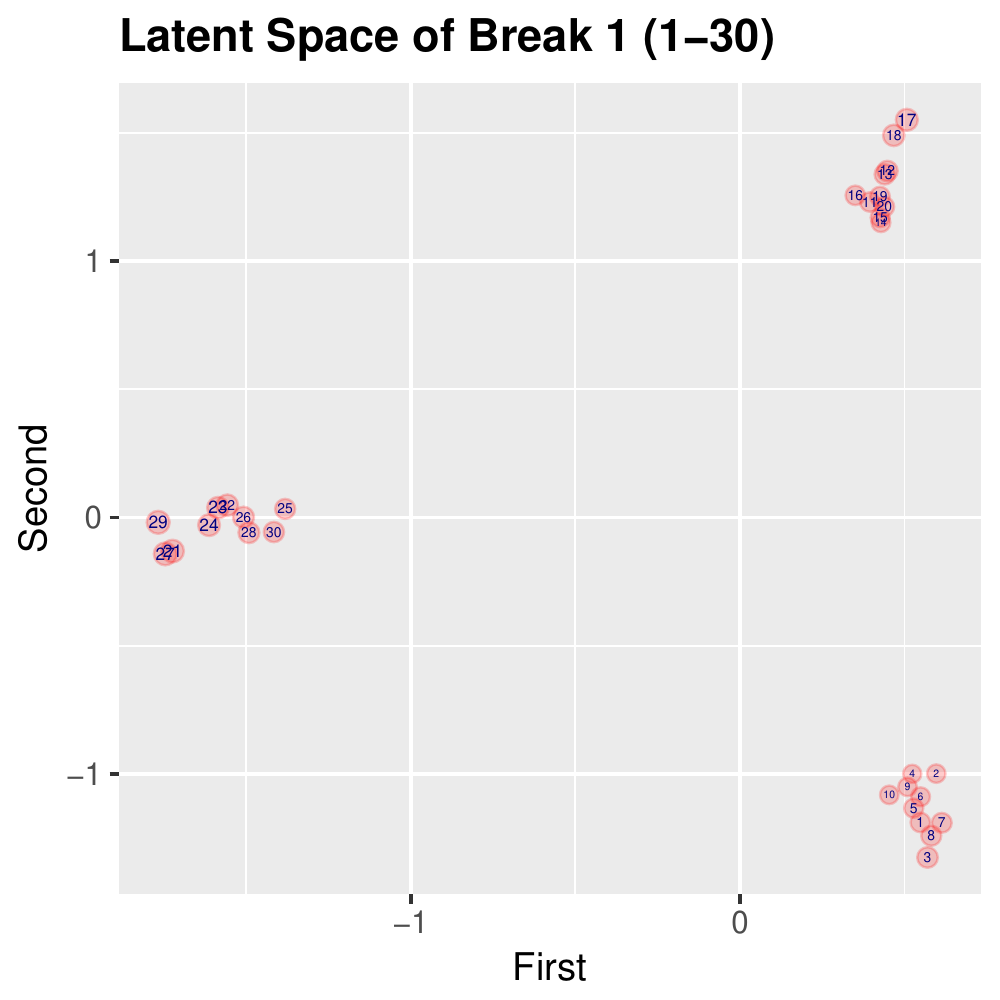}
    \end{minipage}
    &
        \begin{minipage}{.18\textwidth}
      \includegraphics[width=\linewidth]{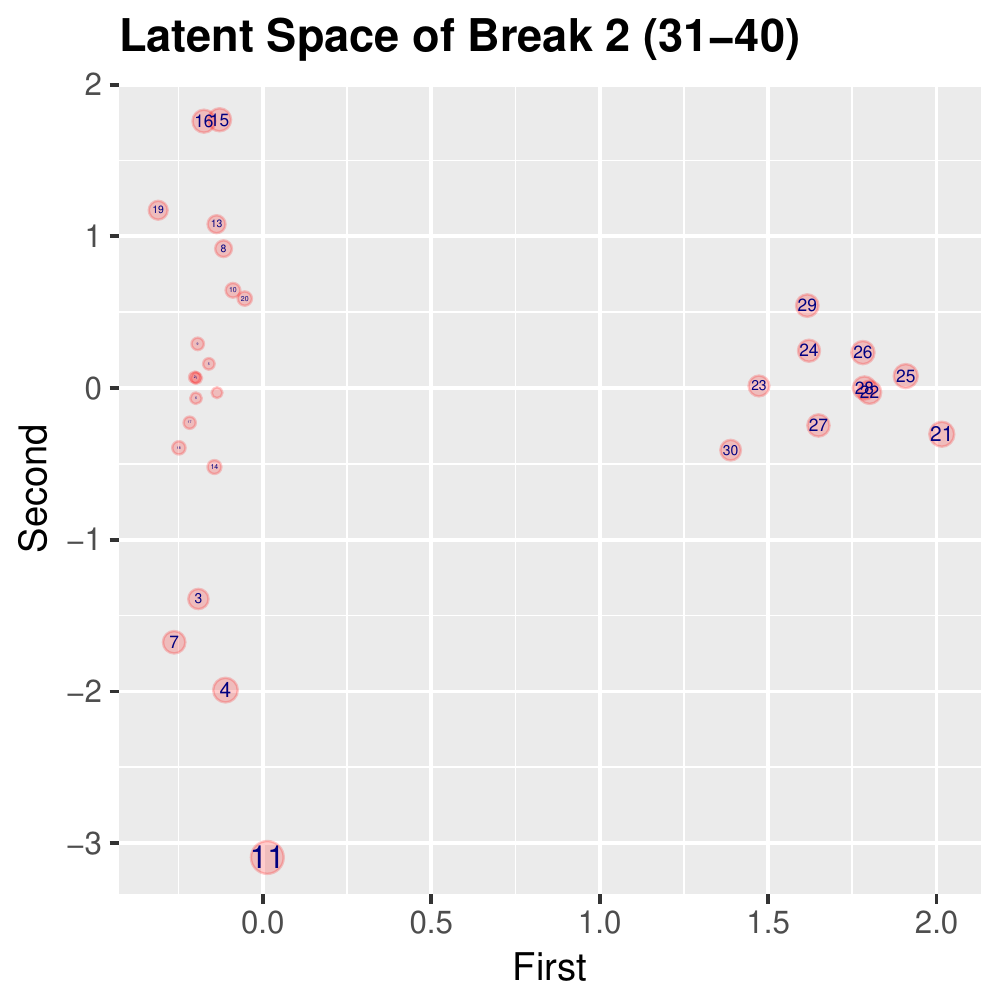}
    \end{minipage}
    &&&
    \begin{minipage}{.18\textwidth}
      \includegraphics[width=\linewidth]{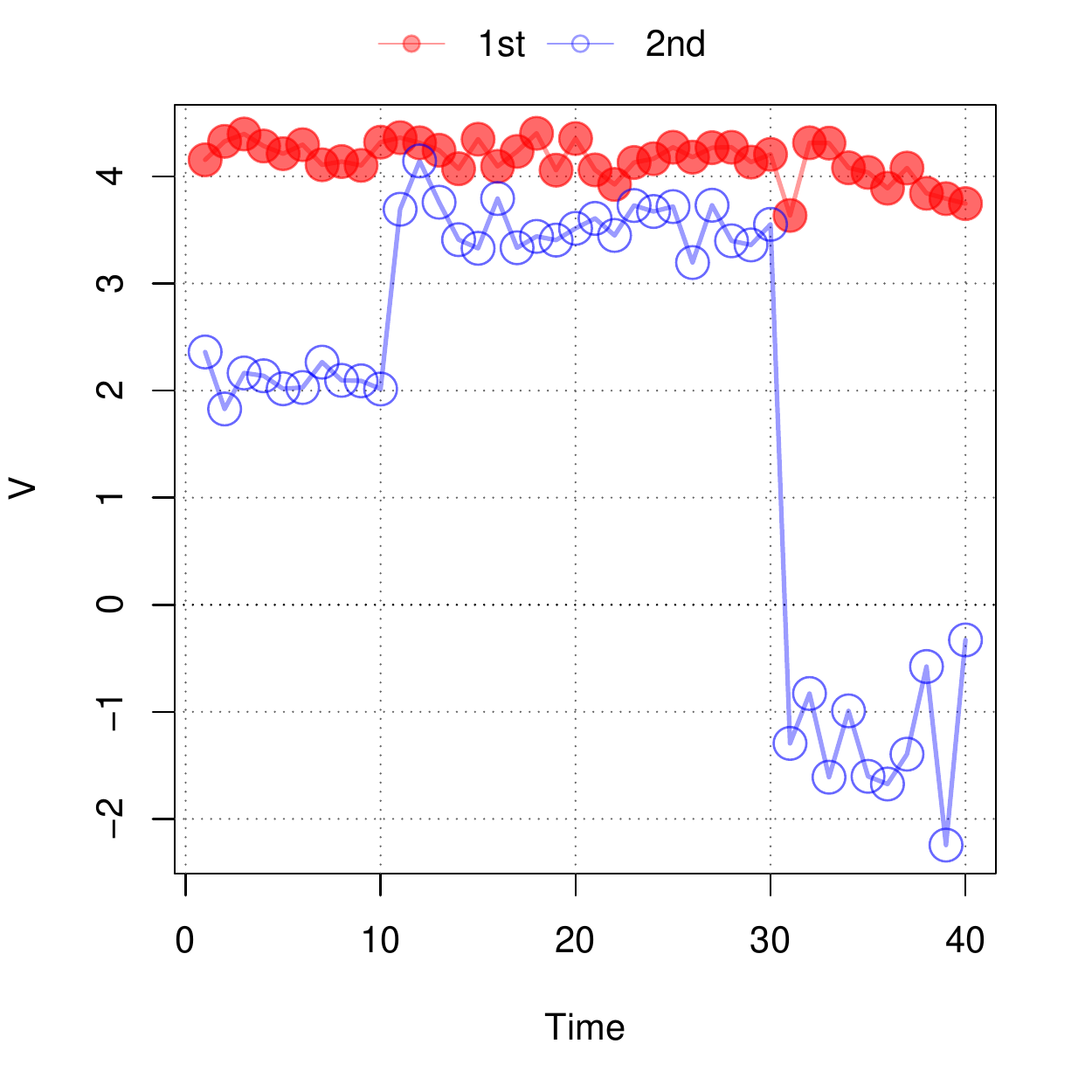}
    \end{minipage} \\
    \begin{tabular}{lcl } Break number &=&2 \\ WAIC &=& 13031\\ -2*log marginal &=&12946  \\ -2*log likelihood &=& 12806   \\ Average Loss &=& 0.00 \end{tabular} 
	&
	 \begin{minipage}{.18\textwidth}
      \includegraphics[width=\linewidth]{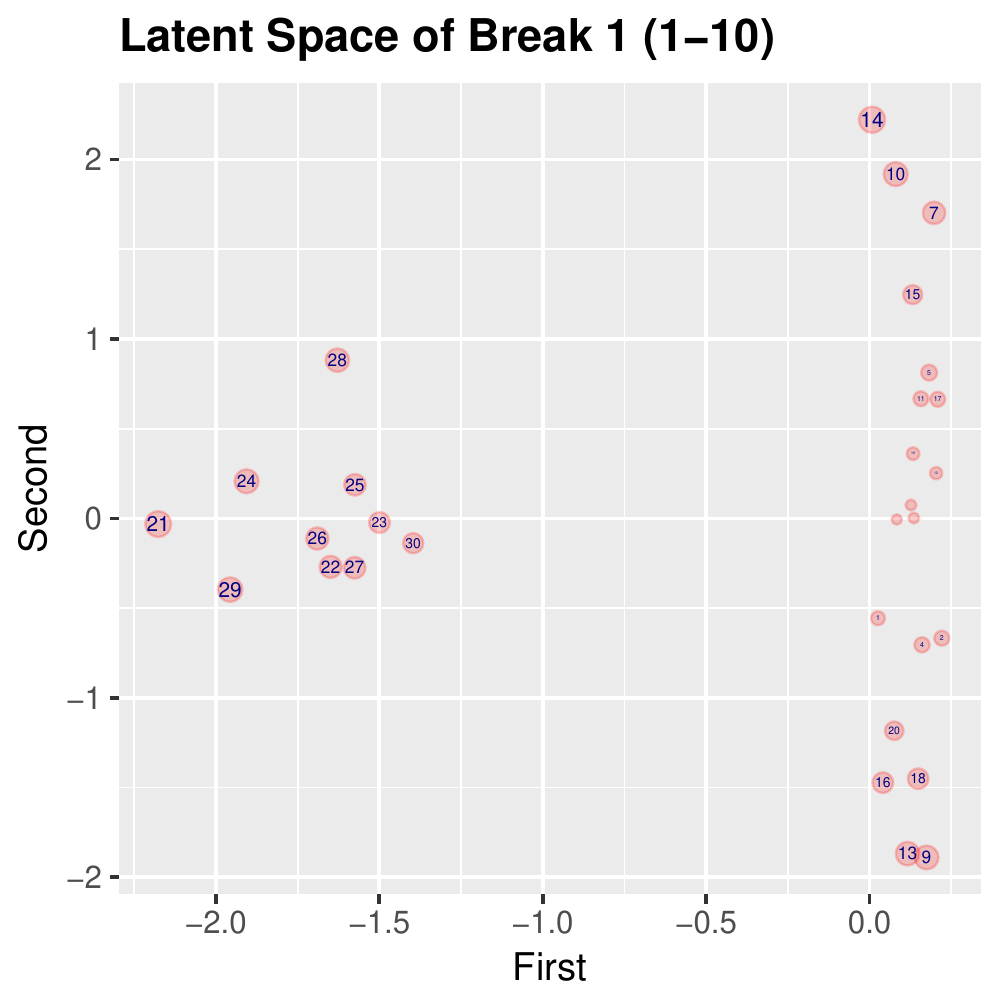}
    \end{minipage}
    &
        \begin{minipage}{.18\textwidth}
      \includegraphics[width=\linewidth]{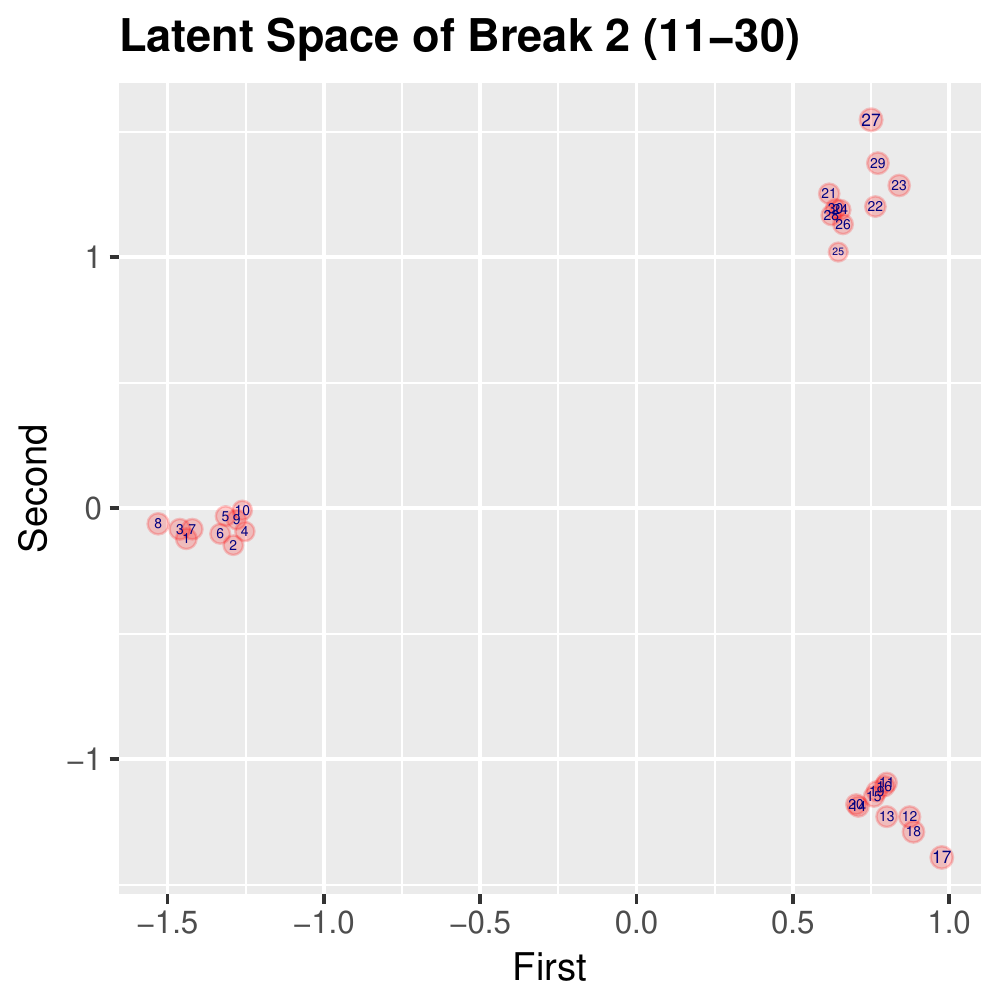}
    \end{minipage}
    &
            \begin{minipage}{.18\textwidth}
      \includegraphics[width=\linewidth]{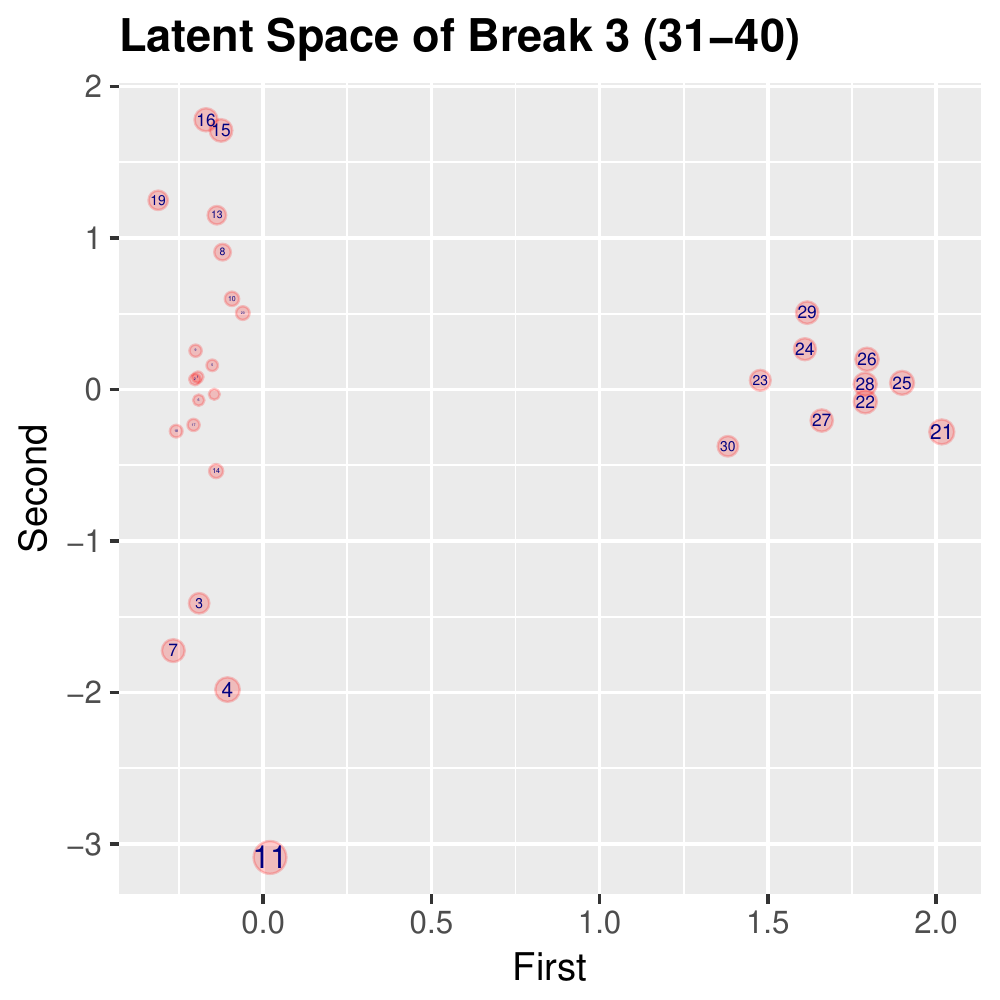}
    \end{minipage}
	&&
    \begin{minipage}{.18\textwidth}
      \includegraphics[width=\linewidth]{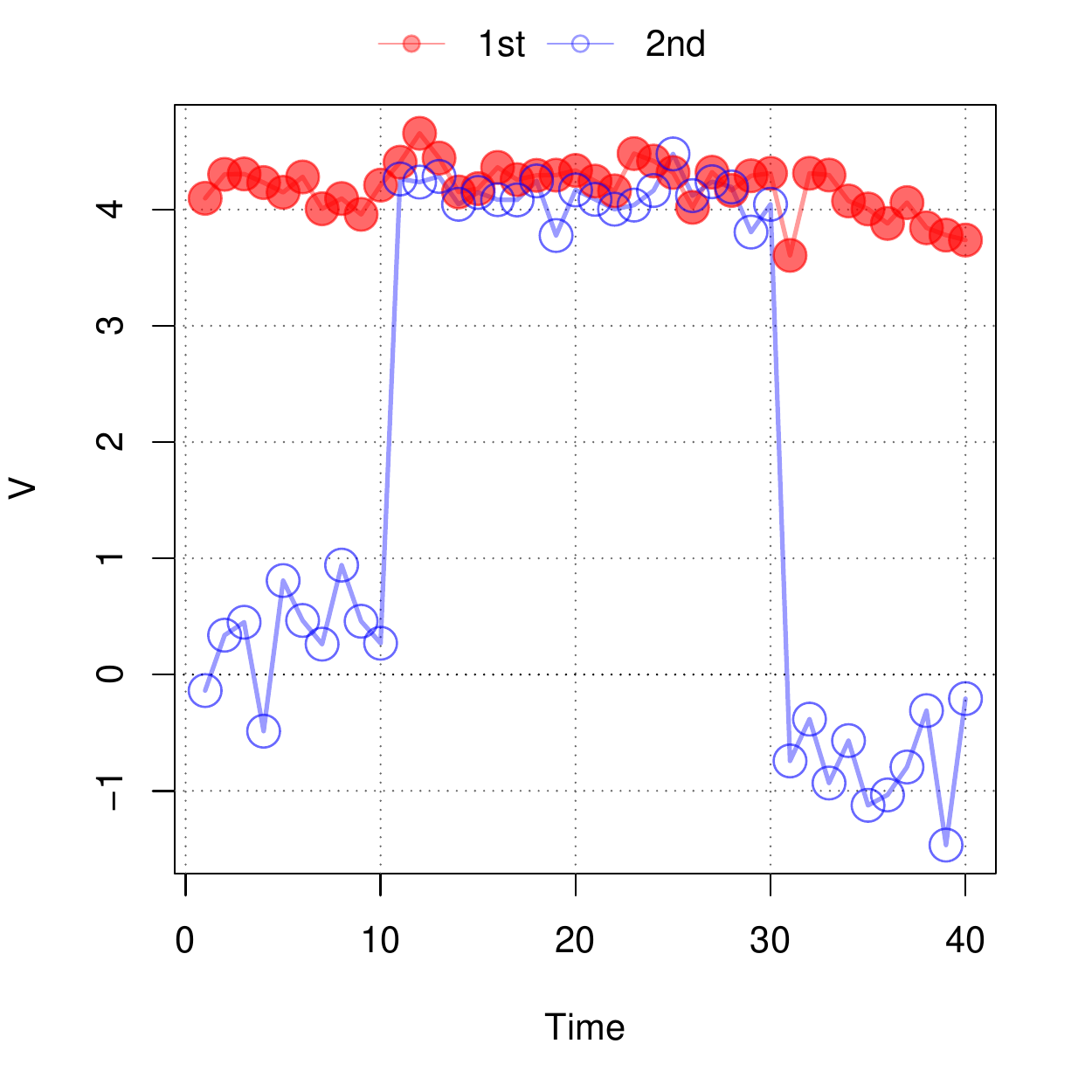}
    \end{minipage} \\
    \begin{tabular}{lcl } Break number &=&3 \\ WAIC &=& 13034 \\ -2*log marginal &=&12890 \\ -2*log likelihood &=& 12714   \\ Average Loss &=& 0.00 \end{tabular} 
  &
	\begin{minipage}{.18\textwidth}
      \includegraphics[width=\linewidth]{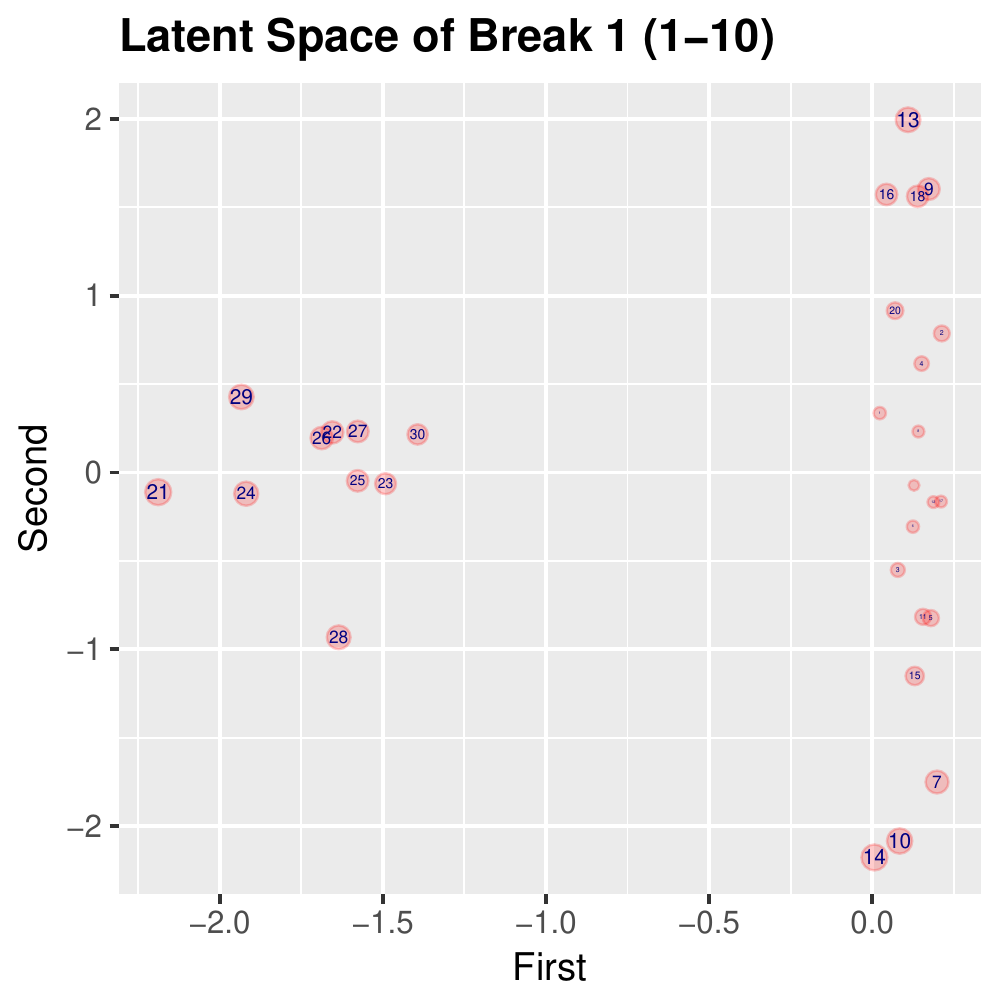}
    \end{minipage}
    &
        \begin{minipage}{.18\textwidth}
      \includegraphics[width=\linewidth]{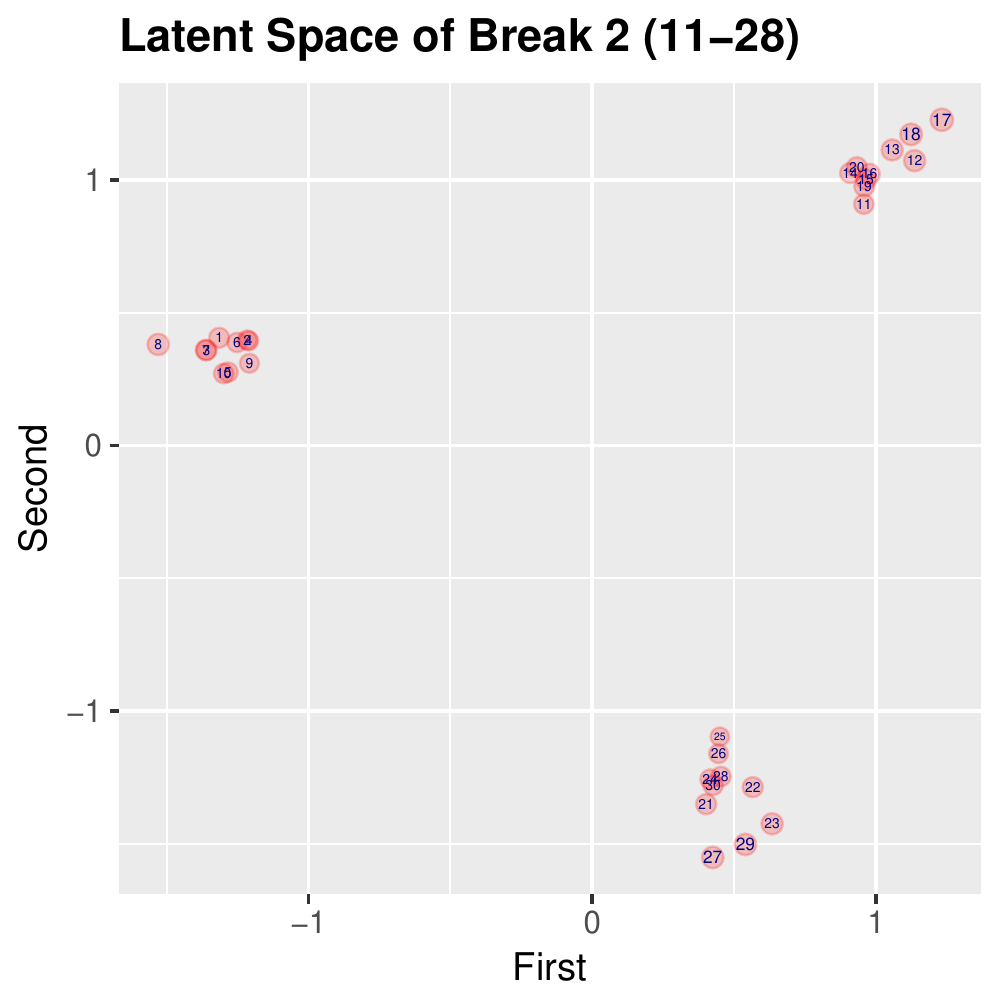}
    \end{minipage}
    &
            \begin{minipage}{.18\textwidth}
      \includegraphics[width=\linewidth]{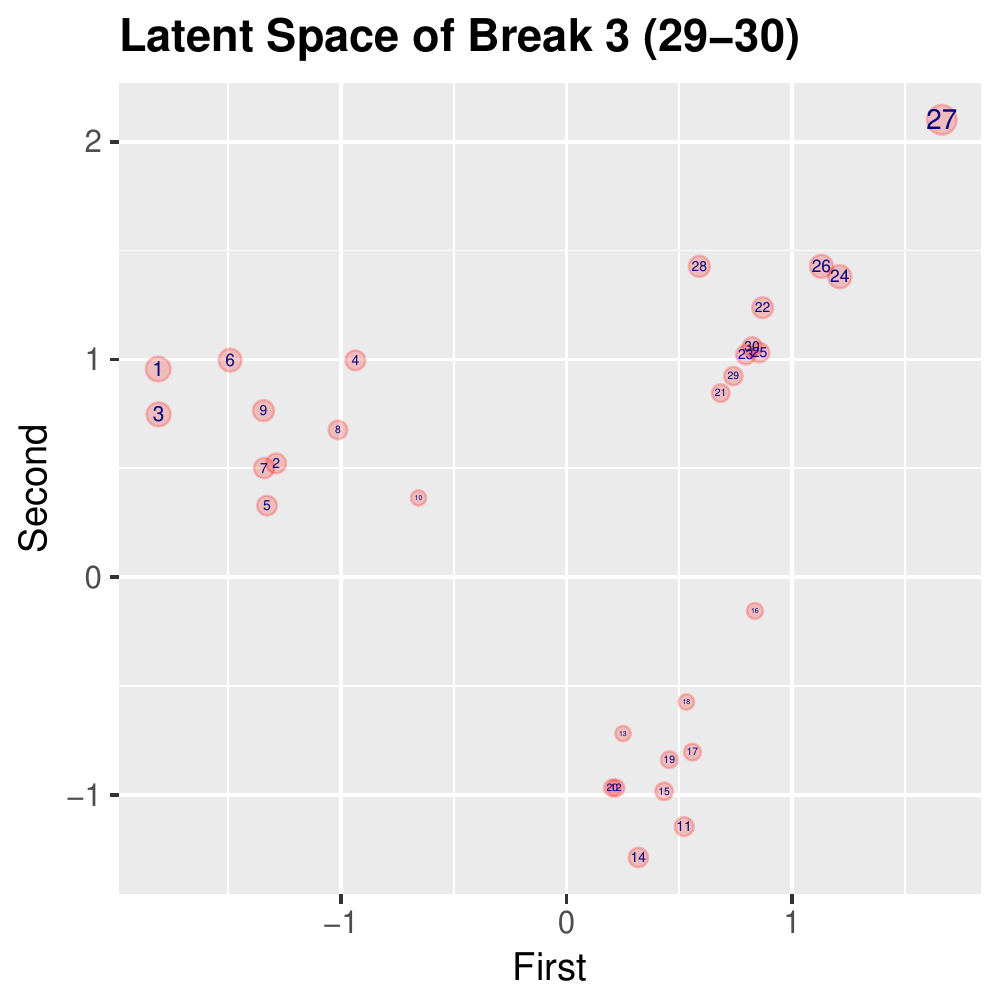}
    \end{minipage}
	&
	           \begin{minipage}{.18\textwidth}
      \includegraphics[width=\linewidth]{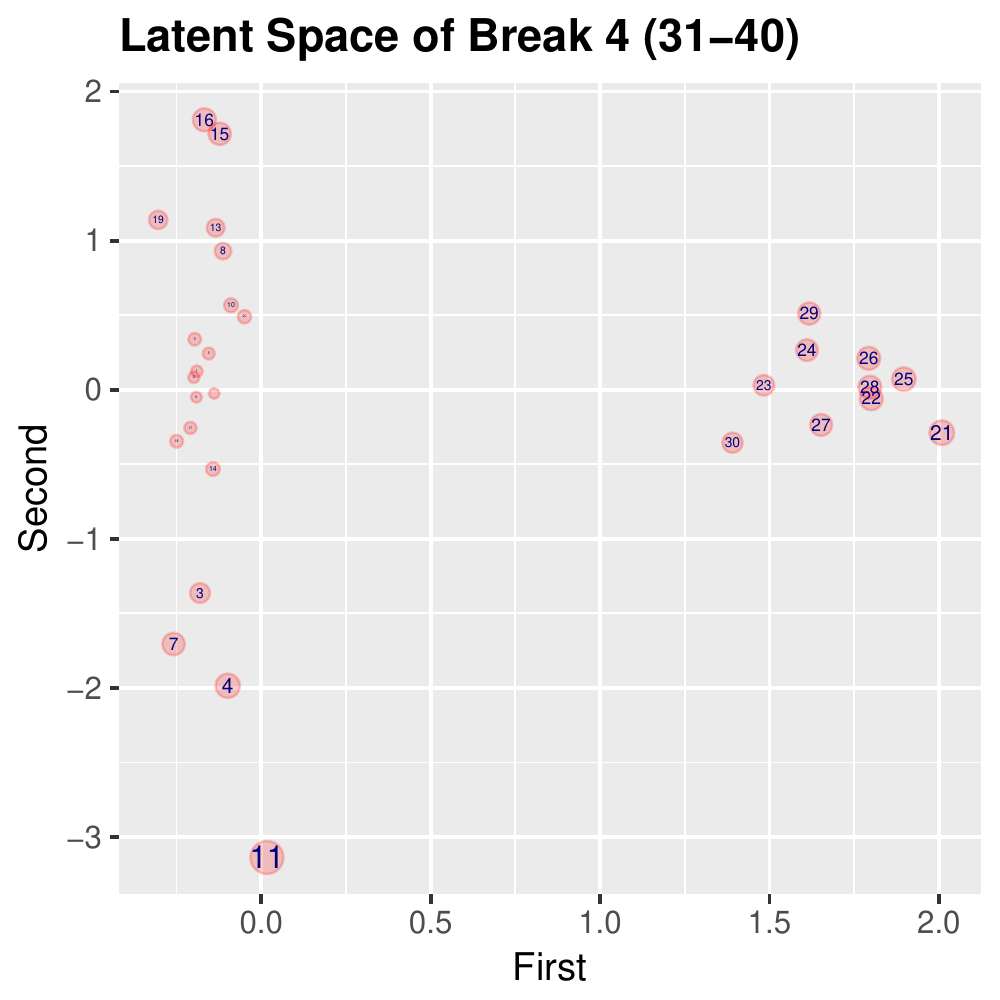}
    \end{minipage}
&
    \begin{minipage}{.18\textwidth}
      \includegraphics[width=\linewidth]{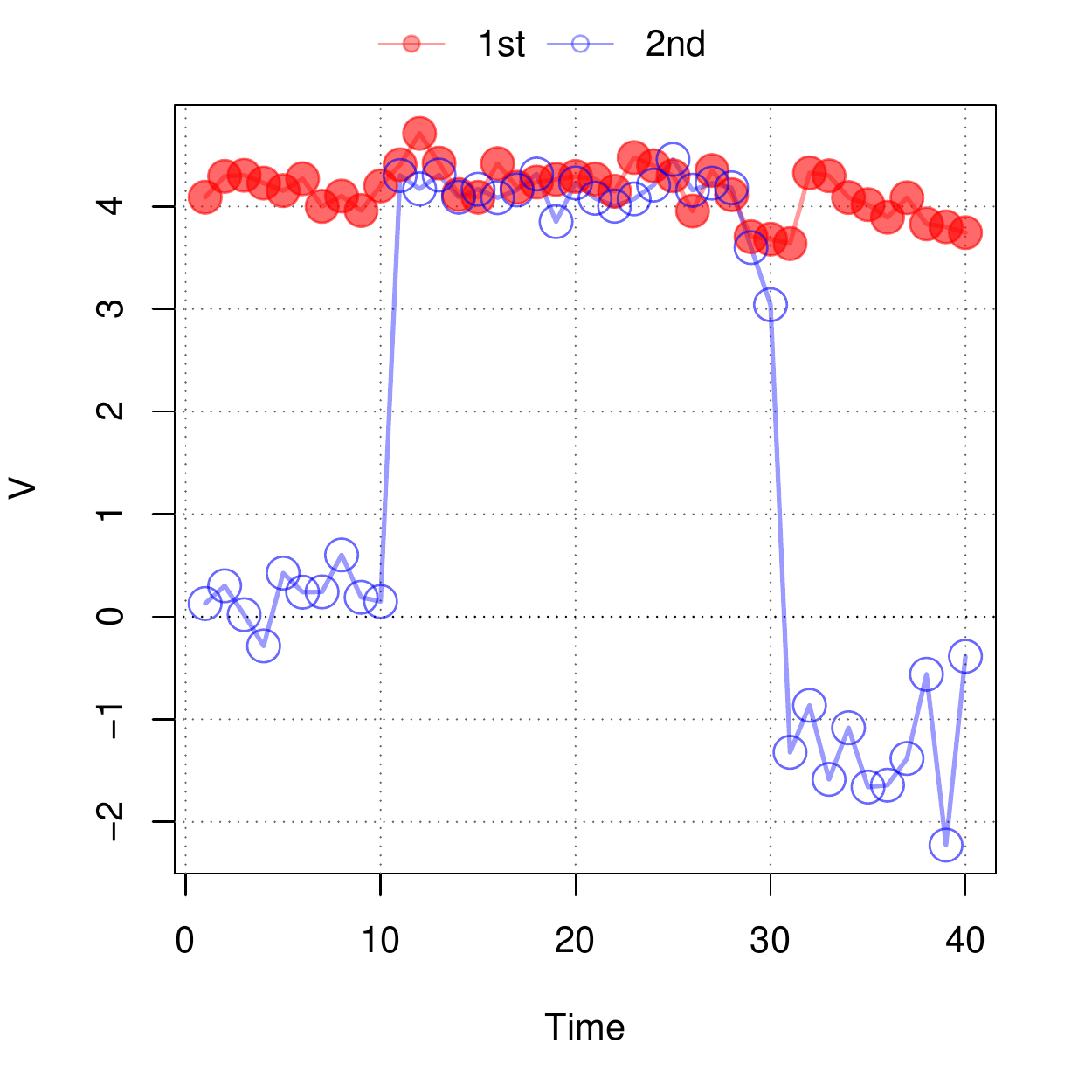}
    \end{minipage} \\
     \bottomrule
  \end{tabular}}
  \caption{Simulation Analysis of Block Split-Merging MTRM. The ground truth is two breaks. All parameters are analogous to the example depicted in Table 3 of the main text except the order of the changes.}\label{sim.split-merge}
\end{sidewaystable}

\clearpage
\section*{Software Implementation}
\begin{verbatim}
require(NetworkChange)
set.seed(1973)

## Generate an array (30 by 30 by 40) with block transitions from 2 blocks to 3 blocks
Y <- MakeBlockNetworkChange(n=10, T=40, type ="split")
G <- 100 ## only 100 mcmc scans to save time

## Fit models
out0 <- NetworkStatic(Y, R=2, mcmc=G, burnin=G, verbose=G, Waic=TRUE)
out1 <- NetworkChange(Y, R=2, m=1, mcmc=G, burnin=G, verbose=G, Waic=TRUE)
out2 <- NetworkChange(Y, R=2, m=2, mcmc=G, burnin=G, verbose=G, Waic=TRUE)
out3 <- NetworkChange(Y, R=2, m=3, mcmc=G, burnin=G, verbose=G, Waic=TRUE)
outlist <- list(out0, out1, out2, out3)

## The true model is out1
WaicCompare(outlist)
MarginalCompare(out)
BreakPointLoss(list(out1, out2, out3))[[1]]

## plot latent node positions
plotU(out1)

## plot layer-specific network generation rules
plotV(out1)

\end{verbatim}

\begin{figure}[!hp]\centering \vspace{0.2cm}
	\caption{Software Ouputs of \texttt{plotU(out1)}}\label{plotU}
	\includegraphics[width= .8 \textwidth]{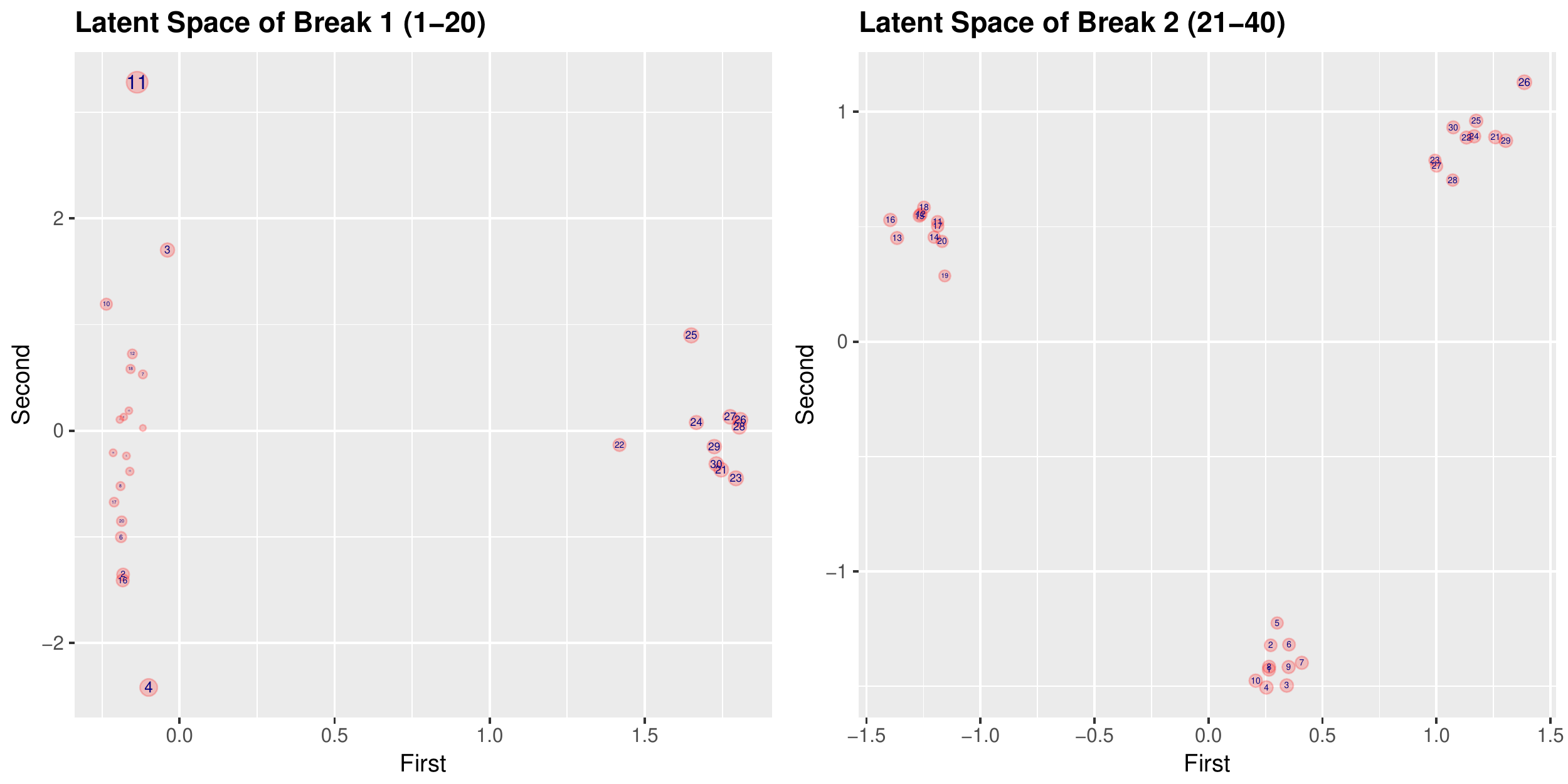}		
\end{figure}

\begin{figure}[!hp]\centering \vspace{0.2cm}
	\caption{Software Ouputs of \texttt{plotV(out1)}}\label{plotV}
	\includegraphics[width=.5 \textwidth]{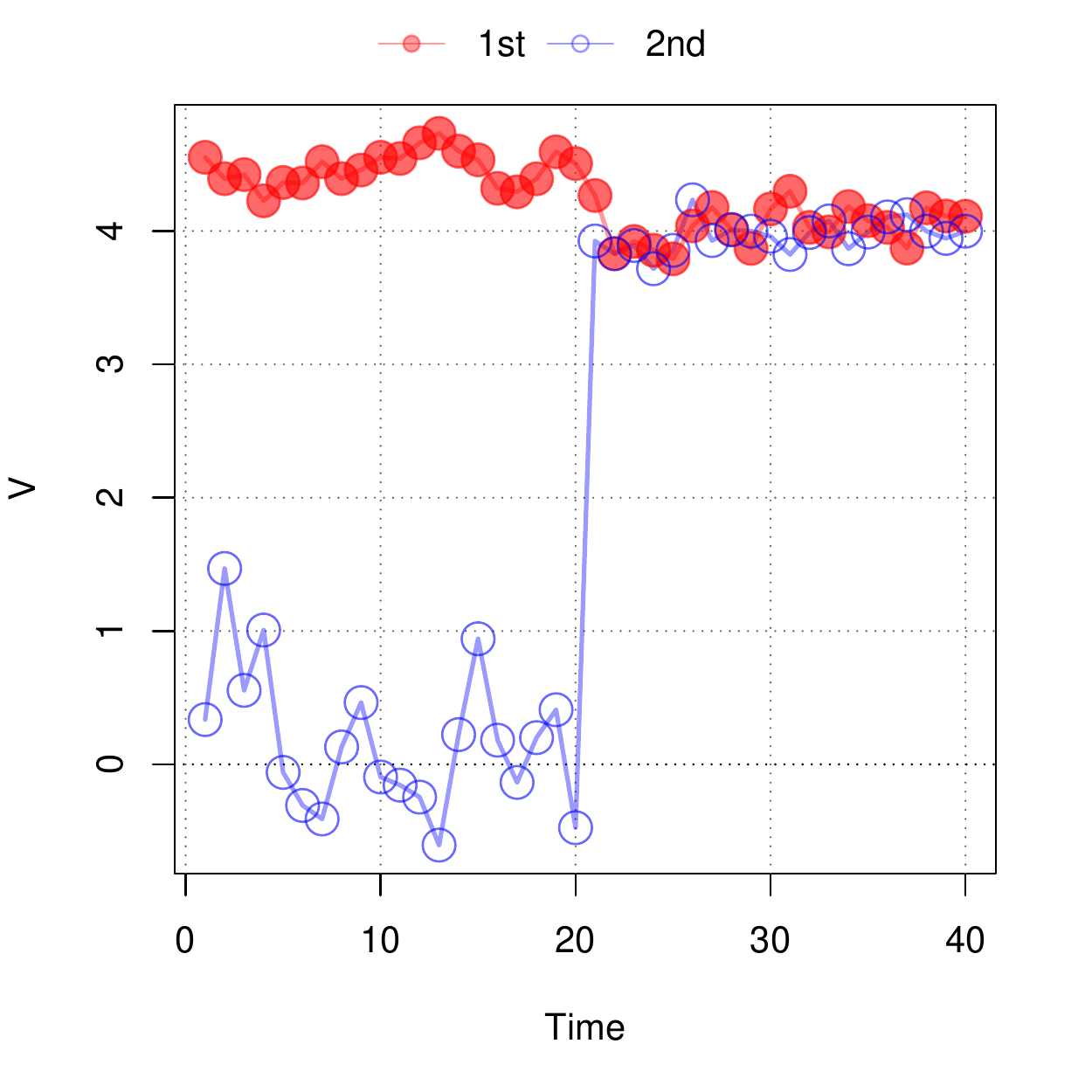}		
\end{figure}

Figure \ref{plotU} shows the output of \texttt{plotU(out1)}, which is the regime-specific latent node positions and Figure  \ref{plotV} shows the output of \texttt{plotV(out1)}, which is time-varying network generation rules. 

\clearpage
\bibliographystyle{apa}
\bibliography{BA}

\end{document}